\documentclass[rmp,aps,twocolumn,nofootinbib,amsmath,floatfix,amssymb]{revtex4}

\usepackage{graphics}
\usepackage{epsfig}
\usepackage{ams}

\begin{document}
\newcommand*{\ket}[1]{$|{#1}\rangle$}
\newcommand*{\beq}{\begin{equation}}
\newcommand*{\eeq}{\end{equation}}
\newcommand*{\beqs}{\begin{equation*}}
\newcommand*{\eeqs}{\end{equation*}}

\title{Atom Interferometers}
\author{Alexander D. Cronin}
 \email{cronin@physics.arizona.edu}
 \affiliation{Department of Physics, University of Arizona, Tucson, AZ 85721}
\author{J\"org Schmiedmayer}
 \email{schmiedmayer@atomchip.org}
 \affiliation{Atominstitut \"Osterreichischer Universit\"aten, TU-Wien, Austria}
\author{David E. Pritchard}
 \email{dpritch@mit.edu}
 \affiliation{Department of Physics, Massachusetts Institute of Technology, Cambridge MA, 02139}

\begin{abstract}
Interference with atomic and molecular matter waves is a rich branch
of atomic physics and quantum optics. It started with atom
diffraction from crystal surfaces and the separated oscillatory
fields technique used in atomic clocks.  Atom interferometry is now
reaching maturity as a powerful art with many applications in modern
science. In this review we first describe the basic tools for
coherent atom optics including diffraction by nanostructures and
laser light, three-grating interferometers, and double wells on
AtomChips. Then we review scientific advances in a broad range of
fields that have resulted from the application of atom
interferometers. These are grouped in three categories: (1)
fundamental quantum science, (2) precision metrology and (3) atomic
and molecular physics. Although some experiments with Bose Einstein
condensates are included, the focus of the review is on linear
matter wave optics, i.e. phenomena where each single atom interferes
with itself.

\end{abstract}
\date{\today}
\maketitle
\tableofcontents

\section{INTRODUCTION}  \label{sec:intro}

Atom interferometry is the art of coherently manipulating the
translational motion of atoms (and molecules) together with the
scientific advances that result from applying this art. We begin
by stressing that \mbox{\emph{motion}} here refers to center of
mass displacements and that \mbox{\emph{coherently}} means with
respect for (and often based on) the phase of the de Broglie wave
that represents this motion. The most pervasive consequence of
this coherence is interference, and the most scientifically
fruitful application of this interference is in interferometers.
In an interferometer atom waves are deliberately offered the
option of traversing an apparatus via two or more alternate paths
and the resulting interference pattern is observed and exploited
for scientific gain.  Atom interferometers are now valuable tools
for studying fundamental quantum mechanical phenomena, probing
atomic and material properties, and measuring inertial
displacements.

In historical perspective, coherent atom optics is an extension of
techniques that were developed for manipulating \emph{internal}
quantum states of atoms. Broadly speaking, at the start of the 20th
century atomic beams were developed to isolate atoms from their
environment; this a requirement for maintaining quantum coherence of
any sort. \textcite{HAN24} studied coherent superpositions of atomic
internal states that lasted for tens of ns in atomic vapors.  But
with atomic beams, Stern-Gerlach magnets were used to select and
preserve atoms in specific quantum states for several ms. A big step
forward was the ability to change atoms' internal quantum states
using RF \emph{resonance} as demonstrated by \textcite{rab38}.
Subsequently, long-lived coherent superpositions of internal quantum
states were intentionally created and detected by \textcite{ram49}.
The generalization and fruitful application of these techniques has
created or advanced a great many scientific and technical fields
(e.g. precise frequency standards, nuclear magnetic resonance
spectroscopy, and quantum information gates).

Applying these ideas to translational motion required the
development of techniques to localize atoms and transfer atoms
coherently between two localities. In this view, localities in
position and momentum are just another quantum mechanical degree of
freedom analogous to discrete internal quantum states. We discuss
these coherent atom optics techniques in Section II and the
interferometers tha result in Section III. Then we discuss
applications for atom interferometers in Sections IV, V, and VI.

\subsection{Interferometers for translational states}

Atom Optics is so named because coherent manipulation of atomic
motion requires that the atoms be treated as waves.  Consequently,
many techniques to control atom waves borrow seminal ideas from
light optics. To make atom interferometers the following components
of an optical interferometer must be replicated:

\begin{enumerate}
    \item State Selection to localize the initial state (generally in
momentum space) \item  Coherent Splitting, typically using
diffraction to produce at least two localized maxima of the wave
function with a well-defined relative phase \item  Free Propagation
so that interactions can be applied to one ``arm",  i.e. one of the
two localized components of the wave function
\item Coherent Recombination so that phase information gets
converted back into state populations \item Detection of a specific
population, so the relative phase of the wavefunction components can
be determined from interference fringes.
\end{enumerate}

\begin{figure*}
\begin{center}
\includegraphics[width = 14cm]{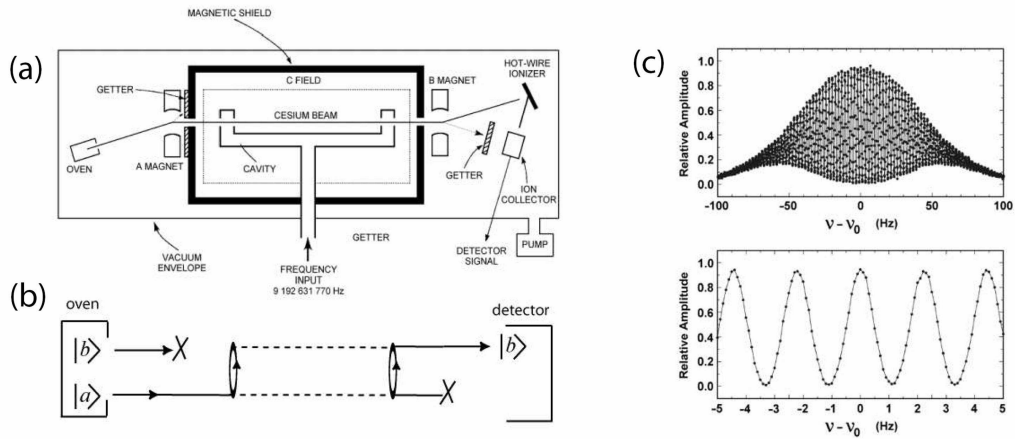}
\caption{(a) Ramsey's separated oscillatory fields experiment
\cite{SBB01}.  (b) The same experiment depicted as an interferometer
for internal states. (c) Interference fringes from the NIST-F1
fountain clock demonstrate the precision obtained with interference
techniques. Fringes result from two separated oscillatory
(microwave) fields exciting atoms in a fountain. On the y-axis is
reported the fraction of atoms in the excited state.  Figures (a)
and (c) are from \cite{SBB01}.} \label{fig:SOF}
\end{center}
\end{figure*}

In hindsight, it is possible to reinterpret much of the work on
internal state resonance as an interferometer.  In particular, the
\emph{separated oscillatory fields} technique \textcite{ram49}
divided a single RF resonance region into two zones that may be
regarded as beam splitters.  In this experiment a Stern-Gerlach
filter (the so-called \texttt{A} magnet in Fig.~\ref{fig:SOF})
selects atoms in state \ket{a}.  The first resonance region
(microwave cavity) then excites atoms into a superposition of states
\ket{a} and \ket{b}.  Atoms then travel through a static
(\texttt{C}) field in a coherent superposition whose relative phase
oscillates freely until the atoms enter the second microwave cavity.
If radiation there is in phase with the oscillating superposition,
then atoms complete the transition to state \ket{b}. But if the
radiation is half a cycle out of phase then atoms are returned to
state \ket{a}. After the final state selector, the detected
intensity oscillates as a function of microwave frequency. Overall,
this method to manipulate the internal states of an atom obviously
maps directly onto the steps listed above and can be regarded as the
first atom interferometer even though it is more frequently
described in terms of resonance of a Bloch vector of an atom moving
classically.

\subsection{Preparation. Manipulation. Detection.}

\emph{Preparation} of position states is hindered by the
uncertainty principle. As soon as free atoms are localized in
position, the attendant momentum uncertainty starts to cause
spatial delocalization. On the other hand, preparation in momentum
space is free of such back action. Therefore in coherent atom
optics, especially with free atoms, it is desirable to reduce the
momentum and its uncertainty for an ensemble of atoms. This is
colloquially referred to as slowing and cooling the atoms,
respectively.

Momentum-state selection can be as simple as two collimating slits
that select atoms with limited transverse momentum. Alternatively,
and preferably, atoms can be concentrated in phase space by
\emph{laser cooling and trapping}\footnote{Original references for
cooling and trapping include: supersonic beams \cite{BEV81,CAM84},
optical molasses \cite{CHB85,ADH86}, optical traps
\cite{ASH70,CBA86,CBA86p,MCH93}, magneto optical traps \cite{RPC87},
magnetic traps \cite{MPP85, PRI83}, atomic fountains \cite{KRC98},
velocity selective coherent population trapping \cite{AAK88},
sideband cooling \cite{WDW78,NHT78,VCK98}, cooling to the ground
state of a trap \cite{JGL92,MMK95} and Bose Einstein Condensation
\cite{AEM95}.}. This is analogous to \emph{optical pumping} for
internal states. In fact, cooling atoms (or ions) in a trap is even
more exactly analogous to optical pumping because trapped atoms are
in discrete translational states and can ultimately be prepared in
the single ground state.

The typical momentum uncertainty achieved with various methods is
summarized in Table \ref{tab:momentum}.  We note that atom
interferometers already work with atoms prepared in beams,
magneto-optical traps, or Bose Einstein condensates.

\emph{Manipulation}.  In most atom interferometers diffraction or
the closely related Raman transitions ``split" atoms into a coherent
superposition of momentum states that typically differ in momentum
by several photon momenta (velocity differences of several cm/sec;
e.g. the recoil velocity for Na atoms due to absorbing a 590 nm
photon is $v_{rec} = \hbar k / m_{Na} =$ 2.9 cm/sec and the velocity
difference between 0th and 1st diffraction orders for Na atoms
transmitted through 100 nm period gratings is $h/(m_{Na}d)$ = 17
cm/sec). As time passes, each atom evolves into a coherent
superposition of spatial positions located a distance $\Delta x =
(p_2 - p_1) t/m$ apart. Moreover, if the initial preparation was
restrictive enough, then the components of each atoms' wavefunction
will be distinctly separated in space. Creating such `separated
beams' in an interferometer invites the experimenter to deliberately
apply different interactions - and hence different phase shifts - to
each component of an atom's wave function.

Observing this phase difference requires recombining the two
components of the superposition. This is generally achieved by using
diffraction or Raman processes again to reverse the momenta of the
two states so they subsequently overlap. When this is done,
interference fringes are observed and the phase $\phi_{int}$ can be
determined from their position.

\begin{table}
\caption{Momentum uncertainty and temperature of atoms prepared
with different techniques. Typical `best case' values for sodium
atoms are tabulated. The momentum uncertainty, \mbox{$\sigma_p = (
\langle p^2 \rangle - \langle p \rangle ^2 )^{1/2}$} is given in
units of 590 nm photon momenta $\hbar k_{ph}$. Temperature is
given by \mbox{$k_BT = \sigma_p^2 / 2 m$} where $k_B$ is the
Boltzmann constant, and $m$ is atomic mass. } \label{tab:momentum}
\begin{ruledtabular}
\begin{tabular}{lcc}
Atomic Sample & $\sigma_p / \hbar k_{ph}$ & T/K \\
\hline
Thermal vapor  &   24,000  &  500  \\
Effusive beam (longitudinal) & 8,000 & 50 \\
Supersonic beam (longitudinal) & 3,000 &  8  \\
Optical molasses or MOT    &  20   & $0.00025$ \\
Collimated beam (transverse) &  1  & $10^{-6}$ \\
Bose Einstein condensate &  0.1  &  $10^{-8}$ \\
\end{tabular}
\end{ruledtabular}
\end{table}

\emph{Detection}.  Once information is transferred from the phase of
a superposition into the population of observable states by using
some kind of beam recombiner, then a state-selective detector is
used to measure the output of an interferometer. In analogy with an
optical Mach-Zehnder interferometer, the fringes can be observed as
atom beam intensity that oscillates between two `output' momentum
states as a function of the interaction-induced phase difference
$\phi_{int}$. Alternatively, fringes can be observed directly in
position space either by moir\'{e}-filtering with a suitable mask or
by directly imaging the atoms. Bragg reflection of laser light can
also be used to detect fringes in atomic density. If the
interferometer manipulates both the internal and (separated)
external states of atoms, then fringes can be detected as
oscillations in population of the internal states after recombining
the atoms, as in Ramsey's experiment.

Historically, alkali atoms were the first to be detected
efficiently, and this was achieved by counting the ions produced as
the atoms ionized on a hot tungsten or rhenium wire\footnote{Various
atom detectors are discussed in \cite{ram85,SCO88,camp00}. For hot
wire detectors see \cite{DML02,LAK25}, for universal detectors see
\cite{DDK00,KKD94}.}. Metastable atoms can be detected directly with
multi-channel plates because of their stored internal energy. More
universal neutral atom detectors use electron bombardment or laser
excitation to produce countable ions. Fluorescence or absorption can
also reveal fringes, especially if a cycling transition is used with
slow atoms.

\subsection{Scientific promise of atom interferometers}

The light interferometers that were developed late in the 19th
century by \textcite{fiz1853} \textcite{mic1881},
\textcite{ray1881}, and \textcite{Fab1899} performed many beautiful
experiments and precise measurements that have had a broad impact in
physics. Recently, the initial idea from de Broglie and
Schr\"odinger that propagating particles are waves has been combined
with technologies to produce interferometers for electrons
\cite{MSA53,MSA54}, neutrons \cite{RTB74}, and now atoms. Even after
the many advances made possible with earlier interferometers,
wonderful further scientific advances from atom interferometers have
long been anticipated. In fact, the concept of an atom
interferometer was patented by \textcite{alt73} and it has been
extensively discussed since.  Early proposals for atom
interferometers were made by \textcite{CDK85},
\textcite{CLA88,Clauser89}, \textcite{KSS88}, \textcite{MOM88},
\textcite{Pritchard89}, \textcite{BOR89} and \textcite{KACS91}.

Even compared to electron-  and neutron-wave physics, interferometry
with atoms offers advantages on several fronts: a wider selection of
atomic properties,  larger cross sections for scattering light,
better characterized environmental interactions, higher precision,
better portability, and far lower cost. Atomic properties like mass,
magnetic moment, and polarizability can be selected over ranges of
several orders of magnitude. For example, Cs has 137 times the mass
and 89 times the electric polarizability of H and is therefore
better suited to measuring inertial effects and detecting weak
electric fields. $^{52}$Cr has a magnetic moment of 6 $\mu_{B}$
while $^4$He has none. Alkali atoms have 10$^{-9}$ cm$^2$ scattering
cross sections for resonant light while electrons have a $10^{-25}$
cm$^2$ cross section for the same light (Compton / Thomson
Scattering). Hence, interactions of atoms and their environment can
be enlarged for better measurements or to study decoherence, or they
can be suppressed to measure something else. Furthermore, atoms
interact with surfaces and other atomic gasses with potentials that
are easily studied by interferometry. Atoms can be manipulated by
lasers whose frequency and wavelength are measured with accuracies
of 10$^{-15}$ and 10$^{-11}$ respectively, offering far better
precision for measurements than the crystals or structures used in
other types of interferometer. Finally, atom sources can be as
simple as a heated container with a small hole in the side or a
pulse of laser light that hits a pellet of the desired material.
These sources are far less expensive than nuclear reactors or even
200 keV electron guns.  In fact atom interferometers on atom chips
can potentially fit in a briefcase.

This richness and versatility is combined with the rewards (and
challenges) that stem from the fact that thermal atomic wavelengths
are typically 30,000 times smaller than wavelengths for visible
light. The power of atom interferometry is that we can measure phase
shifts $\phi_{int} = \hbar^{-1} \int U dt$ due to very small
potential energies. A simple calculation shows that 1000 m/s Na
atoms ($E_\textrm{kin} \sim 0.1$ eV) acquire a phase shift of 1 rad
for a potential of only $U=6.6 \times 10^{-12}$ eV in a 10 cm
interaction region. Such an applied potential corresponds to a
refractive index of $|n-1| = 2.7 \times 10^{-11} $. Measuring the
phase shift $\phi_{int}$ to $10^{-3}$ rad corresponds to an energy
resolution $U/E \sim 10^{-14}$, or a spectrometer with a linewidth
of 10 kHz and spectroscopic precision of Hz/$\sqrt{s}$. This with a
thermal atomic beam, cold atoms can increase the sensitivity
1000-fold!

As we document in this review, atom interferometers have already
measured rotations, gravity, atomic polarizability, the fine
structure constant, and atom-surface interactions better than
previous methods.  Yet atom interferometry itself is just over a
decade old. The realization of such interferometers started with
diffraction gratings that are summarized in Section II of this
review. We catalogue atom interferometer types and features in
Section III. We discuss fundamental issues such as decoherence in
Section IV. Precision measurements are described in Section V, and
atomic and molecular physics applications are described in Section
VI.

\section{ATOM DIFFRACTION} \label{sec:diffraction}

Since half-silvered mirrors do not exist for atoms (solid matter
generally absorbs or scatters atoms), beamsplitters for atom
interferometers are often based on diffraction. Diffraction itself
is an interesting interference effect that has already been cleverly
developed for use with atoms. Hence we discuss atom diffraction now,
and atom interferometers next (in Section III).

Diffraction occurs when a wave interacts with anything that locally
shifts its phase or amplitude (e.g. due to absorption), and is a
hallmark of wave propagation and interference.  It is generally
treated as resulting from the coherent superposition and
interference of amplitudes for wave propagation via different paths
through the diffracting region that have the same starting and
ending points.

A diffraction \emph{grating} is a periodic diffracting region.
Spatial modulation of the wave by the grating generates multiple
momentum components for the scattered waves. The fundamental
relationship between the momentum transferred to waves in the
$n^{th}$ component and the grating period, $d$, is \beq \delta p_n =
n\frac {h}{d} = n \hbar G \eeq where $G=2\pi/d$ is the reciprocal
lattice vector of the grating, and $h$ is Planck's constant. When
the incoming wave has a narrow transverse momentum distribution
centered around $p_{beam}$, this diffraction is generally observed
with respect to angle. Since the de Broglie wavelength is
$\lambda_{dB} = h/p_{beam}$, the resulting diffraction angles (for
nearly normal incidence) are

\beq \theta_n \approx \frac { \delta p_n}{p_{beam}} =
n\frac{\lambda_{dB}}{d} \eeq

To observe the interference a grating must be illuminated at least
in part coherently, i.e. the incident atom waves must have a
well-defined relative phase across several grating periods.  That
means the transverse coherence length must be larger than a few
grating periods, i.e. the transverse momentum distribution must be
small enough to resolve the diffraction orders.   This is usually
accomplished by collimating the incident beam\footnote{The
transverse coherence length is $\ell_{tcoh} \approx
\lambda_\textrm{dB} / \vartheta_\textrm{coll}$, where
$\lambda_\textrm{dB}$ is the de Broglie wavelength and
$\vartheta_\textrm{coll}$ is the (local) collimation angle of the
beam (the angle subtended by a collimating slit). Since for thermal
atomic beams $\lambda_\textrm{dB} \sim 10$ pm a collimation of
$\vartheta_\textrm{coll}<10\mu rad$ is required for a 1 $\mu m$
coherent illumination.}.

\subsection{Early diffraction experiments}

\begin{figure}
\begin{center}
\includegraphics[width = 4cm]{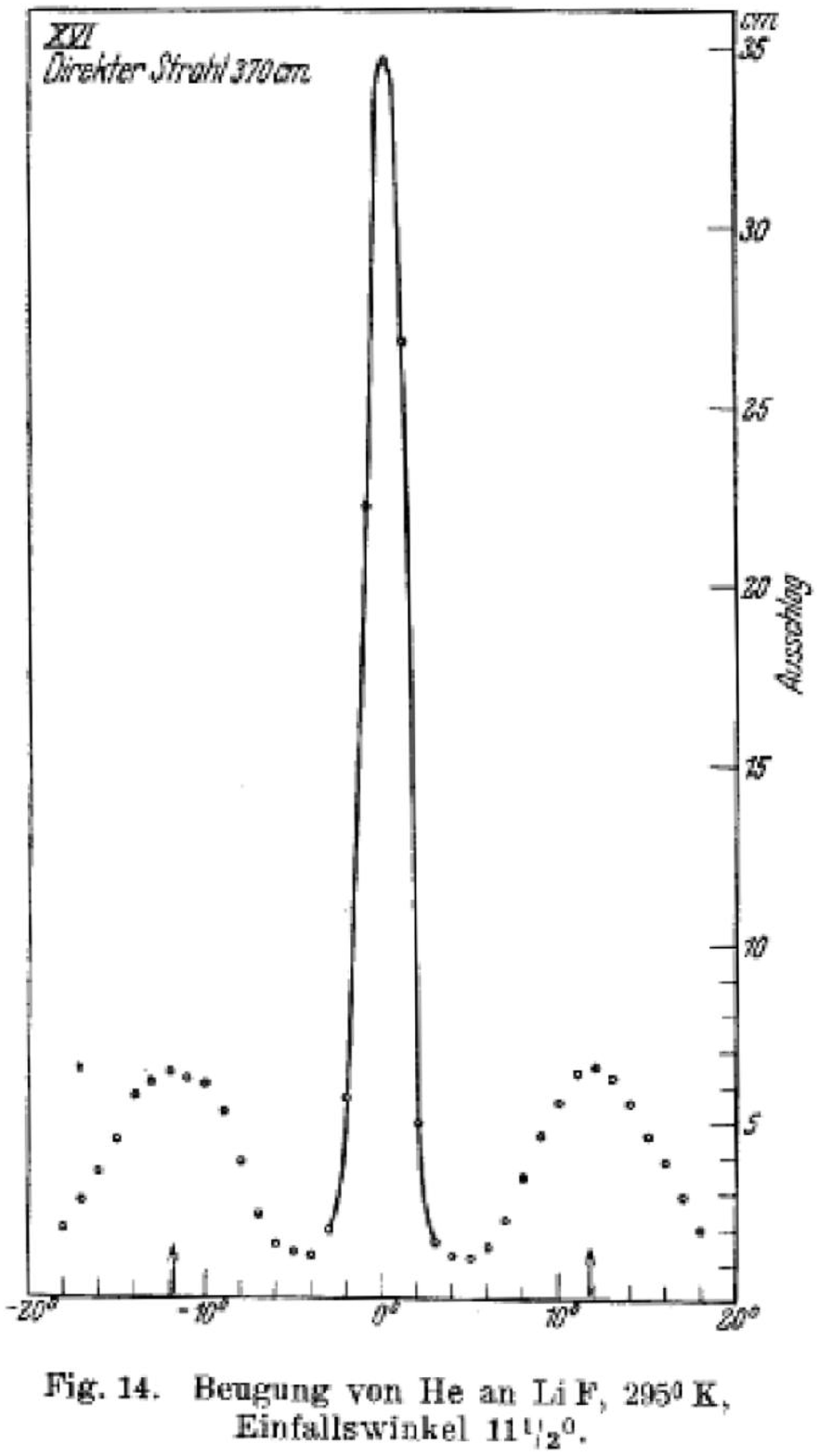}
\caption{Historic data showing diffraction of He atoms from a LiF
crystal surface \cite{ESS30}.  The central peak is due to He atom
reflection.  The side peaks are due to first order diffraction of He
atoms from the LiF crystal lattice.} \label{fig:stern}
\end{center}
\end{figure}

The first examples of atom interference were diffraction
experiments, and the earliest of these was by \textcite{ESS30} just
three years after the electron diffraction experiment by
\textcite{DAG27}.  Figure \ref{fig:stern} shows original data in
which helium atoms were reflected and diffracted from the surface of
a LiF crystal. The small lattice period of the crystal surface (40
nm) gave large diffraction angles and allowed relaxed collimation.
This observation proved that composite particles (atoms) propagate
as waves, but this kind of reflection-type diffraction grating has
not led to a beam splitter suitable for atom interferometry. It did,
however, launch an active field of atom diffraction (both elastic
and inelastic) for studying surfaces.

\subsection{Nanostructures}

One of the first demonstrations of atom diffraction from macroscopic
objects was made by \textcite{LEB69} who observed Fresnel
diffraction from a single 20 $\mu$m wide slit.  With the advent of
modern nano technology it became possible to fabricate elaborate
arrays of holes and slots in a thin membrane that allow atoms to
pass through.  These can have feature sizes of 50 nm or below --
much smaller then typical transverse coherence in well-collimated
atomic beams.  Diffraction from a nanofabricated structure -- a
transmission grating with 200 nm wide slits -- was first observed by
the Pritchard group at MIT \cite{KSS88}.  This led to many beautiful
interference experiments with atoms and molecules.

Nanotechnology has been used to make single slits, double slits,
diffraction gratings, zone plates, hologram masks, mirrors, and
phase shifting elements for atom waves. The benefits of using
mechanical structures for atom optics include: feature sizes smaller
than light wavelengths, arbitrary patterns, rugged designs, and the
ability to diffract any atom and or molecule. The primary
disadvantage is that atoms stick to (or bounce back from) surfaces,
so that most structures serve as absorptive atom optics with a
corresponding loss of transmitted intensity.

\subsubsection{Transmission gratings}

\begin{figure}[t]
\begin{center}
\includegraphics[width = 8cm]{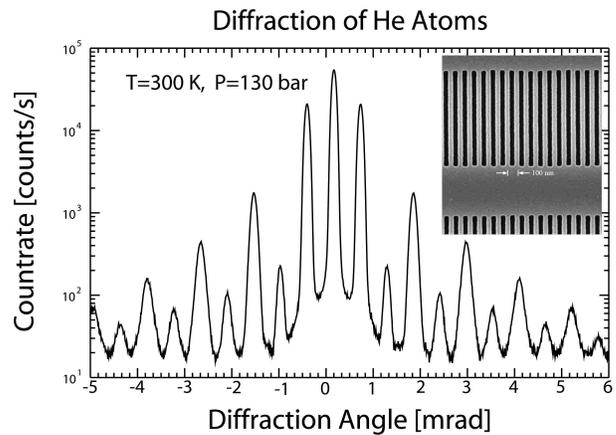}
\caption{Diffraction of He atoms transmitted through a nanostructure
grating. The average velocity and velocity spread of the beam, the
uniformity of the material grating, and the strength of atom-surface
van der Waals forces can all be determined from these data
\cite{GST99}. Figure courtesy of J.P. Toennies, W. Schoellkopf and
O. Kornilov. (Inset)  A 100 nm period grating for atom waves. The
dark regions are slots, and light regions are free-standing silicon
nitride bars. Figure courtesy of T.A. Savas and H.I. Smith at the
MIT NanoStructure laboratory \cite{sava96,SAS90,sss95}. }
\label{fig:He diffn}
\end{center}
\end{figure}

After the demonstration of transmission gratings for atom waves by
\textcite{KSS88}, these gratings have seen numerous applications.
A 100-nm period nanostructure grating made at the MIT
NanoStructures facility and atom diffraction data from this kind
of grating is shown in Figure \ref{fig:He diffn}. Material
structures absorb atoms that hit the grating bars but transmit
atom waves through the slots relatively unperturbed.

Classical wave optics recognizes two limiting cases, near- and
far-field, treated by the Fresnel and Fraunhoffer approximations
respectively.   Both regimes have revealed interesting effects and
led to scientific advance. In the near-field limit the curvature
of the wave fronts must be considered and the intensity pattern of
the beam is characterized by Fresnel diffraction.  Edge
diffraction and the Talbot self-imaging of periodic structures are
examples of near-field atom optics.  In the far-field limit, the
intensity pattern of the beam is characterized by Fraunhofer
diffraction in which the curvature of the atom wave fronts is
negligible and the diffraction orders can be resolved. For a
grating with open fraction $w/d$ and a purely real and binary
valued transmission function the probability for a beam to be
diffracted into the $n$th order is \beq P_n =
\frac{I_n}{I_{inc}}=\left( \frac{w}{d} \right)^2 \left(
\frac{\sin(nw\pi/d)}{nw\pi/d} \right)^2 \label{eq:sinc
diffn_efficiencies} \eeq

Modification of the diffraction patterns due to van der Waals
interaction with the grating bars was first observed by
\textcite{GST99}. This reduces the flux in the zeroth order,
increases flux in most of the higher orders and prevents ``missing
orders" from occurring \cite{CRP04}. Random variations in the
grating bar period can be analyzed as \emph{Debye Waller damping}
which preferentially suppresses higher diffraction orders
\cite{GST00}. Molecular size effects also modify the relative
efficiencies as described by \cite{GST00b} and were used to estimate
the size of the very weakly bound He$_2$ molecule
\cite{SCT94,SHT96,LMK93}.

Molecules such as $^4$He$_2$,  $^4$He$_3$ and other $^4$He clusters,
Na$_2$, C$_{60}$, C$_{70}$, C$_{60}$F$_{48}$, and
C$_{44}$H$_{30}$N$_4$ have been diffracted from similar gratings
\cite{CEH95,ANV99,BAZ03,NAZ03,BHU02,HUH03,SHT96,BGK04}. Scientific
results in this area, such as the study of decoherence and the
formation rate of molecules in beams will be discussed in Section
IV.

\subsubsection{Young's experiment with atoms}

Atomic diffraction from a double slit recapitulates the seminal
Young's double slit experiment in which the diffraction pattern is
created by the interference of waves traversing two cleanly
separated paths.  In that sense it can be seen as a two path
interferometer.  The atomic version by \textcite{car91} used a
mechanical structure with two 1-$\mu$m wide slits separated by 8
$\mu$m to create the interference (Fig.~\ref{fig:Mlykek double
slit}). Diffraction from a single 2-$\mu$m wide slit 62 cm from the
double slit prepared the atom waves ($\lambda_{dB}$= 100 pm) to have
a transverse coherence length larger than the double slit separation
($\ell_{tcoh} = z\lambda_{dB}/2w$ = 15$\mu$m).

\begin{figure}[t]
\begin{center}
\includegraphics[width = \columnwidth]{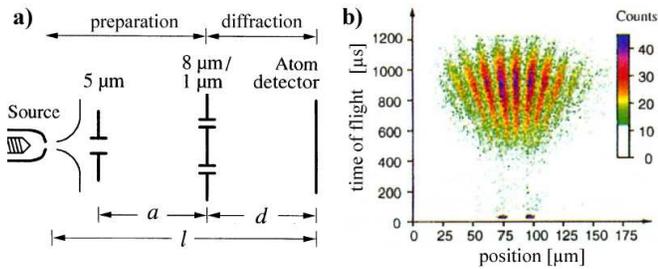}
\caption{(color online) Double-slit experiment with He*. (a)
Schematic. (b) Atom interference pattern with $a=1.05$m and
$d=1.95$m recorded with a pulsed source. \cite{KPM97}.}
\label{fig:Mlykek double slit}
\end{center}
\end{figure}

In the original experiment, a slit was translated in front of the
detector to observe the interference fringes. With a beam
brightness of $B\approx10^{17}$ s$^{-1}$cm$^{-2}$ sr$^{-1}$, the
average count rate was about one atom per second. In a later
version (see Fig Fig.~\ref{fig:Mlykek double slit}) they used a
position sensitive detector to record the whole pattern at once,
giving a larger counting rate.   Time of flight resolution was
added in order to measure the Wigner function of the transmitted
atoms \cite{KPM97}.

A two slit experiment using cold Ne$^*$ atoms was presented by
\textcite{SST92}.  The atoms were dropped from a magneto-optical
trap one meter above a mechanical mask with two slits separated by 6
$\mu$m. At the location of the mask the atoms had a speed of 4.5 m/s
($\lambda_{dB}$=5 nm) and a speed ratio of $v/\sigma_v = 20$. The
mask was equipped with an electrode so that deflection due to an
applied electric field gradient could be measured.

\subsubsection{Charged-wire interferometer}

\begin{figure}[h!]
\begin{center}
\includegraphics[width = 8.5 cm]{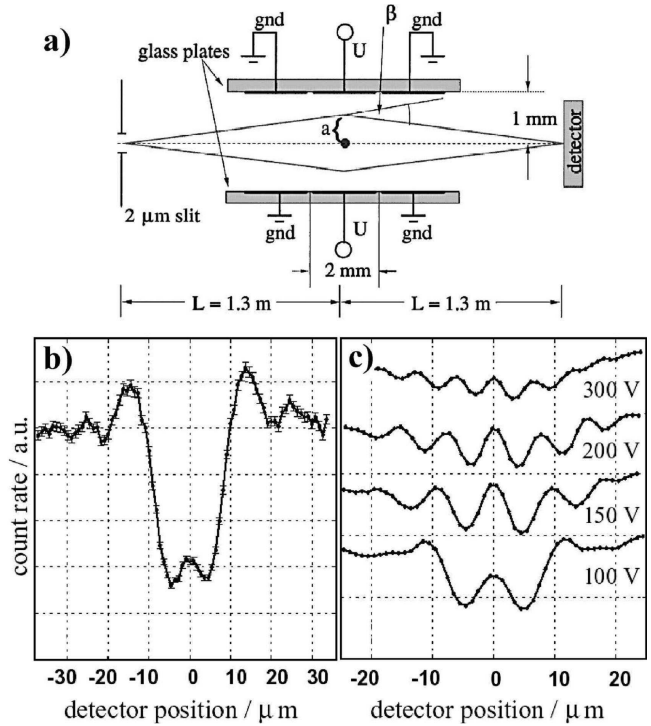}  
\caption{Charged wire interferometer.  (a) Schematic. (b) Measured
diffraction patterns with an uncharged wire. Fresnel fringes and the
Poisson spot are visible.  (c) Interference fringes with different
voltages applied to the electrodes. Figure from \cite{NSP98}.}
\label{fig:nowak1}
\end{center}
\end{figure}

A variation of the atomic Young's experiment was built by
\textcite{NSP98}. A single wire put in a He* beam produces a
near-field diffraction pattern. Charging the wire bends the atom
trajectories passing around it inward, increasing the interference
(Fig.~\ref{fig:nowak1}) analogous to the charged-wire optical
bi-prism interferometer for electrons \cite{MOD55}.

\subsubsection{Zone plates}

\begin{figure}[t]
\begin{center}
\includegraphics[width = 8cm]{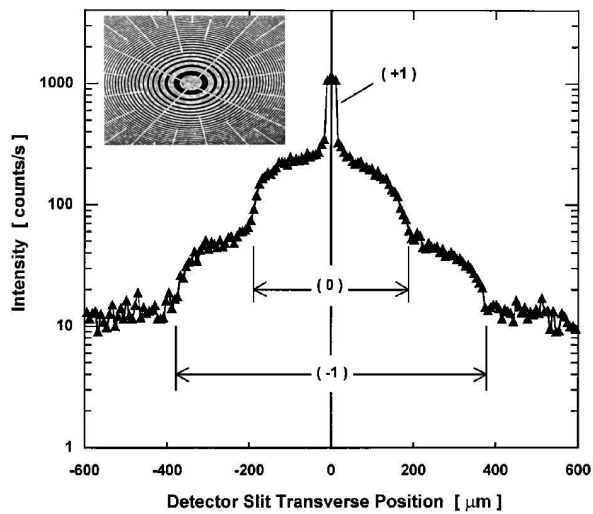}
\caption{A zone plate for focusing atom beams.  The plate (inset)
has free-standing annular rings and radial support struts. The data
shows focused (+1) and defocused (-1) atom beam components. Figure
from \cite{DGR99}.} \label{fig:zone plate}
\end{center}
\end{figure}

Fresnel zone plates have focused atoms to spots smaller than 2
microns (Figure \ref{fig:zone plate}). Zone plates behave locally
like a diffraction grating, therefore the focal length of a zone
plate is $ f = R d_{min} / \lambda_{dB}$ where $R$ is the radius of
outermost zone, and $d_{min}$ is the period of the smallest
features. Focal lengths of $f=450$ mm with $R=0.2$ mm
($\lambda_{dB}=$ 200 pm) \cite{CSS91} and $f=150$ mm with $R=0.3$ mm
($\lambda_{dB}=$180 pm) \cite{DGR99} have been demonstrated.

\subsubsection{Atom holography}

\begin{figure}[b!]
\begin{center}
\includegraphics[width = \columnwidth]{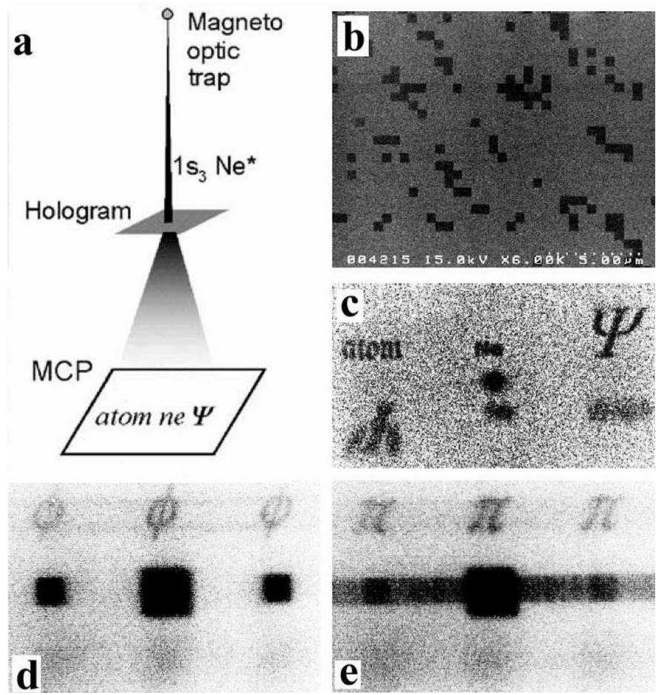}  
\caption{Atom holography.  (a) Experimental setup for image
reconstruction of the hologram by an atom beam. (b) A hologram
designed by computer and realized with a SiN membrane with square
holes. (c) Far field diffraction pattern from the hologram mask.
(d,e) Two different diffraction patterns obtained with a switchable
hologram. \cite{MYK96,FMS00b,FMS00}.} \label{fig:hologram}
\end{center}
\end{figure}

Atom holography with nanostructures can make the far-field atom flux
display arbitrary patterns. Adding electrodes to a structure allows
electric and magnetic fields that cause adjustable phase shifts for
the transmitted atom waves. With this technique,
\textcite{FMK96,FKM99,FMS00} demonstrated a two-state atom
holographic structure that produced images of the letters ``$\phi$"
or ``$\pi$" as shown in Fig.~\ref{fig:hologram}. The different
holographic diffraction patterns are generated depending on the
voltages applied to each nano-scale aperture.

\subsection{Gratings of light}

Laser spectroscopy initially dealt with the internal energy levels
of atoms, and coherent phenomena such as non-linear optics.
Exploiting the momentum transfer accompanying absorption or emission
of light was of little experimental concern until the observation of
quantized deflection (=diffraction) in an atom beam by
\textcite{MGA83} and its subsequent application to a BEC by
\textcite{OMD99}\footnote{Of course, theoretical work on quantized
momentum transfer from light to matter dates back to
\textcite{EIN17} and \textcite{KAD33}.}. Many beautiful experiments
with atoms diffracted from standing waves of light have been
accomplished since these earliest milestones (e.g.
\cite{GML95,GML95b,ROB95,TSK00,DCB02,MOM88,KJD02,SIC99}). Now the
interaction between light and atoms is recognized as a rich resource
for atom diffraction (and interference) experiments; and a unified
view of all possible atom diffraction processes using light beams is
presented in \cite{BER97_Borde}.  Light waves can act as refractive,
reflective and absorptive structures for matter waves, just as glass
interacts with light waves.

In an open two-level system the interaction between an atom and
the light field (with detuning $\Delta = \omega_\textrm{laser} -
\omega_\textrm{atom}$) can be described with an effective optical
potential of the form \cite{OAB96}
 \beq
    U(x) = \frac{\hbar \Omega_1^2}{4\Delta + i2\Gamma}
    \propto  \frac{I(x)}{2 \Delta +  i \Gamma}
    \label{eq:Uopt}
 \eeq
where the (on-resonant) Rabi frequency, $\Omega_1 = \mathbf{d_{ab}}
\cdot \mathbf{E_{optical}}/\hbar$, is given by the atomic transition
dipole moment and the optical electric field, $\Gamma$ is the atomic
decay rate and $I(x)$ is the light intensity. The imaginary part of
the potential comes from the spontaneous scattering processes, and
the real part results from the ac Stark shift. For a more detailed
description we point to the vast literature on mechanical effects of
light\footnote{See for example, \cite{ASH70,ASH80,DAC85,MES99}.}. If
the spontaneous decay follows a path to a state which is not
detected, the imaginary part of the potential in Eq. (\ref{eq:Uopt})
is equivalent to absorption. Therefore on-resonant light can be used
to create absorptive structures. Light with large detuning produces
a nearly real potential and therefore acts as pure phase object.
Near-resonant light can have both roles.

The spatial shape of the potential is given by the local light
intensity pattern, $I(x)$, which can be shaped with all the tricks
of near and far field optics for light, including holography.  The
simplest object is a periodic potential created by two beams of
light whose interference forms a standing wave with reciprocal
lattice vector \beq \vec{G} = \vec{k}_1 - \vec{k}_2.\eeq  This is
often called an \emph{optical lattice} because it is a close
realization of the periodic potentials that electrons experience in
solid state crystals. Thus, Bloch states can be used to understand
atom diffraction \cite{LEM81,CBD01}. Additional points of view that
we shall discuss include the thin-hologram (Raman-Nath
approximation) \cite{mey01}, two-photon Rabi oscillations
\cite{GLC01}, and multi beam interference (dynamical diffraction
theory).

\begin{figure}[h]
\begin{center}
\includegraphics[width = \columnwidth]{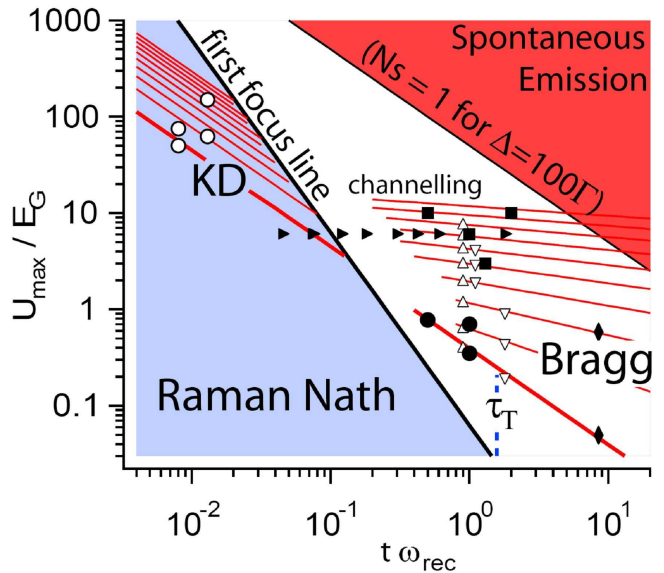}
\caption{(color online) Dimensionless parameter space for atom
diffraction. The vertical axis (optical potential in units of $E_G$)
and horizontal axis (interaction time in units of
$\omega_{rec}^{-1}$) are independent of atomic transition dipole
moment and atomic mass (see Equations \ref{eq:Uopt} and
\ref{eq:EG}). Under the `first focus line' the RNA (Equation
\ref{eq:RNA1}) is satisfied. `KD' labels curves corresponding to
conditions that maximize Kapitza Dirac diffraction into orders 1
through 10 in order from bottom to top (see Equation \ref{eq:Pkd}).
`Bragg' indicates curves that correspond to conditions for complete
($\pi$-pulse) Bragg reflection into orders 1 through 10 (see
Equations \ref{eq:P_Bragg} and \ref{eq:PN_Bragg}). The vertical
dashed line indicates the Talbot time $\tau_T$ (discussed in Section
II.D.). For detuning of $\Delta =100 \Gamma$, the average number of
spontaneously scattered photons per atom is greater than one above
the line marked $N_s=1$. Experiment conditions are shown as points.
$\circ$: Kapitza-Dirac diffraction of an atomic beam \cite{GRP86}.
$\bullet$, $\vartriangle$,$\triangledown$: Bragg diffraction of an
atomic beam \cite{MOM88}\cite{GML95} \cite{KJD02} respectively.
$\blacklozenge$: Bragg diffraction of a BEC \cite{KDH99} (Bragg
spectroscopy of a BEC \cite{SIC99} at $\tau \omega_{rec} = 80$ would
appear to the right of the charted regions, near the first-order
Bragg curve). $\blacktriangleright$ Transition from Kapitza-Dirac
diffraction to oscillation of a BEC in a standing wave light pulse
\cite{OMD99}. $\blacksquare$: Coherent channeling. Figure adapted
from \cite{GLC01} and \cite{KSZ99}. \label{fig:dimensionless}}
\end{center}
\end{figure}

We distinguish different regimes for atom manipulation, for example
(1) thick vs. thin optical lattices, (2) weakly perturbing vs.
strongly channeling lattices, (3) on- vs. off-resonant light, and
(4) static vs. time-dependent optical potentials.  These are
discussed and interrelated in
\cite{GLC01,CBD01,BAK00,KSZ99,OAB99,OAB96,MOO06}. We include a
summary chart (Figure \ref{fig:dimensionless}) that catalogues
different effects caused by gratings of light.

Since light gratings can fill space, they can function as either
\emph{thin} or \emph{thick} optical elements. As in light optics,
for a thin optical element the extent of the grating along the
propagation direction has no influence on the final diffraction
(interference). But in a thick element the full propagation of the
wave throughout the diffracting structure must be considered. In a
grating, the relevant scale is set by the grating period ($d$) and
the atomic de Broglie wavelength ($\lambda_\textrm{dB}$). If a
grating is thicker than $d^2 / \lambda_{dB} $ (half the Talbot
length) it is considered thick, and the characteristics observed are
Bragg scattering or channeling, depending on the height of the
potentials. If the grating is thinner than  $d^2 / \lambda_{dB} $,
it can be analyzed in the Raman-Nath limit, and it produces a
symmetric distribution of intensity into each pair of diffraction
orders of opposite sign ($\pm N)$.  The thin vs. thick transition is
labeled ``first focus line" in Figure \ref{fig:dimensionless}.

The second distinction, mostly relevant for thick gratings, has to
do with the strength of the potential.  One must determine if the
potential is only a perturbation, or if the potential modulations
are larger then the typical transverse energy scale of the atomic
beam or the characteristic energy scale of the grating, \beq
E_\textrm{G} = \hbar^2 G^2/(2 m) = 4 \hbar \omega_{rec},
\label{eq:EG} \eeq associated with one grating momentum unit $\hbar
G$.  ($\hbar \omega_{rec}$ is an atoms `recoil energy' due to
absorbing (or emitting) a photon.) For weak potentials, $U \ll
E_\textrm{G}$, one observes Bragg scattering. The dispersion
relation looks like that of a free particle with avoided crossings
at the edges of the zone boundaries. Strong potentials, with $U \gg
E_\textrm{G}$, cause channeling. The dispersion relations are nearly
flat, and atoms are tightly bound in the wells.

\subsubsection{Thin gratings: Kapitza-Dirac scattering}

\begin{figure}
\begin{center}
\includegraphics[width = 8cm]{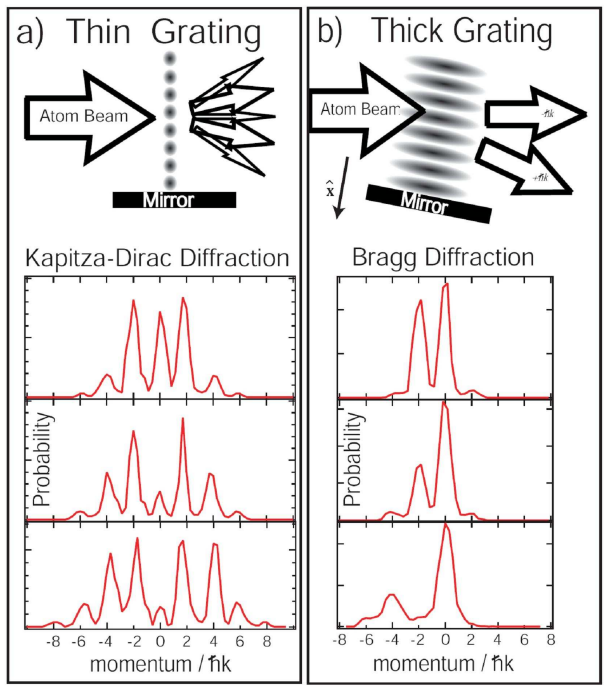}
\caption{(color online) Comparison between diffraction from a thick
and a thin grating. (a) Kapitza Dirac (KD) diffraction, discussed in
Section II.C.1. (b) Bragg Diffraction, discussed in Section II.C.3.
The top row shows the essential difference: thick vs. thin gratings.
The bottom row shows data obtained by the Pritchard group for KD and
Bragg diffraction \cite{GRP86,MOM88}. } \label{fig:standing}
\end{center}
\end{figure}

If atoms are exposed to a standing wave of off-resonant light for a
short time $\tau$, the resulting optical potential due to the
standing wave acts as a thin phase grating with period \mbox{$d =
\lambda_{ph}/2$}. Atom waves are diffracted by this grating so that
many momentum states (each differing by $\hbar G$) are populated as
shown in Figure \ref{fig:standing} (left column).
This is known as \emph{Kapitza-Dirac scattering}\footnote{The
original proposal by \textcite{KAD33} was for Bragg reflection of
\emph{electrons} by a standing wave of light \cite{BAT00,FAB01}.
However, ``Kapitza-Dirac scattering" is now most commonly used in
the literature to describe diffraction of atoms by a thin grating of
light.}, and occurs in the Raman-Nath limit. The probability of
finding atoms in the $N^{\mathrm{th}}$ diffracted state is given by
the Fourier transform of the imprinted phase shift, resulting in the
equation \cite{GLC01}
 \beq
    P_N^{\textrm{K.D.thin}} = J_N^2\left(\frac{\Omega_1^2
    \tau}{2\Delta}\right).
    \label{eq:Pkd}
 \eeq
Here $J_N$ is an $N$th order Bessel function, and $\tau$ is the
duration that the optical intensity is experienced by the atoms. As
defined near Equation \ref{eq:Uopt}, $\Omega_1 = \mathbf{d_{ab}}
\cdot \mathbf{E_{optical}}/\hbar$. Equation \ref{eq:Pkd} is valid
for normal incidence; \textcite{HCA94} considered all angles of
incidence.

The Raman-Nath approximation (RNA) is valid provided the transverse
motion of the atoms remains small. Approximating the potential from
a standing wave as parabolic near a the minimum leads to the
condition for the RNA to be valid:
 \beq
    \tau < \frac{\tau_{osc}}{4} = \frac{1}{2\sqrt{\Omega_R
    E_{G}/\hbar}} \label{eq:RNA1}
 \eeq
where $\Omega_R= \sqrt{|\Omega_1|^2 + \Delta^2}$ is the generalized
Rabi frequency. If the interaction time is longer than this,
Eq.~\ref{eq:Pkd} is no longer valid, and population transfers to
states with the largest momenta (large $N$) are suppressed
\cite{RAN35,MOY78,mey01,KSZ99,WSM91}.

Early attempts to observe the Kapitza-Dirac (KD) effect with
electrons were controversial [see discussion in \cite{BAT00,FAB01}],
and attempts with atoms were unable to eliminate the effects of
spontaneous emission \cite{ALO79}\footnote{In fact, T. Oka made a
wager with DEP that the MIT experiments would continue to show a
maximum at zero deflection, rather than revealing two maxima
displaced from the center as predicted.}. The first observation of
Kapitza-Dirac scattering by \textcite{MGA83}, and \textcite{GRP86}
was therefore a breakthrough: it showed a symmetric double maximum
and also revealed that momentum transfer was quantized in units of 2
$\hbar k_{light}$ thereby indicating a coherent process. Moreover,
these experiments showed that quantized momentum transfer (i.e.
coherent diffraction) is possible even if the interaction time,
$\tau$, is much larger than the atoms' radiative lifetime, provided
that the radiation is detuned from resonance.

With a BEC Kapitza-Dirac scattering was first observed at NIST by
\textcite{OMD99} and has subsequently become an everyday tool for
manipulating BEC's. More recently, a series of light pulses
separated in time [by about one eighth of the Talbot time ($\tau_T=
2d^2 m / h$)] have been used to diffract atoms with high efficiency
into only the $\pm1$ orders \cite{WWD05,WAB04}.

\subsubsection{Diffraction with on-resonant light}

Tuning the light frequency of a standing light wave to resonance
with an atomic transition ($\Delta = 0$) can make an `absorptive'
grating with light.  This is possible when the spontaneous decay of
the excited state proceeds mainly to an internal state which is not
detected. (If the excited state decays back to the ground state,
this process produces decoherence and diffusion in momentum space.)
For a thin standing wave the atomic transmission is given by
 \beq
 T(x) =\exp\left[-\frac{\kappa}{2} [1+\cos (Gx)]\right], \label{eq:t(x)}
 \eeq
where the absorption depth for atoms passing through the antinodes
is  $\kappa$. For sufficiently large absorption only atoms passing
near the intensity nodes survive in their original state and the
atom density evolves into a comb of narrow peaks. Since the
`absorption' involves spontaneous emission such light structures
have been called \emph{measurement induced gratings}. As with all
thin gratings, the diffraction pattern is then given by the scaled
Fourier transform of the transmission function.

\begin{figure}
\begin{center}
\includegraphics[width = \columnwidth]{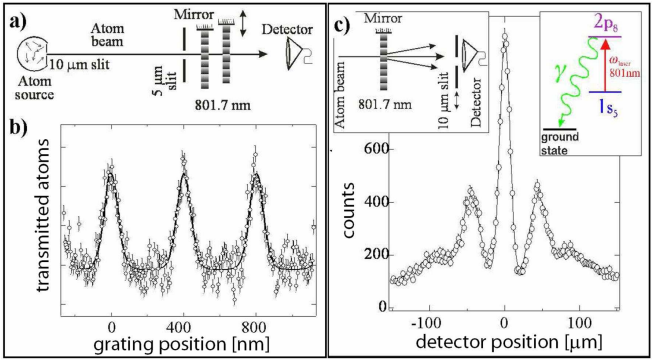}
\end{center}
\caption{(color online) Diffraction from a measurement-induced
grating.  (a) Schematic of two on-resonant standing waves of light.
The first causes atom diffraction.  The second can be translated to
analyze near-field atomic flux. (b) Periodic structure in the
transmitted atomic beam. (c) Far-field atom diffraction from a
measurement induced grating. Figure from \cite{AKB97}.}
\label{fig:OnResDiff}
\end{figure}

Such gratings have been used for a series of near-field (atom
lithography; Talbot effect) and far-field (diffraction;
interferometry) experiments, and an example is shown in Figure
\ref{fig:OnResDiff} \cite{AKB97,JBB96,JTD98,JGS04,ROB95}. These
experiments demonstrate that transmission of atoms through the
nodes of the `absorptive' light masks is a coherent process.



\subsubsection{Thick gratings: Bragg diffraction}

If the standing wave is thick, one must consider the full
propagation of the matter wave inside the periodic potential. The
physics is characterized by multi wave (beam) interference. For two
limiting cases one can regain simple models. For weak potentials,
Bragg scattering; and for strong potentials, coherent channeling.

\begin{figure}[t]
\begin{center}
\includegraphics[width =5.5cm]{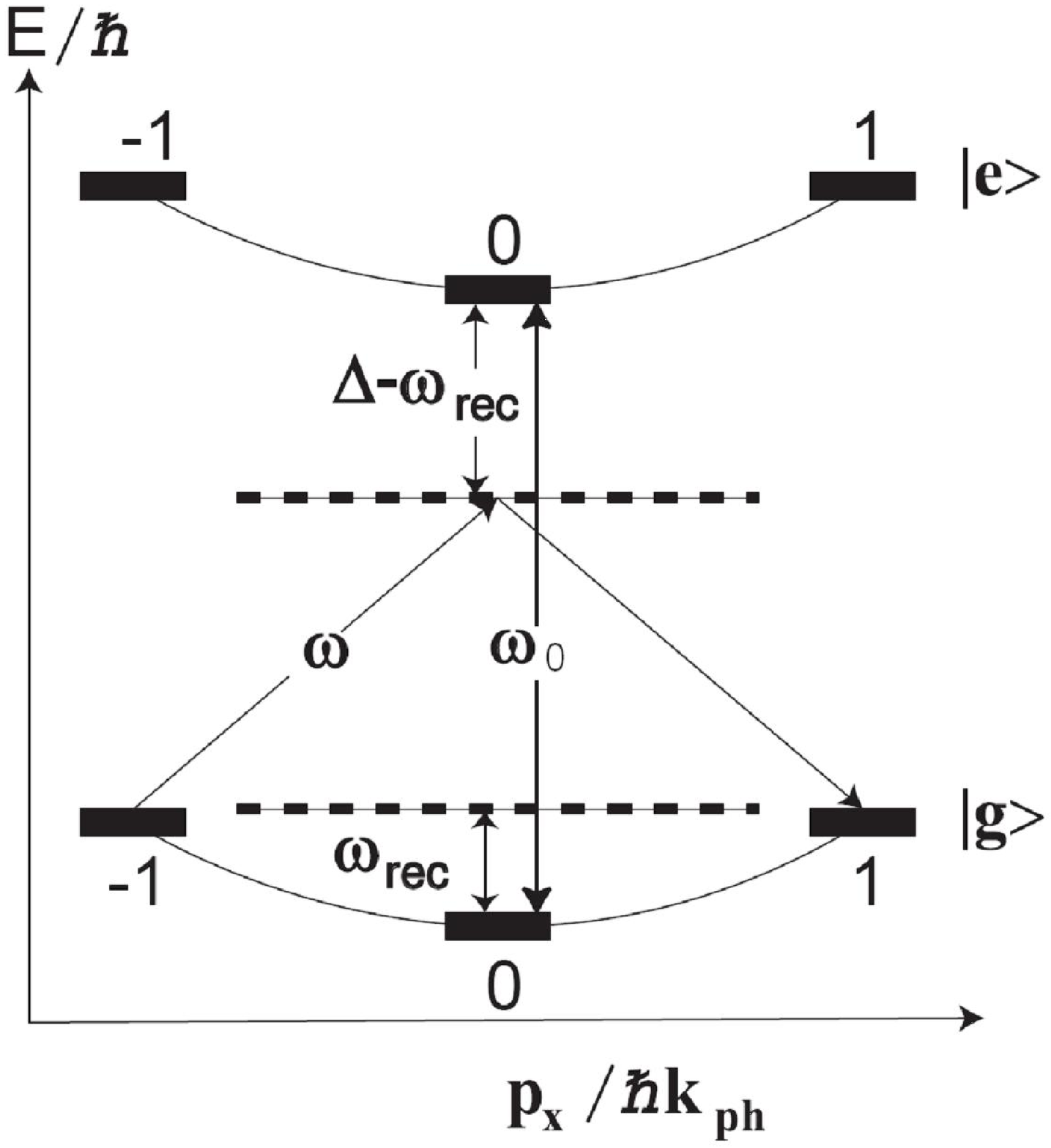}
\end{center}
\caption{Energy states on the energy-momentum dispersion curve
associated with Bragg diffraction. Versions of this classic figure
are found in \cite{MOM88,GML95,KDH99,BER97_Borde,GLC01}.}
\label{fig:braggstates}
\end{figure}

When an atomic matter wave impinges on a thick but weak light
crystal, diffraction occurs only at specific angles, the Bragg
angles $\theta_B$ defined by the Bragg condition
 \beq    N \lambda_{dB} = \lambda_{ph} \sin(\theta_B).  \eeq
Bragg scattering, as shown in Figures  \ref{fig:standing} (right
column) transfers atoms with momentum $-p_x$ into a state with a
single new momentum, $p_x = - p_x + \hbar G$. Momentum states in
this case are defined in the frame of the standing wave in direct
analogy to electron or neutron scattering from perfect crystals.
Bragg scattering was first observed at MIT \cite{MOM88} and first
observed with atoms in a Bose Einstein condensate at NIST
\cite{KDH99}. Higher order Bragg pulses transfer multiples of $N
\hbar G$ of momentum, and this has been demonstrated up to 8th order
with an atomic beam \cite{GML95b,KJD02}. Reviews of Bragg scattering
appear in \cite{GLC01,BAK00,OAB99,DUR99}.

The Bragg scattering process can be understood in terms of
absorption followed by stimulated emission (Figure
\ref{fig:braggstates}). Viewing Bragg scattering as a two-photon
transition from the initial ground state with momentum to a final
ground state (with new momentum) illuminates the close connection
with a Raman transitions \cite{GLC01}.

As a result of the coherently driven 2-photon transition, the
probability amplitude oscillates between the two momentum states
$|g,-\hbar k_{ph}\rangle$ and $|g,+\hbar k_{ph}\rangle$ in a
manner analogous to the Rabi oscillation of atomic population
between two resonantly coupled states.  The probability for Bragg
scattering of atoms from off-resonant standing waves of light is
 \beq
    P_{N=1}^{\textrm{Bragg}}(\tau) = \sin^2\left(\frac{\Omega_1^2 \tau}{4 \Delta}  \right)
    \label{eq:P_Bragg}
 \eeq
The oscillation between the two Bragg-coupled states ($N=0$ and
$N=1$) is known as the \emph{Pendell{\"o}sung} and has been observed
for atoms \cite{MOM88,KJD02,OAB99}, neutrons \cite{SHU68},
electrons, and x-rays. The nice feature with atoms is that the
strength of the grating can be controlled by the intensity of the
light.

The probability for $N$th order Bragg diffraction is
 \beq
    P_{N}^{\textrm{Bragg}}(\tau) =
    \sin^2\left(\frac{\Omega_1^{2N} \tau}{2^{4N-3}[(N-1)!]^2 \Delta^N  \omega_{rec}^{N-1}} \right)
    \label{eq:PN_Bragg}
 \eeq
where we have assumed $\Delta \gg N^2 \omega_{rec}$.

Bragg diffraction of atoms from off-resonant standing waves of light
is often used for studying a BEC's velocity distribution because the
velocity selectivity of the Bragg condition
\cite{SIC99,KDH99,SCG01,CAL00,BLB00}. $\sigma_v$ is improved by
increasing the duration of interaction with the grating, as can be
deduced from the time-energy uncertainty principle, $\sigma_v =
2/(\tau G)$ \footnote{States lie on the energy-momentum dispersion
curve ($E = p^2/2m$) with quantized momentum.  Finite interaction
times ($\tau$) allow states to be populated with a range of energy
$\sigma_{E} = p\sigma_{v} = 2\hbar/\tau $.  For an Nth-order Bragg
process, the state momentum is centered around $p=N\hbar G$.  Hence,
$ \sigma_{v} = 2/(N\tau G) $.}. For first-order Bragg diffraction,
the minimum interaction time required to suppress all but one
diffraction order is $\tau > h/E_{G} \approx 10 \mu$sec; so to
observe Bragg scattering with a 1000 m/s Na atom beam typically
requires standing waves nearly 1-cm thick. However, $\tau$ can be
substantially increased with cold atoms, and $\sigma_v$ less than
1/30 of the recoil velocity has been observed. For higher order
Bragg diffraction the interaction time must be $\tau \gg \pi /
(2(N-1)\omega_{rec})$.

Bragg scattering can be described as a multi beam interference as
treated in the dynamical diffraction theory developed for neutron
scattering. Inside the crystal one has two waves, the refracted
incident `forward' wave ($k_F$) and the diffracted `Bragg' wave
($k_B$). These form a standing atomic wave field, and the
diffraction condition \mbox{($k_{B} - k_{F} = G$)} implies that the
standing atomic wave has the same periodicity as the standing light
wave. At any location inside the lattice, the exact location of
atomic probability density depends on $k_F$, $k_B$ and the phase
difference between these two waves.

For incidence exactly on the Bragg condition the nodal planes of the
two wave fields are parallel to the lattice planes. The eigen-states
of the atomic wave field in the light crystal are the two Bloch
states, one exhibiting maximal ($\Psi_{max}$) the other minimal
($\Psi_{min}$) interaction.

\begin{eqnarray}
 \Psi_{max} &=&
\frac{1}{2} \left[ e^{i\frac{G}{2}x} + e^{-i\frac{G}{2}x}\right]
                = \cos \left( \frac{G}{2}x \right), \nonumber \\
\Psi_{min} &=&  \frac{1}{2} \left[ e^{i\frac{G}{2}x} -
e^{-i\frac{G}{2}x}\right]
                = i \sin \left(\frac{G}{2}x\right) .
 \end{eqnarray}

For $\Psi_{max}$ the antinodes of the atomic wave field coincide
with the planes of maximal light intensity, for $\Psi_{min}$ the
antinodes of atomic wave fields are at the nodes of the standing
light wave. These states are very closely related to the coupled
and non-coupled states in velocity selective coherent population
trapping (VSCPT) \cite{AAK88}.

The total wave function is that superposition of $\Psi_{max}$ and
$\Psi_{min}$ which satisfies the initial boundary condition. The
incoming wave is projected onto the two Bloch states which
propagate through the crystal accumulating a relative phase shift.
At the exit, the final populations in the two beams is determined
by interference and depends on this relative phase (following
equation \ref{eq:P_Bragg}).

\begin{figure}
\begin{center}
\includegraphics[width = \columnwidth]{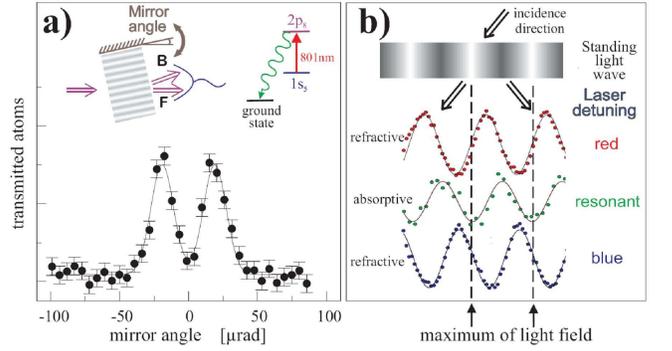}
\end{center}
\caption{(color online) Bragg diffraction of atoms from resonant
standing waves of light.  (a) Atoms entering the light crystal at
the Bragg angle are less likely to emit a spontaneous photon and
therefore survive the on resonant light field (Anomalous
transmission). (b) A resonant standing wave inside a light crystal
serves to measure the atom wave fields inside the crystal. For on-
resonance light crystals one observes the minimal coupled Bloch
state. Figure from \cite{OAB96}.} \label{fig:AnormTrans}
\end{figure}

Bragg scattering can also be observed with absorptive, on-resonant
light structures \cite{OAB96} and combinations of both on and
off-resonant light fields \cite{KOA97}. One remarkable phenomenon is
that the total number of atoms transmitted through a weak
on-resonant standing light wave increases if the incident angle
fulfills the Bragg condition, as shown in Fig.~\ref{fig:AnormTrans}.
This observation is similar to what \textcite{BOR41} discovered for
x rays and called anomalous transmission.

The observed anomalous transmission effect can easily be understood
in the framework of the two beam approximation outlined above.  The
rate of de-population of the atomic state is proportional to the
light intensity seen by the atoms and therefore to the overlap
between the atom wave field with the standing light field.  The
minimally coupled state $\Psi_{min}$ will propagate much further
into the crystal than $\Psi_{max}$. At the exit phase the
propagating wave field will be nearly pure $\Psi_{min}$.  As a
consequence one sees two output beams of (nearly) equal intensity.

Inserting an absorptive mask \cite{AKB97} inside the light crystal
allows one to observe the standing matter wave pattern inside the
crystal \cite{OAB96,OAB99} and verify the relative positions between
the light field and $\Psi_{min}$.

Indeed, tailored complex potentials for atoms can be made out of
combinations of bi-chromatic standing waves \cite{KOA97}. For
example, a superposition of standing waves (one on- and one
off-resonance) with a phase shift $\Delta \varphi = \pm \pi/2$
results in a combined potential of $ U(x)= U_0 e^{\pm i G x}$
which, in contrast to a standing wave, has only \emph{one}
momentum component. Such a potential can therefore only diffract
in one direction. As this predicts, various diffraction orders can
be suppressed by adjusting the phase difference between the
absorptive and the refractive grating. The lack of symmetry is
referred to as a violation of Friedel's law.  The asymmetry in the
observed patterns can also be understood as an interference effect
between diffraction at refractive and absorptive ``subcrystals"
spatially displaced with respect to each other \cite{KOA97}.

\subsubsection{Bloch Oscillations}

Bloch oscillations were predicted by \textcite{Bloch_1929} and
\textcite{Zener_1934} as an interference phenomenon in connection
with the electronic transport in crystal lattices, but can in
general also be observed in any system where accelerated matter
waves move through a periodic potential.  In a simple physical
picture the Bloch oscillations can be viewed as repeated Bragg
reflection from an accelerating grating. To observe high contrast
Bloch oscillations it is desirable to prepare the initial sample
well localized in momentum space, with a width of the momentum
distribution much smaller than the Brilloiuin-Zone.  Therefore a BEC
would be an ideal starting condition.

\begin{figure}[t]
  \includegraphics[angle=0,width=\columnwidth]{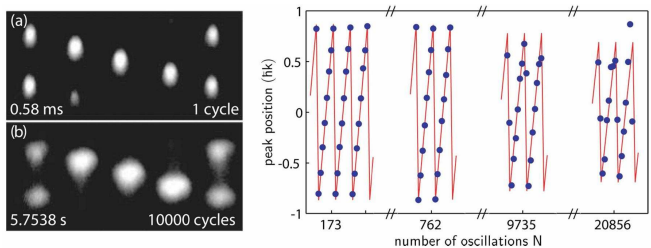}
  \caption{\label{Fig:BlochOsc} Observation of long lasting Bloch
  oscillations in Cs where the interaction was switched off by tuning
  the scattering length close to zero by applying a magnetic field to \mbox{17.12
  G}  (adapted from \textcite{Gustavsson_2007}.}
\end{figure}

The first to observe Bloch oscillations with atomic matter waves was
\textcite{Dahan_1996}, who studied the motion of thermal atoms in an
accelerated lattice.  Since then, because optical lattices can be
precisely controlled, Bloch oscillations have been used for
precision measurements of quantities related to acceleration such as
$g$ or $\hbar / m_{atom}$.

In a real experiments atom-atom interactions damp the Bloch
oscillations (by de-phasing). \textcite{Roati_2004} showed that
Bloch oscillations survive much longer for non-interacting Fermions
($^{40}$K) when compared with Bosons ($^{87}$Rb), and very long
lasting Bloch oscillations were observed both for weakly interacting
Bosons ($^{88}$Sr) by \textcite{FPS06} and especially where the
interaction was switched of by tuning with a Feshbach resonance in
$^{133}$Cs by \textcite{Gustavsson_2007} or $^{39}$K by
\textcite{Fattori_2007}.

\subsubsection{Coherent channeling}

\begin{figure} [t]
\begin{center}
\includegraphics[width = \columnwidth]{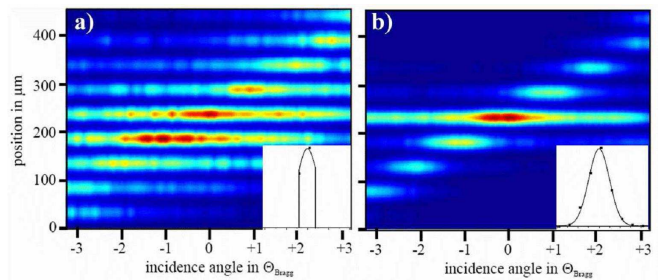}
\caption{(color online) Coherent channeling of atoms through a
strong light crystal.  (a) When the light crystal turns on abruptly
(see inset) many transverse momentum-states are populated, and a
large number of outgoing diffraction orders are observed. (b) atoms
entering the light crystal slowly (adiabatically) only occupy the
lowest energy states, hence only one or two output beams are
observed, as in Bragg scattering. Figure from \cite{KSZ99}.}
\label{fig:channeling}
\end{center}
\end{figure}

When the lattice potential becomes higher then $E_\textrm{G}$
(Eq.~\ref{eq:EG}) the atoms can be localized in the standing light
wave. Atoms impinging on such a strong light crystal are then guided
in the troughs through the crystal, and can interfere afterwards.
Such guiding is called \emph{channeling}. Channeling of electron
beams \cite{JND82} and ion beams \cite{FMP85} in material crystals
is related to channeling of atoms in optical lattices
\cite{KSZ99,HJZ99,SDA87}. If the process is coherent one can observe
a diffraction pattern reminiscent of the KD diffraction from a thin
grating.  See Figure \ref{fig:channeling}.

\subsection{The Talbot effect}

We now turn from far-field atom diffraction to the near-field
region, where a host of different interference effects occur. The
well-known optical self imaging of a grating discovered by Talbot
in 1832 is most important. It has many applications in image
processing and synthesis, photo-lithography, optical testing and
optical metrology \cite{PAT89}, and has proven to be a powerful
tool for interference experiments with matter waves.

Plane waves incident on a periodic structure form a ``self-image" of
the structure at the Talbot distance $L_{T} = 2 d^2 / \lambda_{dB}$
and again at integer multiples of the Talbot length. At half the
Talbot distance a similar self-image is formed but displaced by half
a period.  At certain intermediate distances higher-order Talbot
images are formed.  These have a spatial frequency  that is higher
than the original grating by a ratio of small integers. The position
and contrast of the sub-period images are determined by Fresnel
diffraction as discussed in \cite{PAT89,CLR92,CLL94}. The replica
(Fresnel) images and higher-order (Fourier) images are used in a
Talbot-Lau interferometer \cite{BAZ03}.

\begin{figure}[t]
\includegraphics[width = 8cm]{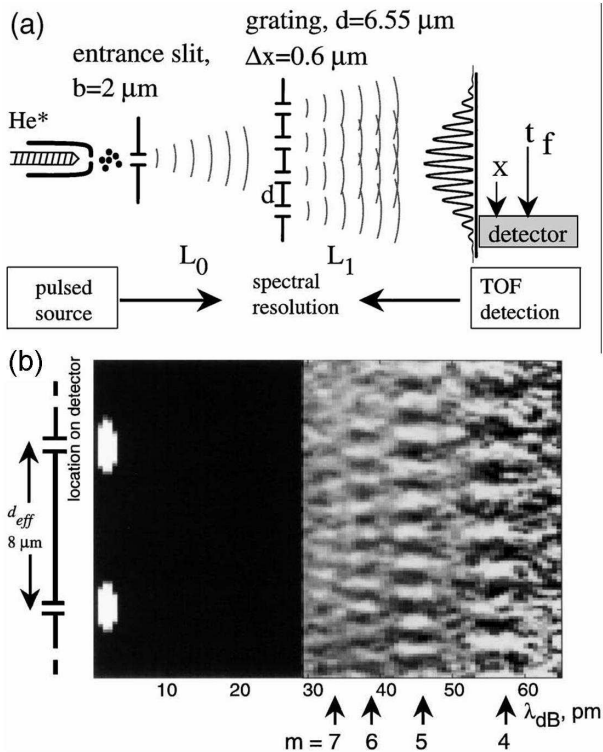}
\caption{The Talbot effect. (a) Schematic of a pulsed source and a
time-resolved detector used to observe near-field diffraction from a
nano-grating with 0.6 $\mu$m diameter windows spaced with a period
of 6.55 $\mu$m.  (b) Higher order Talbot fringes. The spatial atom
distribution vs. de Broglie wavelength is plotted. The arrows
indicate  locations at which Talbot fringes of the $m$th order are
observed. Figure from \cite{NKP97}.}\label{fig:higher Talbot}
\end{figure}

Talbot fringes were first observed with an atom beam and
nanostructure gratings by \cite{SEC93,CEH95t,CLL94} and higher-order
Talbot fringes were observed by \textcite{NKP97} (see Figure
\ref{fig:higher Talbot}). The Talbot effect has also been studied
with on-resonant light \cite{TTS03,TTS05}, and Talbot revivals have
been observed in the time-evolution of atom clouds after pulses of
off-resonant standing waves of light \cite{DHD99,CKS97}. The Talbot
\emph{time} is $\tau_T = L_T / v = 2d^2m/h$.

\textcite{ROH01} proposed detecting the Talbot effect for atoms
trapped in wave guides, and \textcite{RKS01} discussed the formation
of vortices in BEC as a result of the Talbot effect. Proposals to
use the Talbot effect to study the state of electromagnetic fields
in cavities are discussed in \cite{RDZ99,ROS00}. Using the Talbot
effect with multiple phase gratings has been proposed as a way to
make more efficient beam splitters for atom waves
\cite{ROH00,ROH99}, and this is related to the standing-wave
light-pulse sequence described in \cite{WWD05}.

The Lau effect is a closely related phenomenon in which
\emph{incoherent} light incident on \emph{two} gratings causes
fringes on a distant screen, provided that the gratings are
separated by a half-integer multiple of the Talbot length.
References for the Lau effect in light optics include
\cite{Lau48,PAT89,JAL79,BAJ79}. In essence, for the Lau effect the
first grating serves as an array of mutually incoherent sources
and Fresnel diffraction from the second grating makes the pattern
on the screen.  This forms the basis for Talbot-Lau
interferometers which we discuss in Section III.

An especially promising application of Talbot (or Lau) imaging with
atoms is atom lithography as demonstrated in \cite{TBT92,MHP04} and
many others.  For reviews see \cite{MEM03,BBD99}. It is possible to
write smaller gratings and features using the reduced period
intermediate images discussed above. Similar Fourier images have
been used for x-rays to write half-period gratings \cite{FHS79} and
to construct x-ray interferometers \cite{DNS02,MKK03,WND05}. Grating
self-images may also be used in quantum optics experiments to
produce a periodic atom density in an optical resonator.

\subsection{Time-dependent diffraction}

Many new interference effects arise when the diffracting
structures are modulated in time, a situation we have not
considered previously (except for revivals at the Talbot time
after pulsed gratings). These new effects arise with matter waves
because the vacuum is dispersive for atoms - particles with
shorter wavelength (higher energy) propagate faster than those
with longer wavelengths. In contrast, for light in vacuum all
wavelengths propagate at a constant speed, $c$.

Two matter wave components interfering at $(x,t)$ may have
propagated from the same $x'$ but originated from there
at\emph{different} times $t'$ (if they have different velocity)!
Time-dependent boundary conditions can cause matter wave diffraction
phenomena in time that are similar to spatial diffraction phenomena
arising from spatially dependent boundary conditions. This was first
discussed in a seminal paper by \textcite{MOS52}, who argued that
after opening a shutter one should observe a rise in the matter wave
intensity with Fresnel fringes in time, similar to the diffraction
of an edge in space. He called this very general phenomena
\emph{diffraction in time}. Similarly, the opening and closing of a
shutter results in a single slit diffraction in time; two successive
openings makes a double slit; and a periodic change in the opening
of the slit produces a diffraction pattern in time. With diffraction
in time, new frequency (energy) components are created (as in an
acoustic-optic modulator), resulting in components with new momenta.
In analogy to diffraction in space one finds that diffraction in
time has both near-field and far-field regimes. One also observes
Raman Nath and Bragg regimes, depending on the duration of the
interaction and the amount of energy (frequency) transfer.

\subsubsection{Vibrating mirrors}

\begin{figure}[t]
\begin{center}
\includegraphics[width = \columnwidth]{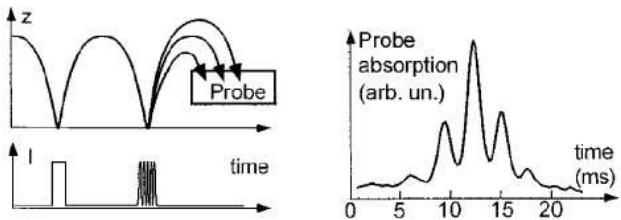}
\caption{Diffraction in time from a pulsed mirror. (a) schematic of
the experiment, showing atom trajectories and a trace indicating
when the mirror was switched on. The first pulse acts as a slit in
time, the second pulse is modulated so that it acts as a grating in
time. (b) the diffraction pattern in time manifests as different
energy components in the resulting atomic beam. Figure from
\cite{SSD95} and \cite{COH98}} \label{diffTimeENS.fig}
\end{center}
\end{figure}

Even though diffraction in time of matter waves was predicted in
1952 the first experimental demonstrations had to wait until the
late 1980's. The experimental difficulty in seeing diffraction in
time is that the time scale for switching has to be faster than the
inverse frequency (energy) width of the incident matter wave. This
condition is the time equivalent to coherent illumination of
adjacent slits in spatial diffraction. The first (explicit)
experiments demonstrating diffraction in time used ultra-cold
neutrons reflecting from vibrating mirrors \cite{HKO87,FMG90,HFG98}.
The side-bands of the momentum components were observed.

A study of diffraction and interference in time was performed by
the group of J. Dalibard at the ENS in Paris using ultra cold
atoms reflecting from a switchable atom mirror
\cite{SSD95,SGA96,ASD96}.  Ultra cold Cs atoms ($T \sim 3.6 \mu
K$) released from an optical molasses fell 3 mm, and were
reflected from an evanescent atom mirror. By pulsing the
evanescent light field one can switch the mirror on and off,
creating time-dependent apertures that are diffractive structures
in the spirit of \textcite{MOS52}.

Even for these ultra cold Cs atoms the energy spread (7MHz) is too
large for the time-diffraction experiment, so a very narrow energy
window was selected by two (0.4 ms) temporal slits. The first slit
was positioned 26 ms after the atoms were released.  Switching on
the mirror a second time, 52 ms later, selected a very narrow energy
slice in the same way as a two-slit collimation selects a very
narrow transverse velocity slice. The arrival time of the atoms at
the final ``screen" was measured by fluorescence induced by a light
sheet.

If the second slit is very narrow ($ < 10 \mu s$) one observes
single slit diffraction in time; if the mirror is pulsed on twice
within the coherence time of the atomic ensemble one observes double
slit interference in time; and many pulses lead to a time-dependent
flux analogous to a grating diffraction pattern as shown in Fig.
\ref{diffTimeENS.fig}. From the measurement of the arrival times the
energy distribution can be reconstructed.  Similar diffraction in
time is observed when a BEC is reflected from a vibrating mirror
\cite{CMP05}.

Because the interaction time between the atoms and the mirror
potential ($< 1 \mu s$) was always much smaller then the modulation
time scale ($> 10 \mu s$), these experiments are in the `thin
grating' (Raman-Nath) regime for diffraction in time.

\subsubsection{Oscillating potentials}

When matter waves traverse a time-modulated potential one can observe coherent
exchange of energy between the oscillating field and the matter wave.  This was
demonstrated in a neutron interference experiment \cite{SHK95} where a
oscillating magnetic potential was applied to one path of an neutron interferometer.
Coherent exchange of up to 5 quanta $\hbar \omega$ was observed in the interference
patterns, even though the transit time through the oscillating potential was much
shorter then the oscillation period.

\subsubsection{Modulated light crystals}

\begin{figure}[t]
\begin{center}
\includegraphics[width = \columnwidth]{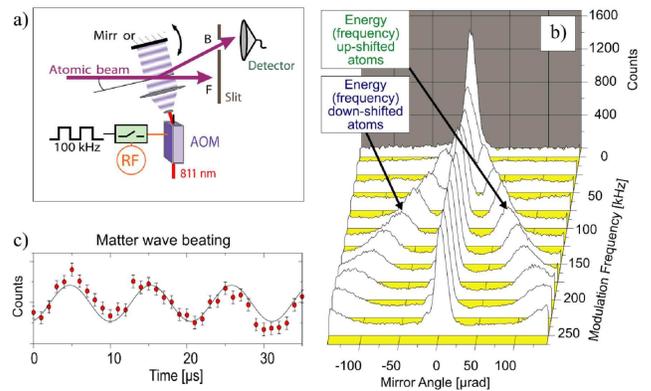}
\caption{(color online) Frequency shifter for matter waves.  (a) A
time-modulated light crystal causes diffraction in time and space.
(b) Rocking curves show how the Bragg angle for frequency-shifted
matter waves is controlled by the grating modulation frequency. (c)
Beating between frequency shifted and unshifted matter waves. Figure
from \cite{BOA96}.} \label{fig:ODBM}
\end{center}
\end{figure}

The time equivalent of spatial Bragg scattering can be reached if
the interaction time between the atoms and the potential is long
enough to accommodate many cycles of modulation. When a light
crystal is modulated much faster then the transit time, momentum
is transferred in reciprocal lattice vector units and energy in
sidebands at the modulation frequency. This leads to Bragg
diffraction at two new incident angles.

Bragg scattering in time can be understood as a transition between
two energy and momentum states. The intensity modulation frequency
of the standing light wave compensates the detuning of the Bragg
angle. The frequency of the de Broglie wave diffracted at the new
Bragg angles is shifted by $ \pm \hbar \omega_{mod}$
\cite{BOA96,BAK00}. Thus, an amplitude modulated light crystal
realizes a coherent frequency shifter for a continuous atomic beam.
It acts on matter waves just as an acousto-optic modulator acts on
photons, shifting the frequency (kinetic energy) and requiring an
accompanying momentum (direction) change.

In a complementary point of view, the new Bragg angles can be
understood from looking at the light crystal itself. The modulation
creates side bands $\pm \omega_\textrm{{mod}}$ on the laser light,
creating moving crystals which come from the interference between
the carrier and the side bands.  Bragg diffraction from the moving
crystals occurs where the Bragg condition is fulfilled in the frame
co-moving with the crystal, resulting in diffraction of the incident
beam to new incident angles.

The coherent frequency shift of the Bragg diffracted atoms can be
measured by interferometric superposition with the transmitted
beam. Directly behind the light crystal the two outgoing beams
form an atomic interference pattern which can be probed by a thin
absorptive light grating \cite{AKB97}.  Since the energy of the
diffracted atoms is shifted by $\hbar \omega_\textrm{{mod}}$, the
atomic interference pattern continuously moves; this results in a
temporally oscillating atomic transmission through the absorption
grating (Figure \ref{fig:ODBM}).

Starting from this basic principle of frequency shifting by
diffraction from a time dependent light crystal many other time
dependent interference phenomena were studied for matter waves
\cite{BAK99,BAK00}. Light crystals are an ideal tool for these
experiments since one can easily tailor potentials by controlling
the laser intensity and frequency and create more complex
structures by superimposing different independently controlled
crystals.

\begin{figure}[t]
\includegraphics[width = 8 cm]{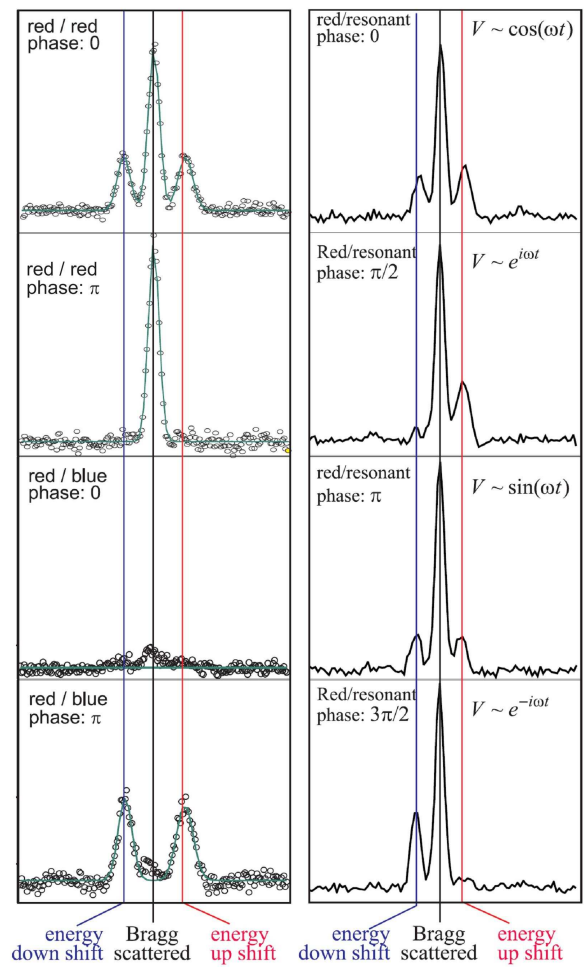}
\caption{(color online) Diffraction in time from two superimposed
light crystals with a controlled relative phase between the
modulations. (Left) two off-resonant light crystals are
superimposed. The relative phase of the temporal modulation controls
the intensity of the frequency shifted and unshifted Bragg beams.
(Right) an on-resonant and an off-resonant crystal are superimposed.
The relative phase controls the time-dependent potential.  For phase
$\pi$/3 (3$\pi$/2) only frequency up (down) shifted components
appear. Figure adapted from \cite{BAK00}.} \label{fig:Mod2}
\end{figure}

For example using light from two different lasers one can create two
coinciding light crystals generated in front of the retro-reflection
mirror.  Varying detuning and phase between the two modulated
crystals creates situations where diffraction is completely
suppressed, or where either the frequency unshifted or the frequency
shifted order is suppressed (Figure \ref{fig:Mod2}). The combination
of real and imaginary potentials can produce a driving potential of
the form $U(t) \sim e^{ \pm i \omega_m t}$ which contains only
positive (negative) frequency components respectively.  Such a
modulation can only drive transitions up in energy (or down in
energy).

\subsection{Summary of diffractive Atom Optics}

To summarize, in Section II we have reviewed atom diffraction from
nanostructures and standing waves of light. Nanostructures absorb
atoms, can be arbitrarily patterned (e.g. holograms) and affect all
atomic and molecular species.  Standing waves of light can make a
phase (or in some cases amplitude) grating for a particular species
of atom in a specific state. Light gratings can be thick or thin,
strong or weak, and can be modulated in time. Both types of grating
exhibit interesting and useful interference phenomena in both the
near- and far-field regimes.

\begin{figure} [t!]
\includegraphics[width = \columnwidth]{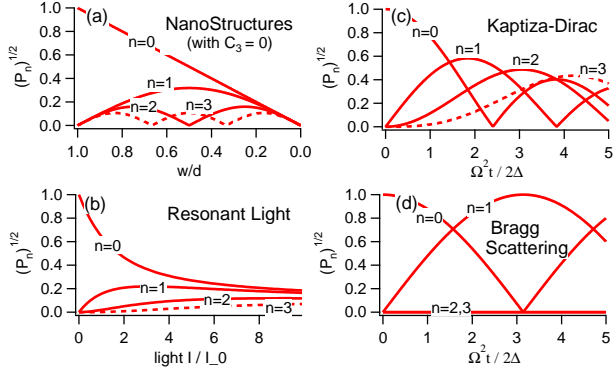}
\caption{Summary of diffraction efficiency for atoms
$|\psi_n/\psi_{inc}|$ from different types of gratings. (a)
nanostructures with $C_3$=0, (b) standing waves of on-resonant
light, (c) Kapitza-Dirac (thin phase mask) diffraction, and (d)
Bragg (thick crystal) scattering.  The x-axis is proportional to
the intensity of the light (or the open fraction in the case of
nanostructures).\label{fig:efficiency}}
\end{figure}

Figure \ref{fig:efficiency} summarizes the diffraction efficiency
of four different kinds of time-independent gratings: two
absorbing gratings (nanostructures and standing waves of
on-resonant light) and two non-dissipative gratings (in the
Kapitza Dirac and Bragg regimes). These efficiencies are given by
equations \ref{eq:sinc diffn_efficiencies}, \ref{eq:Pkd},
\ref{eq:P_Bragg}, and the Fourier transform of equation
\ref{eq:t(x)}.

\begin{figure}[h]
\includegraphics[width = \columnwidth]{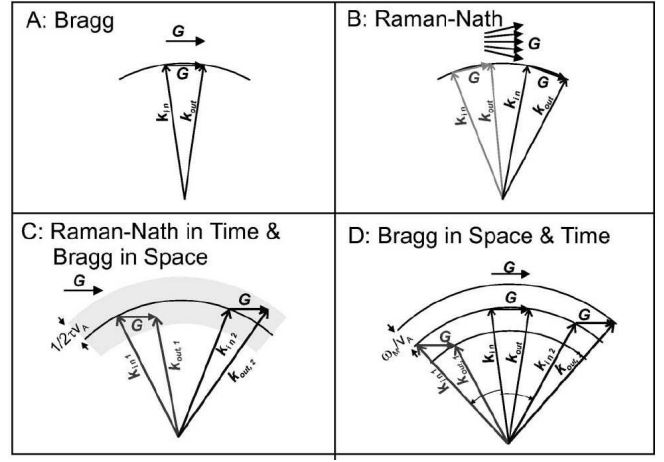}
\caption{Momentum diagrams for cases: (A) A thick grating, (B) A
thin grating, (C) A thick pulsed grating (D) A thick harmonically
modulated grating. \cite{BAK00}\label{fig:RNB}}
\end{figure}

Figure \ref{fig:RNB} summarizes thick and thin gratings in space and
also in time with Ewald constructions to denote energy and momentum
of the diffracted and incident atom waves. The diffraction from
(modulated) standing waves of light can also be summarized with the
Bloch band spectroscopy picture \cite{BAK00,CBD01}.

\subsection{Other coherent beam splitters}

Whereas diffraction occurs without changing the atom's internal
state, another important class of beam splitters uses laser or RF
transitions that do change atoms' internal state while transferring
momentum. Therefore, as they coherently split atomic wavefunctions
into two (or more) pieces they cause entanglement between the atomic
motion and the internal atomic states. Important examples that are
used in atom interferometry include absorption from a traveling wave
of light \cite{BOR89,BER97_Borde}, stimulated Raman transitions
\cite{KAC92} and longitudinal RF spectroscopy \cite{GKR01}. The
longitudinal Stern-Gerlach effect \cite{MPV91} also causes
entanglement between motion and internal degrees of freedom.  We
discuss these in Section III on atom interferometry.

Potentials that change slowly from single-well to double-well
represent an entirely new type of beam splitter that is more
applicable to trapped atoms than propagating light.  We discuss this
tool for coherent splitting of atomic wavefunctions in the next
chapter on atom interferometry.

Reflecting surfaces have been used for atom diffraction, atom
holography, and Young's experiment with atoms, e.g. in
\cite{LCH97,CSH98,GKK05,GKZ07,DEW93,CSS94}, \cite{SHF02}, and
\cite{KSF03,ESA04} respectively. The challenges of using
reflection-type atom optical elements include low reflection
probability and strict requirements for flatness in order to
maintain atom wave coherence. Still, the toolkit for coherent atom
optics is expanded by quantum reflection, in which atom waves
reflect from an attractive potential,  and also classical
reflection, where repulsive potentials can be formed with evanescent
waves of blue-detuned light or engineered magnetic domains. Various
mirrors for atoms are discussed in
\cite{HWW99,KLL96,SHF02a,SHI01,BLS89, MCS00,HMK97,SSE02,FOZ07}.

Other beam splitters for atoms that have not been used for atom
interferometer experiments will not be discussed in this review.

\section{ATOM INTERFEROMETERS} \label{sec:interferometers}

\subsection{Introduction}

The essential features of interferometers generally and atom
interferometers in particular are listed in the succession of five
steps: (1) prepare the initial state, (2) split the wavefunctions
coherently into two or more states,  (3) apply interactions that
affect the two states differentially, generally due to their
different spatial location, (4) recombine these components
coherently, and (5) measure the phase shift of the detected fringes.

The crucial step of coherent splitting (2) has been accomplished for
atom interferometers using diffraction gratings, photon absorption,
Raman transitions, longitudinal Stern-Gerlach magnets and even
physical separation of confined atoms into multiple potential wells.
In the following we discuss these in the framework of the
interferometers in which they have been used, and review the basic
features of several atom interferometer designs. A detailed survey
of scientific research with atom interferometers is given in
Sections IV, V, and VI.

\subsubsection{General design considerations}

When designing and building interferometers for atoms and molecules,
one must consider key differences between matter waves and light.
The dispersion relations, the coherence properties, and our tools to
control the two different kinds of waves are among the important
differences.

One striking difference is the fact that matter waves have short
deBroglie wavelengths ($\sim$ 10 pm  for thermal atoms up to $\sim$1
$\mu$m for ultracold atoms), and also have a very short coherence
lengths ($\sim$100 pm  for thermal atomic beams, and seldom larger
then 10 $\mu m$ even for atom lasers or BEC). This requires that the
period and the position of the interference fringes must be
independent of the deBroglie wavelength of the incident atoms. In
optical parlance this is a property of \emph{white light}
interferometers.

A second concern with atoms is that they interact strongly with each
other. Therefore matter waves are often non-linear, especially in
the cases where the atoms have significant density as in a BEC or
atom laser.

A third distinguishing feature is that atoms can be trapped. This
leads to a new class of interferometers for confined particles,
which we discuss at the end of this section.

\subsubsection{White light interferometetry}

The challenge of building a white light interferometer for matter
waves is most frequently met by the 3-grating Mach Zehnder (MZ)
layout. This design was used for the first \emph{electron}
interferometer by \textcite{mar52}, for the first \emph{neutron}
interferometer by \textcite{RTB74}, and for the first atom
interferometer that spatially separated the atoms by
\textcite{kei91}.  In the MZ interferometer the role of splitter and
recombiner is taken up by diffraction gratings. They also serve as
the mirrors that redirect the separating atom waves back together.
(In fact simple mirrors won't serve this purpose if the initial
state is not extremely well collimated.  This is because most
interferometer designs that employ simple mirrors will make the
fringe phase strongly correlated with input beam position and
direction.) In their seminal work \textcite{SIM54}, noted that with
grating interferometers
\begin{quote}``the fringe spacing is independent of wavelength.
This `achromatic' behavior ... appears to be characteristic of
instruments using diffraction for beam splitting.''\end{quote}

The explanation is that diffraction separates the split states by
the lattice momentum, then reverses this momentum difference prior
to recombination.  Faster atoms will diffract to smaller angles
resulting in less transverse separation downstream, but will produce
the same size fringes upon recombining with their smaller angle due
to their shorter deBroglie wavelength.  For three evenly spaced
gratings, the fringe \emph{phase} is independent of incident
wavelength, surprisingly also for asymmetric designs (such as that
in Fig \ref{fig:marton}a) where the intensity maximum for different
wavelengths occurs at different distances off the symmetry
axis\footnote{The popular design in figure \ref{fig:marton}a is
asymmetric because the interferometer paths are formed by
diffraction orders 0 and 1 for one arm, and orders 1 and -1 for the
other.}.

Many diffraction-based interferometers produce fringes when
illuminated with a source whose transverse coherence length is much
less than its (large) physical width, or even the grating period.
Under such conditions, the different diffraction orders will not be
separated, so diffraction can not be resolved and it will not be
possible to exploit the physical separation of the orders to apply
an interaction to only one arm of the interferometer.  Nevertheless,
high contrast fringes will still be formed.

The three grating interferometer produces a ``position echo" as
discussed by \textcite{CDK85}.  Starting at one grating opening, one
arm evolves laterally with $\hbar G$ more momentum for some time,
the momenta are reversed, and the other arm evolves with the same
momentum excess for the same time, coming back together with the
first arm at the third grating.  If the gratings are registered, an
atom's trapezoidal pattern starts at a slot on the first grating, is
centered on either a middle grating slot or groove, and recombines
in a slot at the third grating. Not surprisingly, spin-echo and
time-domain echo techniques (discussed below) also offer
possibilities for building an interferometer that works even with a
distribution of incident transverse atomic momenta.

\subsubsection{Types and categories}

A large variety of atom and molecule interferometers have been built
since 1991. The list includes Mach Zehnder, Talbot-Lau, optical
Ramsey-Bord\'{e}, and stimulated Raman transition interferometers.
There are also longitudinal Stern-Gerlach, and longitudinal RF
interferometers. Some of these designs render the interference
fringes in position space. Some make fringes in momentum space.
Still other designs make the interference fringes observable in
internal atomic state space. We catalogue these interferometers
briefly here according to their features before examining each in
detail throughout this section.

(1) Internal state changing interferometers are one broad category.
Some beam splitters change atom's internal state, analogous to a
polarizing beam splitter in light optics.  For example, stimulated
Raman transitions entangle internal and external states, so atoms in
these interferometers are in a coherent superposition of different
momentum-spin states.

(2) Time-domain vs. space-domain is another broad classification.
In a time-domain interferometer, the beam splitters are pulsed so
all atoms interact with the gratings and the interferometer for
the same amount of time.

(3) Near-field (Talbot-Lau) and far-field (Mach Zehnder)
classification applies for diffractive atom optics. Near-field
interferometers can function even with poorly collimated beams,
but the gratings in a Talbot-Lau interferometer (TLI) must be
separated by precise multiples of the Talbot length or else the
contrast degrades.

(4) Separated path interferometers are a special category of atom
interferometer in which the paths are sufficiently physically
separated that the atom wave in one arm can be isolated and
interactions can be applied to it alone.

(5) Freely propagating cold atoms can have long times of flight
($\sim\frac{1}{2}$ s) as compared to thermal atom beams ($\sim$1
ms).  Confinement in a trap \emph{during} the interferometer
operation may soon provide even longer interaction times.

(6) Atom traps and waveguides offer the possibility of making
\emph{confined atom} interferometers in which the atom wavefunction
is split in coordinate space rather than momentum space. Obviously,
the ability to interfere atoms that are spatially confined in all
three dimensions throughout the entire interferometer is
unprecedented with light interferometers.  Additional topologies
such as multiple wells and ring-traps and longitudinal waveguides
have also been demonstrated.

Finally, we distinguish single atom interferometers from those in
which (non-linear) collective effects are significant.   Even
atoms launched from a  magneto-optical trap are generally not
dense enough to cause significant non-linear effects.
Interferometry with Bose Einstein condensates (or atom lasers) on
the other hand can show non-linear atom optics phenomena that
range in significance from phase noise to number squeezing and
Josephson oscillations.

These distinctions -- (1) internal state changing vs. state
preserving, (2) space-domain vs. time-domain, (3) near-field vs.
far-field, (4) separated path or not, (5) trapped or freely
propagating, (6) dilute vs. dense -- all affect the performance of
atom interferometers for different applications.

Since the first atom interferometers were built for Na, Cs, Mg, and
He* in 1991, others have been made for Ar*, Ca, H* He, K, Li, Ne*,
Rb atoms, and ${\rm He}_{2}$, ${\rm Li}_{2}$, ${\rm Na}_{2}$, ${\rm
I}_{2}$, C$_{60}$, C$_{70}$, C$_{60}$F$_{48}$, and
C$_{44}$H$_{30}$N$_4$ molecules. Interferometers starting with
trapped atoms have been made for Ca, Cs, He*, Mg, Na, and Rb and
interferometers using Bose-Einstein condensates have been
demonstrated with Na and Rb, .  These lists are still growing.

\subsection{Three-Grating Interferometers}

The simplest white light interferometer is a Mach Zehnder
interferometer built from 3 diffraction gratings.  The first grating
acting as a beam splitter, the second as a redirector, reversing the
(transverse) momentum of the beam, and the 3rd as a recombiner or
analyzer of the interference.

\begin{figure*}[t]
\begin{center}
\includegraphics[width = 16cm]{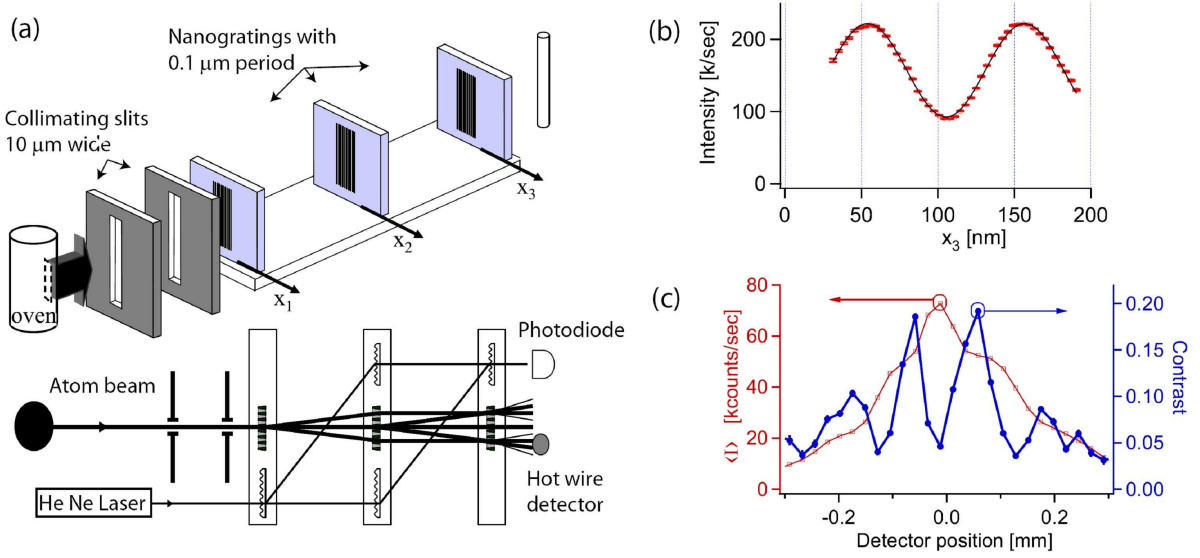}
\caption{(color online) Three grating Mach-Zehnder atom
interferometers. (a) Atom Interferometer setup used in
\textcite{kei91}. (b) Interference fringe data and best fit with
$\langle I \rangle$ = 157,000 counts per second and $C = 0.42$. A
total of 5 seconds of data are shown and the uncertainty in phase
calculated by equation \ref{eq:sigma_phi} is $\sigma_{\phi}= 2.7
\times 10^{-3}$ radians. (c) Average intensity $\langle I \rangle$
and contrast $C$ as a function of detector position [under different
conditions than (b)].} \label{fig:marton}
\end{center}
\end{figure*}

\subsubsection{Mechanical gratings}

The first 3-grating Mach-Zehnder interferometer for atoms was built
by \textcite{kei91} using three 0.4-$\mu$m period nano fabricated
diffraction gratings. Starting from a supersonic Na source with a
brightness of $B\approx10^{19}$ s$^{-1}$cm$^{-2}$sr$^{-1}$ the
average count rate  $\langle I \rangle$, in the interference pattern
was 300 atoms per second. Since then, gratings of 100 nm period have
been used to generate fringes with up to 300,000 atoms per sec.

We use this design (shown in Fig.~\ref{fig:marton}) to illustrate
how a standing wave interference pattern is formed by the two
running waves.  Starting with a common beam that is incident on two
gratings (G1 and G2), one wave is formed by 0th and 1st order
diffraction, while the other is formed by -1st and +1st order
diffraction. The difference in momentum is thus one unit of $\hbar G
\hat{x}$.  So we describe the incident running waves by the
functions $\psi_{1}$ and $\psi_{2} e^{iGx} e^{i\Delta \phi_{int}} $.
These running waves differ in momentum explicitly by $\hbar G
\hat{x}$ due to diffraction.  The waves also differ in phase by
$\Delta \phi_{int}$ due to different interactions along the two
paths.

In the zone where these coherent running waves overlap, the atom
beam intensity is
 \begin{eqnarray}
    I(x) &=& \left|\psi_{1}+ \psi_{2} e^{i\Delta
    \phi_{int}} e^{i G x}\right|^2 \nonumber\\ \nonumber\\
    I(x) &=& \langle I \rangle + \langle I \rangle C \cos (\Delta
    \phi_{int} + Gx ) .
    \label{eq:iire}
 \end{eqnarray}
This interference pattern is a standing wave in space but is
unchanging in time.  The fringes have a period of $d=G/(2\pi)$ (just
like the gratings), and a spatial offset in $\hat{x}$ (i.e. a phase)
that depends on the location of the two gratings G1 and G2 as well
as the interaction phase $\Delta \phi_{int}$.  Equation
\ref{eq:iire} is a general result, and the fringes can be detected
in many different ways.

The intensity pattern has a mean intensity and contrast
 \begin{eqnarray}
    \langle I \rangle &=& |\psi_1 |^2 + |\psi_{2} |^2 \nonumber \\
    C &=& \frac {I_{\textrm{max}}-I_{\textrm{min}} } {  I_{\textrm{max}} +I_{\textrm{min}} }
       = \frac{\psi_1^{*} \psi_{2} + \psi_2 ^{*} \psi_{1}  } {|\psi_1 |^2 + |\psi_{2}|^2}.
       \label{eq:Czone}\nonumber
 \end{eqnarray}

\noindent If one of the interfering beams is much stronger then the
other, for example $|\psi_1 |^2 \gg |\psi_{2} |^2$, then the
contrast of the interference pattern scales like
 \beq
    C \sim \frac{2 |\psi_2 |}{|\psi_1 |} = 2 \sqrt{\frac{I_2}{I_1}}.
 \eeq
Consequently one can observe $20 \%$ ($2 \%$) contrast for an
intensity ratio of 100:1 ($10^4$:1) in the interfering beams.

If the waves are not perfectly coherent, then the incoherent part
adds to the overall intensity, and the contrast is diminished. If
more than two components overlap, the situation is somewhat more
complicated.

The spatial oscillations in intensity can be detected, for example,
by measuring the atom flux transmitted through a third (absorbing)
grating (G3). In this case G3 acts as a mask to transmit (or block)
the spatially structured matter wave intensity. By translating G3
along $x$ one obtains a moir\'{e} filtered interference pattern
which is also sinusoidal and has a mean intensity  and contrast
\begin{eqnarray}
    \langle I \rangle &=& \frac{w_3}{d} \langle \tilde{I} \rangle, \label{eq:IIp} \\
    C &=& \frac{\sin(G w_3/2)}{(G w_3/2)} \tilde{C}. \label{eq:CCp}
\end{eqnarray}
where $\tilde{I}$ and $\tilde{C}$ refer to the intensity and
contrast just prior to the mask. The phase of the filtered
interference pattern is given by \beq \phi = G(x_1 - 2 x_2 + x_3)
+ \Delta \phi_{int} \label{eq:3g std phase}\eeq where $x_1$,
$x_2$, and $x_3$ are the relative positions of gratings 1, 2 and 3
with respect to an inertial frame of reference  \cite{BER97}.


\begin{figure*}
\includegraphics[width = 18cm]{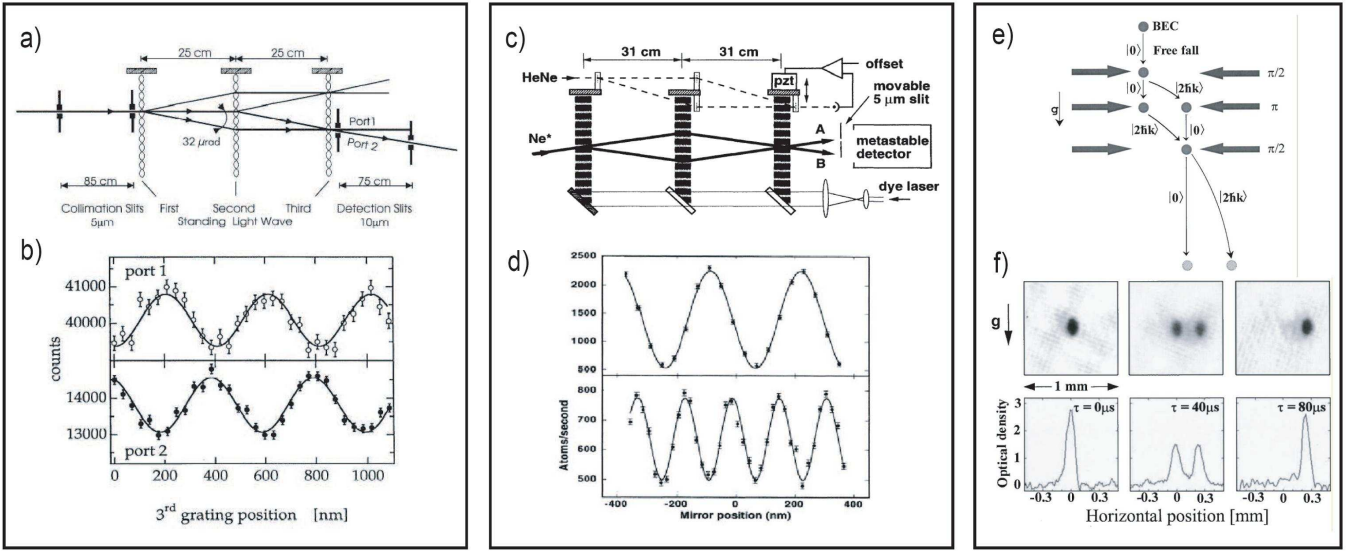}   
\caption{Atom interferometers based on three standing waves of
light.  (a) Atom beam and three Kapitza-Dirac gratings.  (b) Atom
interference patterns for both output ports demonstrate
complementary intensity variations. This is a consequence of atom
number conservation. Figures a and b reproduced from \cite{ROB95}.
(c) Interferometer based on three Bragg gratings. Dashed line shows
the path of auxiliary optical interferometer used for stabilization.
(d) Intensity fluctuations in beam A vs. position of the Bragg
gratings.  For second order Bragg diffraction, fringes of half the
period are formed. Figures c and d reproduced from \cite{GML95}. (e)
Schematic of the $\pi/2-\pi-\pi/2$ Bragg interferometer for atoms in
a BEC falling from a trap. (f) Absorption images and density
profiles demonstrating different outputs of the interferometer.
Figures e and f reproduced from \cite{TSK00}. \label{fig:rob1} }
\end{figure*}

The phase $\phi$ will have a statistical variance,
$\sigma_{\phi}^2$, in the simplest case due to shot noise
(counting statistics) \cite{LHS97,BER97} given by
 \beq
    (\sigma_{\phi})^2 \equiv \left\langle (\phi - \langle \phi \rangle )^2 \right\rangle = \frac{1}{C^2 N}
    \label{eq:sigma_phi}
 \eeq
where $N$ is the total number of atoms counted. For discussion of
how phase fluctuations depend on atom-atom interactions within the
interferometer see \cite{SCD93,SEM03,PES06}. To minimize the
uncertainty in measured phase we therefore seek to maximize $C^2 N
\propto C^2 \langle I \rangle$ by choosing the open fractions
$w_i/d$ for the three gratings, where $w_i$ is the window size for
the $i^\textrm{th}$ grating and $d$ is the grating period.  The open
fractions that maximize $C^2\langle I \rangle$ are $(w_1/d, w_2/d,
w_3/d) = (0.56, 0.50, 0.37)$.   With these open fractions, the
theoretical value of $C=0.67$ and $\langle I \rangle /
I_\textrm{inc}$= 0.015.  If vdW interactions between atoms and
gratings are included, then open fractions of the first two gratings
should be increased for best performance \cite{CWP05}.

There are in fact several different interferometers formed by the
gratings. For example, the 1$^{\mathrm{st}}$ and 2$^{\mathrm{nd}}$
orders can recombine in a skew diamond to produce another
interferometer with the white fringe property. Additional mirror
images of these interferometers make contrast peaks on either side
of the original beam axis, as shown in Fig.~\ref{fig:marton}. All
those interferometers can have fringes with the same phase, and
consequently one can therefore build interferometers with wide
uncollimated beams which have high count rate, but lower contrast.
(The contrast is reduced because additional beam components such as
the zeroth order transmission through each grating will also be
detected.)

Mechanical gratings with much larger periods have been used to
make interferometers in the extreme limit of non-separated beams.
We discuss these in the Talbot-Lau interferometer section ahead.

For well-collimated incoming beams, the interfering paths can be
separated at the 2nd grating. For example in the interferometer
built at MIT the beams at the second (middle) grating have widths of
30 $\mu$m and can be separated by 100 $\mu$m (using 100-nm period
gratings and 1000 m/s sodium atoms ($\lambda_{dB} =$ 16 pm). Details
of this apparatus, including the auxiliary laser interferometer used
for alignment and the requirements for vibration isolation, are
given in \cite{BER97}.

This geometry was used in the first atom interferometer with
physical isolation of the spatially separated paths.  Isolation
was provided by inserting a 10 cm long metal foil between the two
paths, so that the electric or magnetic field or gas pressure
could be varied on the left or right arm separately. This resulted
in measurements of atomic polarizability \cite{ESC95}, the index
of refraction due to dilute gasses \cite{SCE95,RCK02}, contrast
interferometry using magnetic rephasing \cite{SEC94}, and
diffraction phases induced by van der Waals interactions
\cite{PEC05,PEC06}. In experiments not explicitly needing
separated beams, this apparatus has been used to measure phase
shifts due to rotations \cite{LHS97} and to study decoherence due
to scattering photons and background gas \cite{CHL95,KCR01,UPC05}.
This apparatus was also used to perform the first separated beam
experiments with molecules (Na$_2$) \cite{CEH95}.

An interferometer with similar nanogratings was developed at the
MPI in G\"{o}ttingen and used to measure the polarizability of He
and He$_2$ \cite{TOE01}.

\subsubsection{Interferometers with light gratings}

One can also build MZ interferometers with gratings made from
light (Fig.~\ref{fig:rob1}).  These light gratings are generally
near-resonant standing waves that make species-specific phase
gratings.  Because they transmit all the atoms, light gratings are
more efficient than material gratings.

The third grating in a light interferometer can function in many
ways to enable detection of the fringes.  It can recombine atom
waves so their relative phase dictates the probability to find
atoms in one output port (beam) or another.  Alternatively,
fringes in position space can be detected with fluorescence from a
resonant standing wave.  Another detection scheme uses backward
Bragg scattering of laser light from the density fringes.  This
can be used in multi-path interferometers where phase shifts
affect the contrast of the fringes. (We discuss such
\emph{contrast interferometry} in the next section.). Detecting
the direction of exiting beams requires that the incident beams
must be collimated well enough to resolve diffraction, and may
well ensure that the beams are spatially separated in the
interferometer.


\textcite{ROB95} used light gratings in the Kapitza-Dirac regime
with a $5\mu$m-wide collimated beam.  Many different interferometers
were formed, due to symmetric KD diffraction into the many orders.
Two slits after the interferometer served to select both the
specific interferometer, and the momentum of the outgoing beam
(ports 1 and 2 in Figure \ref{fig:rob1}.) Fringes with 10\% contrast
show complementary intensity variations, as expected from particle
number conservation in a MZ interferometer with phase gratings.

It is even more efficient to use Bragg diffraction because no atoms
are lost to `unwanted' orders. \textcite{GML95} used Bragg
diffraction and a Ne$^*$ beam and obtained contrast of $C$=63\%.
Higher order Bragg diffraction was also used to demonstrate smaller
period interference fringes shown in Figure \ref{fig:rob1}.  A Bragg
scattering interferometer for Li atoms with a contrast of 0.84 and a
count rate of 17 kc/s was recently used to measure the
polarizability of Li atoms \cite{DCB02,MJB06,MJB06b}.

\subsubsection{Time domain and contrast interferometers}

Light gratings can easily be turned on and off, allowing one to
control the interaction times of atoms with the three gratings.
Thus, independent of initial longitudinal momentum, all atoms will
see an equal interaction and will subsequently separate equally
(since they have the same momentum transferred by the grating).
Such interferometers are especially valuable in precision
experiments since time is so easily measured accurately. This
consideration also applies to optical Raman pulses, which will be
discussed in a later section. \textcite{TSK00} made a 3-grating
interferometer for ultra cold atoms by pulsing gratings thrice in
time.  These pulsed gratings are turned on for a duration long
enough to produce Bragg diffracted momentum states.  This duration
does not affect the precise timing between interactions, but is
long enough that diffraction in time is unimportant. Their fringes
were read out in momentum space by measuring the atom cloud
position in absorption images taken shortly after the third
grating pulse. Atoms released from a BEC were used insuring that
the momentum spread of the cloud was smaller than a photon recoil
momentum $\hbar k_{ph}$ thus allowing resolution of the output
states.  More examples of time-domain interferometers based on
three diffraction gratings include \cite{WAB04,GDH02}.

\begin{figure}[t]
 \begin{center}
   \includegraphics[width = 7cm]{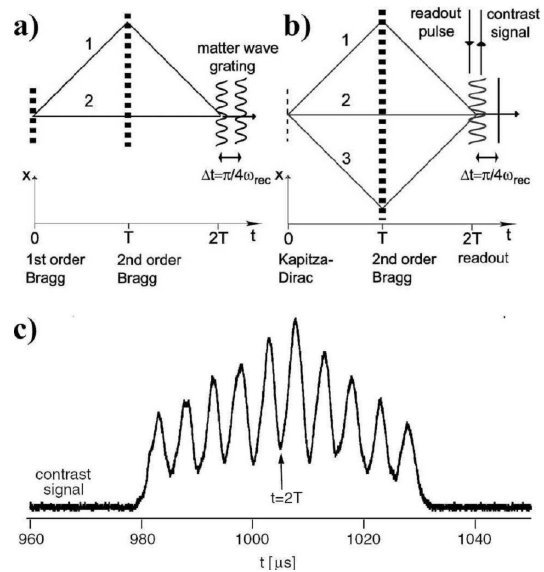}
   \caption{Contrast interferometry. (a) Space-time representation of a
   two-path interferometer that is sensitive to  the photon recoil phase.
   (b) The three-path geometry. The overall fringes have
   large contrast at $2T$ and zero contrast at $2T + \pi / 4
   \omega_{rec}$.  Bottom:  Typical single-shot signal from the
   contrast interferometer. \cite{GDH02}.} \label{fig:contrast1}
 \end{center}
\end{figure}

\textcite{GDH02} used one Kapitza-Dirac pulse followed by a second
order Bragg pulse to make an interferometer with \emph{three}
paths as shown in Fig.~\ref{fig:contrast1}. One can understand
this arrangement as two separate two-path interferometers whose
density fringes overlap. Because the phase of each two-path
interferometer changes in time in opposite directions, the two
density gratings move in and out of register as time (and phase)
increase, hence the contrast oscillates rapidly with time. This
interferometer has been used to measure $h/m_{Na}$ to a precision
of 7 ppm.  This demonstrates the utility of \emph{contrast
interferometry} in which measurements of contrast, not phase, are
used.  Contrast interferometry was pioneered in \textcite {SEC94}
where interference patterns from atoms with different magnetic
sublevels moved in and out of register.

This contrast interferometer design offers several advantages
compared to phase measurements made with a regular interferometer.
First, the fringe phase can be accurately determined in a single
``shot", eliminating effects of shot to shot atom intensity
fluctuations.  Second, most experimental sources of phase noise
affect each two-path interferometer in the same way they move the
fringes but don't change the contrast.  For example, measurements
with the interferometer in \cite{GDH02} were nearly insensitive to
vibrations, rotation, accelerations, and magnetic field gradients. A
relative phase shift between the two interferometers can be caused
however, by diffraction phases \cite{BDM03}.  If the Kapitza-Dirac
pulse causes a phase shift between the 0th and 1st diffraction
order, then the contrast does not peak at exactly $2T$, where $T$ is
the time between diffraction grating pulses. Hence, fluctuations in
the intensity of light used for the KD pulse can then lead to
fluctuations in the time at which the total contrast peak is
visible.

\textcite{GDH02} detected the contrast of the fringes in space by
measuring the intensity of reflected (Bragg diffracted) light
probing the fringes in space.  The intensity of reflected light can
be continuously monitored as the two sets of interference fringes
pass through each other in time. This causes oscillations in the
intensity of reflected light as shown in Fig.~\ref{fig:contrast1}.

\subsubsection{Talbot-Lau (near field) interferometer}

\begin{figure}[b]
\begin{center}
\includegraphics[width = 8cm]{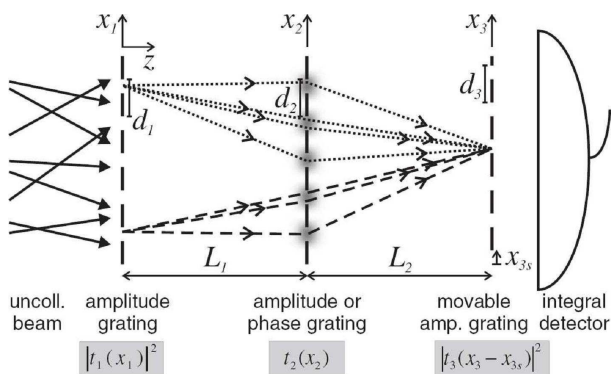}
\caption{A sketch of the Talbot-Lau interferometer setup consisting
of three gratings.  The first grating is illuminated by an
uncollimated molecular beam.  Still, coherent interference occurs
between all paths that originate from one point at the first grating
and meet at the a point on the third grating.  By varying the
grating position $x_{3}$, a periodic pattern in the molecular
distribution can be detected. Figure from \cite{BAZ03}.
\label{fig:TLIsketch}}
\end{center}
\end{figure}

We now turn to near-field interferometers.  As discussed in
Section II D, a high degree of spatial coherence is needed to
create recurring self-images of a grating due to near-field
diffraction (the Talbot effect).  But completely incoherent light
can still produce fringes downstream of a grating pair (the Lau
effect).  When two gratings with equal period ($d$) are separated
by a distance $L_1$, the Lau fringe contrast is maximum at a
distance beyond the second grating of \beq L_2 = \frac{L_1 L_T
\frac{n}{2m}}{L_1 - L_T \frac{n}{2m}} \eeq where
$L_T=2d^2/\lambda_{dB}$ is the Talbot length and the integers $n$
and $m$ refer to the $n$th revival of the $m$th higher-order
Fourier image. The fringe period is then \beq d' = d \frac{L_2 +
L_1}{m L_1}. \eeq If a 3rd grating is used as a mask to filter
these fringes, then a single large-area integrating detector can
be used to monitor the fringes. This 3-grating arrangement is a
Talbot-Lau Interferometer (TLI). A typical TLI uses three
identical gratings and $L_1 = L_2 = L_T/2$ with $n=1$ and $m=2$.

The first grating can be regarded as an array of small but mutually
incoherent sources of diverging waves.  Shortly after the second
grating near-field diffraction causes any shadow effects to become
blurred out.  At a distance $L_2$ from the second grating, spatial
structure in the intensity starts to reemerge. The intensity
oscillations observed with a TLI are not a ray-optics phenomenon;
they are due to wave interference for the multiple paths shown in
Figure \ref{fig:TLIsketch}. Evidence for this is that $L_2$ depends
on $\lambda_{dB}$ (and hence a fairly monochromatic velocity
distribution is needed for optimum contrast). The second grating can
be a phase grating, but the first and third gratings must be
amplitude gratings.  The theory of this interferometer is discussed
in \cite{CLR92,CLL94,BER97-BBO,BER97-CLL,BAZ03}.

A famous feature of a TLI is that the contrast is unaffected by the
beam width. A large transverse momentum spread in the beam is also
tolerated.  Hence much larger count rates can be obtained with a
TLI.

Furthermore, in a TLI the relationship $L_1=L_2=L_T/2$ means that
the maximum grating period is $d < \sqrt{L_1 \lambda_{dB}} \sim
M^{-1/4}$ where $M$ represents mass for a thermal beam.  In
comparison, for a MZI with resolved paths the requirement is $d<
\lambda_{dB} L / (\Delta x) \sim M^{-1/2}$ where $\Delta x$ is the
width of the beam and $L$ is the spacing between gratings. Thus
the TLI design is preferable for massive particles.

A Talbot-Lau interferometer was first built for atoms by
\textcite{CLL94} using a slow beam of potassium atoms. The
experiment used gratings with a period of $d$=100 $\mu$m, and a
count rate of $\langle I \rangle = 4 \times 10^7$ atoms/sec was
achieved. The source brightness was 2500 times weaker than in the
3 grating Mach Zehnder interferometer of \textcite{kei91}, but the
signal was about 3000 times stronger.  Because of its attractive
transmission features, and the favorable scaling properties with
$\lambda_{dB}$, the TLI has been used to observe interference
fringes with complex molecules such as C$_{60}$, C$_{70}$,
C$_{60}$F$_{48}$, and C$_{44}$H$_{30}$N$_4$ \cite{BHU02,HUH03}. Of
course, the TLI does not separate the orders - indeed components
of the wave function are only displaced by one grating period at
the Talbot length. Even though the TLI interferometer cannot offer
separated paths, it is still sensitive to inertial forces,
decoherence, and field gradients \cite{CLL94b,HHB04,HUB03}.

\textcite{CKS97} used the phrase ``time domain interferometry" to
describe a Talbot-Lau interferometer that consists of two (Kapitza
Dirac) gratings pulsed in time, and renders interference fringes in
position space.  A third pulse of light, a traveling wave, was Bragg
reflected by the atomic density pattern and thus served as the
detection scheme for fringes.  The atom fringe contrast (and
backscattered probe light) oscillates with the characteristic Talbot
time $\tau_T = \L_T / v = 2d^2 m / h$, and this readout mechanism
demonstrates 100 percent contrast even with an `uncollimated cloud'
of atoms. Talbot Lau interferometry with (pulsed) light gratings has
also been explored by \cite{CDB00,DHD99,TTS03,TTS05,WSP05,WCB06}.

\subsection{Interferometers with path-entangled states}

In some interferometers, the internal state of the atoms depends on
the path through the interferometer.  Hence the state of the atom is
entangled with the path.  This usually occurs when the RF or laser
photons that cause a transition between internal states also impart
momentum, thus creating such entanglement.

Such entanglement has implications both for what the interferometer
can measure, and for how the interference can be detected. Detection
is the more obvious; if recombination results in oscillations
between two internal states, then state-sensitive detection can
reveal the fringes without need for the atom paths to be spatially
resolved. The influence of having different internal states in the
middle of the interferometer is more subtle.  Many atomic properties
such as polarizability and scattering lengths depend on the state;
hence such interferometers naturally measure the difference of that
property between the states which is generally less informative than
the property in one state.

\subsubsection{Optical Ramsey-Bord\'{e} interferometers}

\begin{figure}[t]
\begin{center}
\includegraphics[width = \columnwidth]{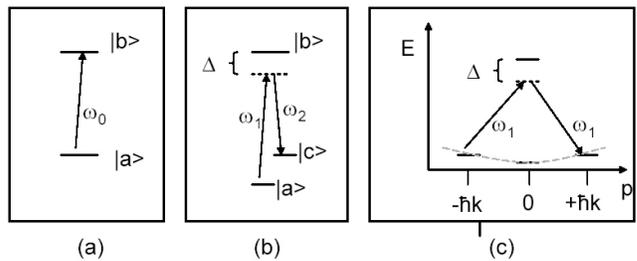}
\caption{Different schemes used to place atoms in a superposition of
momentum states. (a) superposition with a meta-stable state using a
$\pi/2$ pulse. (b) Stimulated Raman transition with two light
fields. (c) Bragg scattering with monochromatic light. $\Delta$ is
the detuning from resonance.  The dashed curve is the kinetic energy
$\frac{p_{light}^2}{2m}$. } \label{fig:beamsplitterdiagram}
\end{center}
\end{figure}

\begin{figure*}[t]
\begin{center}
\includegraphics[width = 9cm]{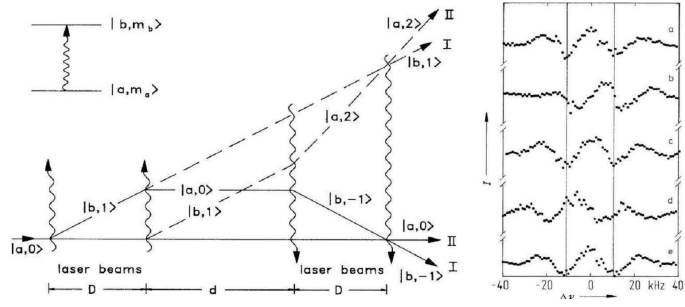}
\caption{Ramsey-Bord\'{e} Interferometer.  (Left) In the first
interaction zone the matter wave is coherently split into two
partial waves with internal states \ket{a,m_a} and \ket{b,m_b}
corresponding to energy levels $a$ and $b$ respectively, and the
number $m$ of photon momenta transferred to the atom. (Right) Fringe
shifts due to rotation at different rates. Figures and caption
reproduced from \textcite{RKW91}. \label{fig:borde1}}
\end{center}
\end{figure*}

When a traveling wave of resonant light intersects a two-level atom,
the atom is put into a superposition of ground and excited states in
which the photon absorbed in promoting the atom to the excited state
has added its momentum to that of the ground state, resulting in a
differential momentum of $\hbar k_{ph}$ between ground and excited
state (Fig \ref{fig:beamsplitterdiagram}a).
 \beq
    |a,p \rangle \rightarrow \sin(\theta)|a,p \rangle + \cos(\theta) |b,p+\hbar k_{ph} \rangle
 \eeq
Bord\'{e}'s seminal 1989 paper that Optical Ramsey spectroscopy by
four traveling laser fields is an atom interferometer when taking
the momentum transfer in the excitation process into account
\cite{BOR89}.  Such an experiment is now often called a
Ramsey-Bord\'{e} interferometer. In comparison, the classic
Chebotayev paper  \cite{CDK85} focused on Kapitza-Dirac or Bragg
diffraction gratings that preserve atoms' internal state. A unified
description of these cases is found in \cite{BER97_Borde}.

If the excitation is on resonance, the fraction of amplitude that
is deflected by the transition is determined by the \emph{pulse
area} $\theta = \int \Omega_1 dt $ where $\Omega_1 =
\mathbf{d_{ab}} \cdot \mathbf{E_0}/\hbar$ is the bare Rabi
frequency. A Ramsey-Bord\'{e} $\frac{\pi}{2}$-pulse (named for the
condition $\theta = \pi/2$) results in an equal splitting of the
amplitude between states \ket{a} and \ket{b} by resonant light. If
the excitation is detuned by $\Delta$
 \beq
    P_b(t) = \frac{1}{2}\left(\frac{\Omega_1}{\Omega_R}\right)^2[1-\cos(\Omega_R t)],
 \eeq
where $\Omega_R = \sqrt{\Omega_1^2+\Delta^2}$ is the generalized
Rabi frequency. When the detuning grows the oscillations become
more rapid and less complete.

For an optical Ramsey-Bord\'{e} interferometer to work, the lifetime
of the excited state must be comparable to the transit time through
the interferometer in order to avoid coherence-destroying
spontaneous decay of state $|b \rangle$ (see Section IV.B. on decoherence).
Consequently optical Ramsey-Bord\'{e} interferometers are generally
used with long-lived, metastable excited states such as the 1S-2S
transition in H, or the lowest-lying intercombination lines of Mg or
Ca \cite{GHN98,OBF99,RPH98,MRI89}.

In the 4 zone Ramsey-Bord\'{e} interferometer atoms passing
through the first laser beam are put in a superposition of
internal states \ket{a} and \ket{b}. Several possible paths exit
this apparatus, but only the paths shown in Fig.~\ref{fig:borde1}
cause interference fringes in the populations (outputs I and II of figure 26).
Oscillations in the state \ket{b} population are controlled by the
phase of the laser at each of the 4 zones, therefore the simplest way to
produce fringes is to adjust the laser frequency. Additional phase
shifts in the fringes can be caused by any interaction that
affects the internal states differentially, for example magnetic
fields. Because of the photon recoil, the two paths are also
separated in space and are therefore sensitive to field
\emph{gradients} and inertial displacements.

This 4 zone design of a Ramsey-Bord\'{e} interferometer was
realized by \textcite{RKW91} who also demonstrated the linear
increase of phase shift with rotation rate $\Omega$. The data
shown in Fig.~\ref{fig:borde1} are the first demonstration of the
Sagnac effect for atom interferometers.

Since then many Ramsey-Bord\'{e} interferometers were employed for
H, Mg, and Ca atoms and I$_2$ molecules for precision experiments
such as high resolution spectroscopy
\cite{HGW98,HMN02,GHN98,KZR94,SSH93,KDM05,SDS04,DSL05,WBD02,OBF99}
and fundamental studies such as geometric phases, and light shift
potentials \cite{MBR95,MHW00,YKM02}, transition Stark shifts
\cite{RSS93,MTI93} and multiple beam high-finesse atom
interferometry \cite{WHH96,HPR97,HRL99,RPH98} and molecule
interferometry \cite{BCD94}.

\subsubsection{Raman interferometry}

\begin{figure}[b]
\begin{center}
\includegraphics[width = 8cm]{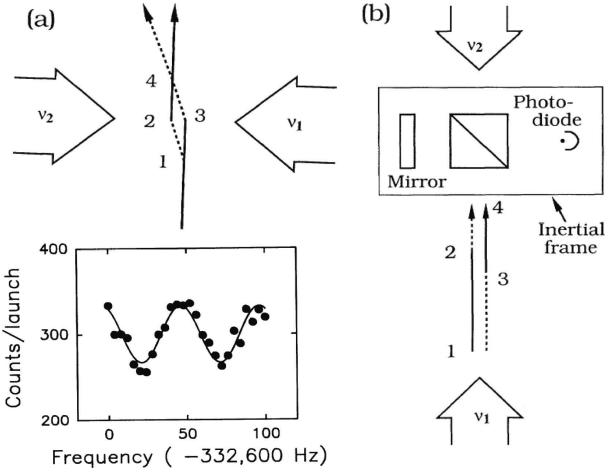}
\caption{Raman pulse interferometer.   (a) Transverse splitting and
(b) longitudinal splitting of atoms with a $\pi/2 - \pi - \pi/2$
pulse interferometer. The mechanical recoil from the first $\pi/2$
pulse (at position 1) coherently splits the atomic wave packet. The
$\pi$ pulse (positions 2 and 3) redirects each wave packet's
trajectory. By adjusting the phase of the second $\pi/2$ pulse
(position 4), the atom can be put into either \ket{1} or \ket{2}. In
the experiment, the atoms were prepared in the \ket{1} state (solid
lines) and detected in the \ket{2} state (dashed lines). (Bottom)
Interferometer fringes are observed by scanning the frequency of the
Raman laser beams. Figures reproduced from \textcite{KACS91}. }
\label{fig:kasevich1}
\end{center}
\end{figure}

A similar beam splitter can be implemented using \mbox{\emph{Raman
transitions}} between two low-lying (e.g. hyperfine) states in a
three-level atoms (Fig.~\ref{fig:beamsplitterdiagram}b).  The
superposition is now between two long-lived states and can be driven
with lasers tuned off-resonant from the excited state so that
spontaneous emission is no obstacle to coherence time.  For building
an atom interferometer one has to transfer momentum during the Raman
transition.  Consequently two counter propagating running
waves\footnote{Counter-propagating light beams make Doppler
sensitive transitions that are highly selective for atomic velocity;
co-propagating light beams make Doppler insensitive transitions.
Doppler sensitive Raman transitions can prepare atoms with a
momentum uncertainty of less than a photon recoil \cite{KWR91}.}
($\omega_1$ and $\omega_2$) of light with frequencies tuned to Raman
resonance ($\hbar \omega_1 - \hbar \omega_1 =
E_{|a\rangle}-E_{|c\rangle} = \Delta E_\textrm{hf}$) are used to
stimulate Raman transitions between two hyperfine states \ket{a} and
\ket{c} (Fig.~\ref{fig:beamsplitterdiagram}b). Absorption from one
light beam and stimulated emission into the other gives atoms a
momentum kick of $\hbar \Delta k = \hbar k_1 + \hbar k_2 \sim 2\hbar
k_\textrm{ph}$.  Since the hyperfine splitting $\Delta E_\textrm{hf}
\ll \hbar \omega_{1,2}$ is much smaller then the energy of either of
the photons ($\hbar \omega_{1,2}$) the momentum transfer can be
approximated by $2 \hbar k_\textrm{ph}$ (reduced by the cosine of
the half-angle between light beams).

Transfer of amplitude from \ket{a} to state \ket{c} mimics the
dynamics of a driven 2 level system with coupling frequency equal
to the product of the individual Rabi frequencies divided by
$\Delta$ (see Fig.~\ref{fig:beamsplitterdiagram}b).

An alternative to Raman transitions is Stimulated Adiabatic Rapid
Passage \mbox{STIRAP} described by \textcite{GRS90} and
\textcite{BTS98}. This process is more controllable since it does
not depend so critically on laser power.  The method is based on
adiabatic change of a ``dark state" and has the disadvantage that
only one superposition of the two states survives (the other
decays spontaneously) hence its application to interferometry
gives only one output state.

Starting with laser-cooled sodium atoms launched from a trap,
\textcite{KACS91,KAC92} demonstrated an interferometer based on
stimulated Raman transitions by employing a $\pi/2 - \pi - \pi/2$
sequence (Fig.~\ref{fig:kasevich1}).  The $\pi/2$ pulses act as beam
splitters, and the $\pi$ pulse acts to completely change the state
and reverse the differential momentum in each arm of the
interferometer in essence a three-grating interferometer. Similar to
the Ramsey-Bord\'e interferometer, the paths have internal state
labels. The interference is detected as oscillations in the
population of the different internal states after the
interferometer, as measured with state-sensitive fluorescence or
ionization.  Since the gratings are pulsed in time such an
arrangement is a \emph{time domain} atom interferometer.  These
experiments employed atomic fountains for Na \cite{KACS91,KAC92} or
Cs atoms \cite{PCC99} to permit longer interaction times. In the
first experiments (with Na) a time delay between pulses of 100 msec
gave a wavepacket separation of 6 mm (cf. 66 $\mu m$ for thermal
beams with fabricated gratings \cite{kei91}). Chu and coworkers have
refined this technique to get high contrast ($C$=65\%) fringes with
a count rate of $\langle I \rangle = 10^6$ atoms per second. This
allowed measurements of $g$ at the part-per-billion level
\cite{PCC99,PCC01}. The theory of this interferometer is discussed
in detail by \textcite{BER97-YKC} and \textcite{KAC92}. Higher order
Raman transitions can be stimulated with multiple pulses, and
momentum differences of 60 $\hbar k_{ph}$ have been used for
interferometry \cite{WYC93}.

A beam experiment using the same kind of Raman transitions was used
by \textcite{GBK97,GLK00} to measure rotation rates, and achieved
short-term sensitivity to rotations of $6 \times 10^{-10}
$(rad/s)/$\sqrt{\mathrm{Hz}}$ as discussed in Section V.A..  In this
configuration, the gratings were not pulsed, so this was a
\emph{space domain} interferometer.

We discuss several other applications of this kind of
interferometer, like precise measurements of gravity gradients
\cite{SMB98,MFF02}, Newton's constant $G$, and the value of
$\hbar/M$ \cite{PYC97,WYC93} in Section V on precision
measurements.

\subsection{Longitudinal interferometry}
The standard description of Ramsey's separated fields experiment
treats the longitudinal motion classically and as being the same for
both states.  This is obviously incorrect if the states have
different magnetic moments and the beam passes into a region with a
different magnetic field - the field gradient puts a different force
on components with different magnetic moments, and could even
reflect one state but not the other.  Another source of longitudinal
energy shift is excitation by RF radiation whose frequency is below
(or above) resonance: the remaining energy to excite the atom comes
from (or goes into) the kinetic energy of the excited state
component. In fact, the transition can be made by a gyrating field
with zero temporal frequency, especially if the beam is moving fast
so that the spin can't follow the field as it passes.  We discuss
these cases below.

\subsubsection{Stern Gerlach interferometry}

\begin{figure}[b]
\begin{center}
\includegraphics[width = 8cm]{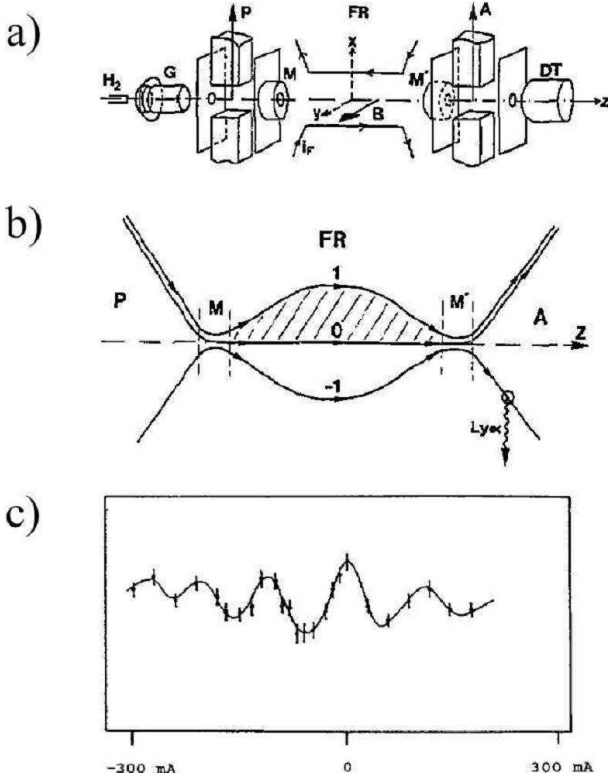}
\caption{Longidudinal Stern Gerlach interferometer.  (a)
Experimental set-up: source G, polarizing and analyzing magnetic
fields P and A, mixers M and M', frame FR with current $i_F$ is
creating a magnetic field $B$, detector DT. (b) The energy landscape
for the Zeeman states (-1,O,1) of H* (2s$_{1/2}$, $F=1$) along axis
$Z$. (c) Interference pattern obtained with a transverse magnetic
field in region FR. Figure from \cite{CWM93}.} \label{fig:sg1}
\end{center}
\end{figure}

While a Stern-Gerlach magnet can entangle an atom's spin and
momentum \emph{transverse} to the beam velocity, it is difficult
to redirect and recombine amplitudes along these two paths
\cite{REI99,SES89,ESS88,SSE88}. In a different geometry, atoms in
a beam can be split \emph{longitudinally}, so that components of
each atom are separated along the direction of the beam velocity.
This is easy to accomplish, and has the advantage (for
interferometry) that the two paths overlap
\cite{RMB91,MPV91,DDS95}.

A longitudinal Stern-Gerlach interferometer from \textcite{RMB91} is
shown in Fig.~\ref{fig:sg1}. A partially polarized beam of
metastable hydrogen atoms  in the state 2s$_{1/2}$, F=1
($\lambda_{dB} = $ 40 pm) is prepared in a linear superposition of
magnetic sub-levels by a non-adiabatic passage (projection on the
new eigen-states) through a magnetic field perpendicular to the
atomic beam. The magnetic field gradient along the beam shifts the
longitudinal momentum of different atomic center of mass wave
packets proportionally to their magnetic state. Next, the different
magnetic sub-levels enter a constant magnetic field region, and
after 10 cm are recombined again in a region identical to the one
used as a beam splitter. Finally, an analyzing magnetic field
selects a particular magnetic polarization, whose intensity is then
measured by detecting Lyman-$\alpha$ photons emitted in the decay of
the 2p$_{1/2}$ state to the ground state.  A typical interference
pattern is shown in Fig.~\ref{fig:sg1} \cite{CWM93,RMG92}.

Interference fringes are obtained in the beam intensity by
changing the magnetic field strength, and arise from the different
potentials experienced by the magnetic sublevels in the region of
constant magnetic field. The longitudinal Stern-Gerlach
interferometer was applied to demonstrate the effect of
topological phases on the atomic wavefunction for a non-adiabatic
cyclic evolution \cite{MRG92}.

\subsubsection{Spin echo}

\begin{figure}
\begin{center}
\includegraphics[width = 7cm]{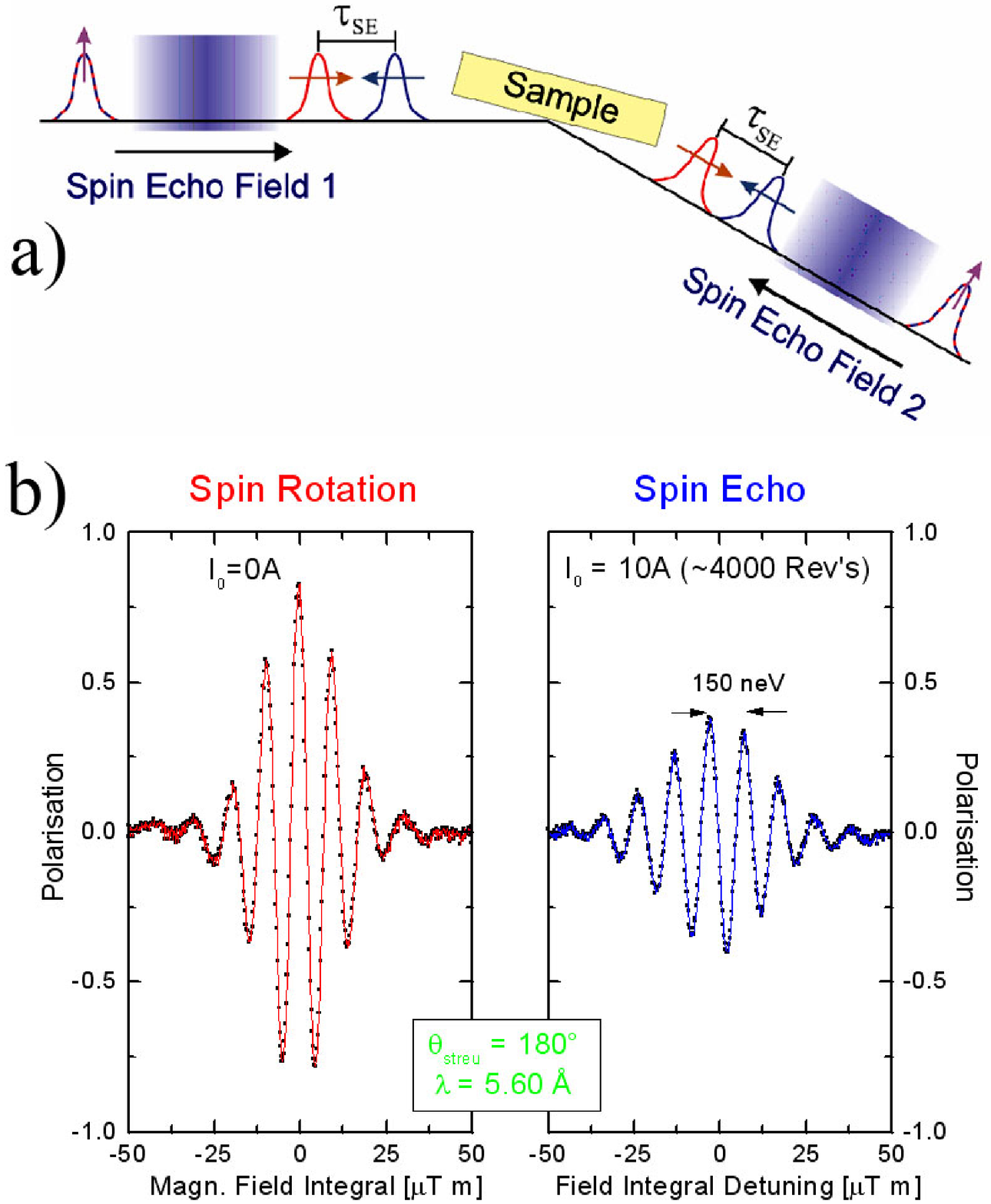}    
\caption{(color online) Atomic Beam Spin Echo interference technique
for a spin-1/2 particle. (a) Schematic of setup: upon entering spin
echo field 1, the linearly polarized wave packet \ket{\uparrow} is
split into two polarizations \ket{\rightarrow} and \ket{\leftarrow},
having different energies in the longitudinal magnetic field. By
inverting the direction of the spin echo field 2 with respect to the
first one, the Zeeman states \ket{\rightarrow} and \ket{\leftarrow}
exchange roles (like a $\pi$-flip). At the end they overlap and
coherently add up to \ket{\downarrow} or \ket{\uparrow} depending on
the phase shift. The initial linearly polarized wave packet
reappears as an echo. (b) Experimental ABSE data using a 4 K beam of
$^3$He atoms. Plotted is the beam averaged linear polarization as a
function of the spin echo field. (spin rotation): when spin echo
field is off, the interference pattern is generated through
Stern-Gerlach interferometry. (spin echo): when the same (but
inverted) current is applied through both spin echo coils an echo
appears. Figure curtesy of M. DeKieviet} \label{fig:ABSE}
\end{center}
\end{figure}

Along similar lines, \textcite{DDK97} developed an atomic beam
spin-echo (ABSE) interferometer with $^3$He atoms.   Following the
Stern-Gerlach arrangement described above one can apply a reversed
field (or a $\pi$-pulse) and extend this type of interferometer
with an ``echo", in complete analogy to the spin echo technique
used for neutrons \cite{MEZ93}.  The $^3$He ABSE has the advantage
that $^3$He can reflect from a surface at grazing incidence, and
therefore can be applied as interferometric probe of surfaces
\cite{ZSS98,DDS95,DDH00,DDK97}.

In a quantum mechanical picture, the Larmor precession can be viewed
as a magnetic birefringence (Fig.~\ref{fig:ABSE}a). Note that the
Zeeman states \ket{\rightarrow} and \ket{\leftarrow} arrive with
some time delay $\tau_{SE}$ (spin echo time) at the scattering
center, which allows time-correlation spectroscopy of the sample.
The contrast in the measured echo signal depends then on the degree
to which the Zeeman states are scattered coherently. For non-static
samples this will depend on $\tau_{SE}$ (see Fig.~\ref{fig:ABSE}b).
The interference contrast directly measures the correlation function
$I(\mathbf{q},\tau_{SE})$ in the time domain, which is the Fourier
transform of the scattering function $S(\mathbf{q}, \tau_{SE})$
($\mathbf{q}$ is determined by the scattering geometry). ABSE with
$^3$He atoms has been successfully applied in surface science as an
appealing alternative to time-of-flight experiments. The spin echo
experiment is much more sensitive, with an energy resolution
extending into the sub-neV-range.

ABSE is not restricted to longitudinal interferometry; depending on
the direction of the magnetic field gradient, the paths of the
magnetic sub-states may diverge perpendicular to the beam direction.
Using atomic hydrogen with a de Broglie wavelength of around 100 pm,
\textcite{LAN98} measured a transverse spin echo interference signal
for path separations exceeding 100 nm.

In an entirely optical setup a spin echo was demonstrated through
hyperfine pumping a thermal beam of lithium atoms \cite{ZSS98}.
Here, the spin echo is induced via a ``virtual magnetic field", by
applying a short pulse of intense, far detuned photons. The light
causes a shift in the hyperfine levels that depends linearly on
the quantum number $m_F$, just like Zeeman splitting
\cite{RSM90,COD72}.

\subsubsection{Longitudinal RF interferometry}

\textcite{DKR97} showed that a detuned radiofrequency field
constitutes a beamsplitter in longitudinal momentum space for
atoms.  If an atom makes a transition to an excited quantum state
by absorbing a quantum of off resonant RF radiation, then its
longitudinal velocity is changed such that total energy is
conserved.

\begin{figure}[h!]
\begin{center}
\includegraphics[width = 7cm]{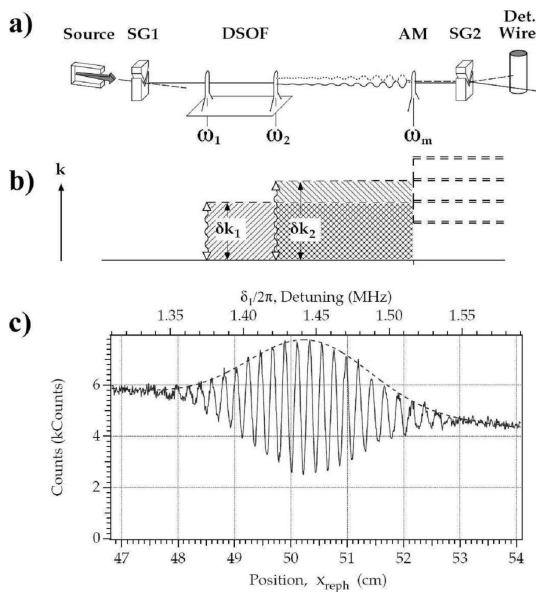}
\caption{Longitudinal RF interferometer (a) Schematic. Coils at
longitudinal positions $x_1$ and $x_2$ with oscillatory fields at
$\omega_1$ and $\omega_2$, respectively, make the differentially
tuned separated oscillatory fields (DSOF). The amplitude modulator
coil is located at $x_m$. The ground state is selected by upstream
Stern-Gerlach magnet SG1, and the excited state by SG2. (b) Wave
number $k$ versus the longitudinal position $x$ for states that are
detected. Dashed lines indicate the excited internal state, and
hatched areas denote the differential phases accrued by atoms
excited at $x_1$ ($x_2$). (c) Fringes demonstrated with the DSOF
system and an additional AM modulator \cite{SDK98}.}
\label{fig:long1}
\end{center}
\end{figure}

Using two such beam splitters \textcite{SDK98} constructed a
longitudinal atom interferometer in a generalization of Ramsey's
separated oscillatory fields (SOF) configuration. This technique is
referred to as differentially tuned separated oscillatory fields, or
DSOF. Oscillations in excited state population both in time and
space occur after an atom beam passes the two DSOF regions. To
measure the phase and amplitude of these oscillations, a third
oscillatory field and a state selective detector were used as shown
in Fig.~\ref{fig:long1}.

This interferometer is well suited to studying the longitudinal
coherence properties of matter-wave beams. Scanning the position of
the third oscillating field demonstrates that the DSOF system can
produce or detect coherent momentum superpositions.

The envelope of the fringes in space Fig.~\ref{fig:long1}
indicates the velocity width of the atom beam was 36 $\pm$ 4 m/s
and the fringe period in space indicates the most probable beam
velocity was 1080 $\pm$ 3 m/s.  An argon seeded supersonic source
of sodium atoms was used.

The same DSOF arangement was used to demonstrate the absence of
off-diagonal elements in the densiy matrix in a supersonic atom
beam, thus showing that there are no coherent wave packets
emerging from this source. \cite{RDK99}.  In a further
demonstration, the DSOF longitudinal interferometer was used to
measure the complete longitudinal density matrix of a deliberately
modulated atom beam \cite{DKR97,RKD99}. A fully quantum mechanical
treatment of this system was developed for this analysis
\cite{KDH98}, and these experiments are summarized by
\textcite{KBB00}.

\subsubsection{St\"{u}ckelberg interferometers}

St\"{u}ckelberg oscillations occur when a level-crossing for
internal states acts as a beam splitter.  For example, if an atom
can change its internal state on the way either to or from a
reflecting surface, then two amplitudes for making a transition will
interfere. Oscillations in the probability for state-changing atomic
reflection can thus be regarded as longitudinal interferometry. One
application is to survey the van der Waals potential near surfaces
\cite{CSH98,MCS00}.

\subsection{Coherent reflection}

\begin{figure}[h]
\begin{center}
\includegraphics[width = 7cm]{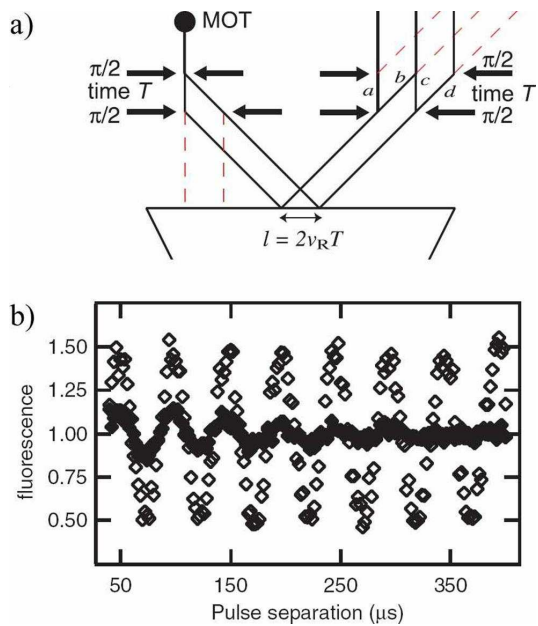}
\caption{An atom mirror inside an interferometer. (a) Diagram of the
interferometer. The arrows represent Raman $\pi/2$ pulses which
create superpositions of different internal states and momenta. The
atomic mirror is an evanescent wave at the surface of a glass prism
represented by the trapezoid. The letters a, b, c and d, label the 4
possible paths. (b) Fringes obtained by scanning the pulse
separation T with the mirror(filled symbols) and without the mirror
(open symbols).  Figures from \cite{ESA04}.} \label{fig:westbrook}
\end{center}
\end{figure}

Here we briefly list more experiments in which \emph{reflected} de
Broglie waves are demonstrably coherent. Shimizu demonstrated
reflection mode holograms \cite{SHF02} and a reflection-mode double
slit experiment \cite{KSF03}.  Westbook used a Raman-pulse atom
interferometer to study coherent reflection from an evanescent light
field \cite{ESA04} as shown in Figure \ref{fig:westbrook}. Dekieviet
used a longitudinal Stern-Gerlach interferometer to study quantum
reflection of $^3$He. \cite{DRD03}.  Zimmermann used a
chip-integrated magnetic grating to diffract and interfere reflected
BEC's \cite{GKK05,GKZ07,FOZ07}.

\subsection{Confined Atom Interferometers with BEC's}

In this section we discuss a different type of interferometer, where
the atoms are confined in a 3-dimensional potential well during the
splitting of their wave function and application of the interaction.
In this new type of splitting process, the single trap holding the
ultra cold gas of atoms (or BEC) is continuously deformed into two
adjacent potential wells, each containing a part of the wave
function.  Thus the splitting step in the interferometer occurs in
position space.

This splitting in position space is in sharp contrast to previously
discussed atom and most optical interferometers, in which the
splitting process occurs in momentum space. Using diffraction
gratings or pulses of light transfers momentum; similarly a
partially reflecting surface changes the momentum of the reflected,
but not the transmitted, beam.  The two maxima then separate to a
varying extent in position space only because the wave is split in
momentum space.  In the trapped atom interferometers discussed here
the atom waves remain confined and are separated by moving the
potential wells apart.

Important advantages of confined atom interferometers are manifold.
The confinement can support the atoms against gravity, offering
potentially unlimited experiment times with obvious advantages for
precision experiments. The location of the atom wave can be known
very precisely.  This is essential in experiments studying
atom-surface interactions like the Casimir potential, or for
studying spatially varying fields or interactions with small objects
that are coupled to the atoms via an evanescent wave. If a BEC is
confined, the large scale coherence allows new ways to measure the
relative phase of two condensates using just a small sample of the
atoms. Additionally, confined atom interferometers, especially those
using atom chips, can be small and portable.

Confined atom experiments differ qualitatively from the many
experiments that have been carried out using BEC's as a bright
source of cold atoms propagating in free space \cite{GDH02,TSK00}.
In those the physics is dominated by single-particle dynamics and
does not exploit the particular coherence properties of BECs.  In
the interferometers described here, the intrinsic properties of
the BEC allow novel measurements, and create new problems to be
overcome.

Confined atom interferometers naturally operate with significant
density to achieve the advantages of large signals, from which
several disadvantages follow. First of all, the matter wave optics
becomes non linear.  The atom-atom interactions lead to a mean field
potential (the chemical potential in a BEC) that can cause a
relative frequency shift between atoms in the two wells. In addition
the potential wells have to be controlled very accurately in
stiffness and depth, to prevent additional sources of systematic
frequency shifts.  (In waveguide interferometers where the atoms are
confined only in two directions, any residual potential roughness
gives additional problems.)

Splitting a condensate coherently produces a state whose relative
phase is specified at the expense of a superposition of number
states with different relative populations because of the
(approximate) number-phase uncertainty relation. Knowing the
relative phase of two condensates requires an uncertainty in the
relative number of atoms in each well, even though the total number
may be certain. The wave function in each well is therefore a
coherent superposition of states with different relative mean field
interactions (different relative chemical potentials) and therefore
evolve at different rates. The resulting dephasing limits the
coherence time to less than 50 ms for a typical million-atom BEC
(with diluteness parameter, $na^3\approx10^{-4}$).

In addition one has to carefully consider the collective
excitations of the condensate (e.g. sound or shape oscillations)
which may arise if the potential changes too suddenly.  This can
be overcome by applying techniques from coherent control as shown
in \textcite{HRB07}.

Recombining the split double well into a single trap allows in
principle the readout of the relative phase as a relative population
difference between ground state and first excited state
\cite{ACF02,HVB01}. In the recombination, the non linear
interactions lead to creation of (fast moving) solitons. These can
enhance the sensitivity \cite{NH04} of phase measurements, but are
much harder to control. Consequently the experiments recombine the
split waves by releasing them from the trap, then free expansion
reduces the non-linearity and facilitates the overlap.

Confined atom interferometers have so far come in two types: BEC's
confined to waveguides (i.e. in two dimensions) which are
described in the next sub section, and those confined in traps
(i.e. in all three dimensions) using focused light beams
(subsection 2) and magnetic fields generated by atom chips
(subsection 3). Finally in the last subsection we describe an
example where it is possible to establish and read out the
relative phase of two condensates that do not overlap during the
entire process and discuss whether this can be seen as a type of
interferometry involving two classical objects.

\subsubsection{Interference with guided atoms}

Given the existence of optical interferometers using fiber optical
wave guides, and the success in confining and guiding ultra cold
atoms, it is natural to consider similar designs for atoms. While
preliminary theoretical study shows that special designs should
allow multi-mode interferometers \cite{ACF02}, no interferometer
devices involving atom waveguide beam splitters have been
demonstrated.

\begin{figure}[t]
\begin{center}
\includegraphics[width = \columnwidth]{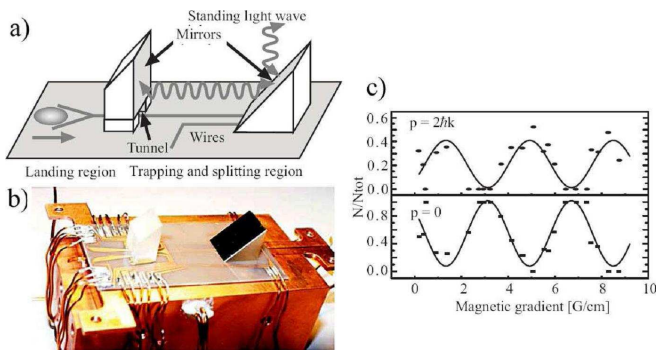}    
\caption{(color online) Michelson Atom Interferometer. (a) Schematic
drawing of the atom chip (not to scale). The prism-shaped mirrors
are integrated with micro fabricated wires on an aluminum nitride
substrate. The dimensions of the whole chip are 5 cm by 2 cm. (b)
Photo of the atom chip on its copper holder. (c) Interference
fringes after 1 ms propagation time in the waveguide with the
magnetic gradient turned on for 500 $\mu$s while the average
separation of clouds is 8.82 $\mu$m. Figures and caption reproduced
from \cite{WAB04}. } \label{fig:michelson1}
\end{center}
\end{figure}

The first waveguide atom interferometer, by \textcite{WAB04} and
improved on by \textcite{GDH06}, was designed to test coherent
propagation in atom waveguides, not waveguide beamsplitters.  It was
a familiar three grating interferometer in which pulsed light
gratings split and recombined a BEC confined in a weakly confining
(magnetic) guide along the axis.  The BEC is split at $t=0$ into two
momentum components $\pm 2 \hbar k_L$ using a double pulse of a
standing light wave. A Bragg scattering pulse at $t=T/2$ then
reverses the momentum of the atoms and the wave packets propagate
back.  At $t=T$ the split wave packets overlap and a third
recombining double pulse completes the interferometer. The output
port is given by the momentum of the atoms as detected by imaging
(typically 10 ms) after release from the guide. To apply a phase
shift between the two arms of the interferometer, a magnetic field
gradient was turned on for a short (500$\mu$s) time while the atom
clouds were separated. In the original experiment \cite{WAB04} the
propagation time in the interferometer was varied from $T=1$ ms to
$T=10$ ms. The contrast of the fringes was as high as 100\% for
$T=1$ ms, but droped to 20\% for $T=10$ ms. The degradation of the
contrast is mainly due to the non linear term coming from the
interaction between the atoms.  By reducing the transverse
confinement and consequently the non linear interaction
\textcite{GDH06} reached much longer coherent propagation up to 180
$\mu m$ and times up to 50 ms.

There is ample optical precedent for waveguide interferometers using
2-dimensional confinement since there is wide application of optical
fiber interferometers both scientifically and commercially.  On the
other hand, interferometry with 3-dimensionally trapped atoms has no
precedent in light optics\footnote{One could argue that a
Fabry-Perot is an (imperfect) trap for photons and that the LIGO
interferometer which uses Fabry-Perot interferometers nested in a
Michelson interferometer is not far from this precedent.}.

\subsubsection{Coherent splitting in a double well}

Three dimensional trapped atom interferometers are a qualitatively
new type of interferometer without precedent in optics since it is
not possible to trap photons, move the trap around, and then
somehow recombine the photons.  A trapped atom interferometer does
just that.

Coherent splitting of the wave function by slowly deforming a single
trap into a double well is the generic trapped atom beam splitter,
achieving physical separation of two wavefunction components that
start with the same phase.  When the two wells are well separated,
an interaction may be applied to either.  Finally the split atoms in
the two wells are recombined to observe the interference.

\begin{figure}[t]
\begin{center}
\includegraphics[width = 8cm]{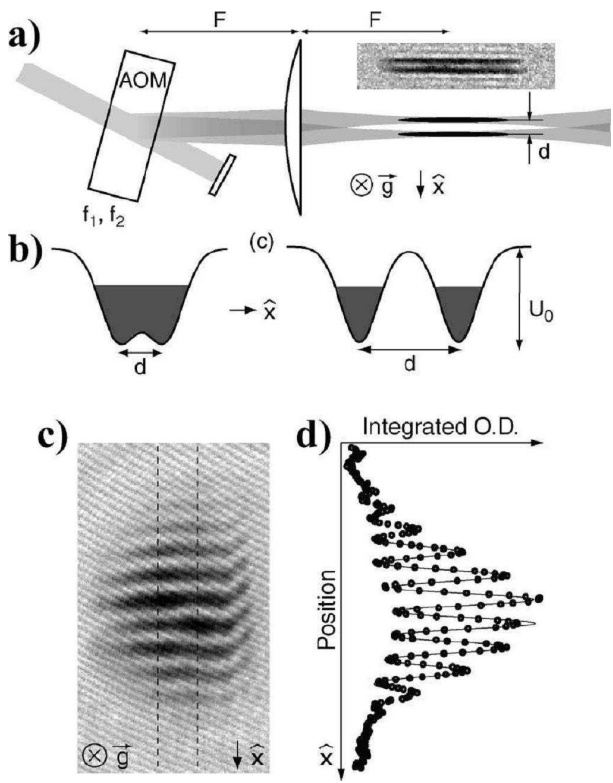}
\caption{Top:  Optical double-well potential. (a) Schematic diagram
of the optical setup for the double-well potential. The insert shows
an absorption image of two well-separated condensates in the
double-well potential (he field of view is 70 $\times$ 300 $\mu$m).
(b) Energy diagram, including the atomic mean field, for the initial
single-well trap with $d$=6 $\mu$m and for the final double-well
trap with $d = 13 \mu$m ($U_0 = 5$kHz, atomic mean field energy  $
\sim 3$kHz, potential ``barrier''). (c)  Absorption image of fringes
created by condensates released from the double-well potential
immediately after splitting (30 ms of ballistic expansion, field of
view $600 \times 350 \mu$m). (c) Density profile obtained by
integrating the absorption signal between the dashed lines. Figure
and caption adapted from \cite{SSP04}.} \label{fig:doublewell}
\end{center}
\end{figure}

Such coherent splitting was first demonstrated by \textcite{SSP04}
who split a BEC by deforming an optical single-well potential into a
double-well potential.  A BEC was first loaded into the single trap
and allowed 15 sec to damp its excitations.  The splitting was done
over ~ 5 ms, slowly enough compared to a 600 Hz transverse
oscillation frequency in the trap not to excite substantial
transverse excitation of the two new condensates, but not slowly
enough that the mean field interaction would cause the atom to
divide exactly evenly between the two wells (with exactly $N/2$ on
each side there would be no number uncertainty and hence the
relative phase would have been indeterminate).

The interferometer was completed by releasing the trapped
separated BEC's and determining their relative phase from the
resulting fringes.  Releasing the condensates dramatically lowers
the mean field interaction prior to overlap, hence averting
problems arising from the non-linearity of atom optics. Another
big advantage is that overlapping two BEC's produces high contrast
fringes, enabling an accurate determination of the phase from each
``shot" of the interferometer.

Observing the fringes in repeated experiments, starting with fresh
condensates each time, addressed the key question: is the relative
phase between the split condensates random or consistent from shot
to shot? There had been some theoretical controversy on this
subject. The fringes observed when the load, split and immediate
release sequence was repeated were in the same place, showing that
the relative phase between the two condensates was consistent,
i.e. that is can be controlled deterministically.  It was also
shown that the phase evolved coherently for up to 5 ms.

The condensates were separated by 13 $\mu m$ in these experiments,
and the single atom tunneling rate between the two wells was
estimated to be ~ $5 \times 10^{-4}$ s$^{-1}$, sufficient to uncouple the
BEC's in separated wells and let their phases evolve
independently. It was verified that each condensate evolved phase
independently and was phase shifted as expected by a local Stark
shift.

This experiment showed definitively that splitting the well led to
BEC's with a common phase, introduced a new method to determine
the phase that was not affected by mean field interactions, and
showed that coherence could be maintained for several oscillation
periods of transverse condensate motion.

\subsubsection{Interferometry on atom chips}

The combination of well established tools for atom cooling and
manipulation with state-of-the-art micro fabrication technology has
led to the development of atom chips \cite{FKS02}. Atoms are
manipulated by electric, magnetic and optical fields created by
micro-fabricated structures containing conductors designed to
produce the desired magnetic and electric fields. Technologically,
atom chip based atom interferometers promise to be relatively
inexpensive and presumably are relatively robust. Atom chips have
been demonstrated to be capable of quickly creating BEC's and also
of complex manipulation of ultra cold atoms on a micro scale. We
trace here the development of techniques to coherently split the
condensate and perform atom interferometry.

Many basic interferometer designs and beam splitters on an atom chip
were conceived and tested \cite{FKS02}. Most of them rely on
splitting a magnetic potential in multi-wire geometry. The first
experiments demonstrating splitting, but not coherence, were carried
out in Innsbruck 1996-1999 with splitting a guide with a Y-shaped
wire \cite{CCF00,CHF00} and a trap with a 2-wire configuration
\cite{FKS02}.

\begin{figure}[t]
\begin{center}
\includegraphics[width = 8cm]{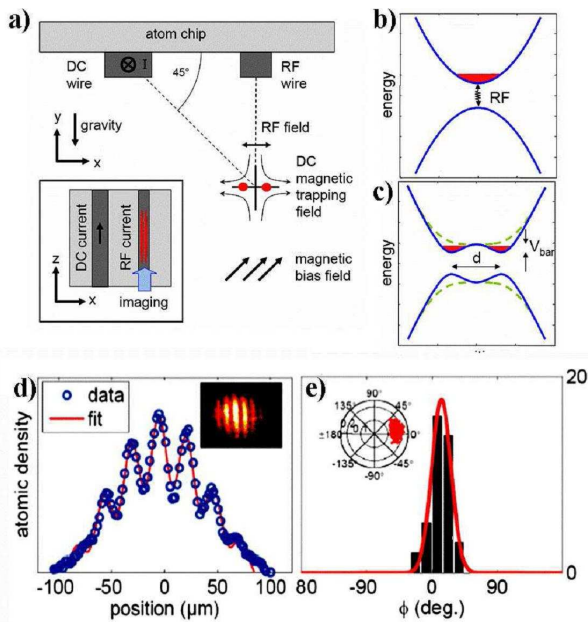}    
\caption{(color online) Coherent splitting with an RF induced double
well on an atom chip.  (a) A wire trap is split by coupling the
magnetic substates by RF radiation. To achieve the correct
orientation (splitting orthogonal to gravity) the trap is rotated
and placed directly over the RF wire. (b,c)  The energy landscape
before and after splitting. (d) Interference is observed by
switching the trap of, and letting the atomic cloud overlap in time
of flight. The image integrates over the length of the condensate.
(e) Observed distribution of fringe phase and contrast obtained from
multiple experiments \cite{SHA05}.} \label{fig:AtomChipDW}
\end{center}
\end{figure}

At MIT interference with random phase using such a two wire setup
was observed by \textcite{SSJ05}. Simultaneously the first coherent
splitting of trapped micro manipulated atoms on atom chip was
acieved by \textcite{SHA05} at Heidelberg, using radio frequency
induced adiabatic potentials \cite{ZOG01,CKM04,LSH06,LHS06}.
Analyzing interference patterns formed after combining the two
clouds in time-of-flight expansion, demonstrated that the splitting
is coherent (i.e. phase preserving) Figure \ref{fig:AtomChipDW}.

The splitting using radio frequency induced adiabatic potentials
as developed in Heidelberg overcomes the disadvantages of the 2
wire setup: weak confinement during the splitting, and extreme
sensitivity to magnetic field fluctuations. The new method allows
very well controlled splitting over a large range of distances -
from 2 to 80 $\mu m$ - thus accessing the tunneling regime as
well as completely isolated sites.

The Heidelberg experiments \cite{SHA05,HLF06} are remarkable since
they were performed with 1d BEC (chemical potential $\mu < \hbar
\omega_\perp$), much longer then the phase coherence length.
Nevertheless the interference patterns persist for as long as the
condensate.  All different regimes from physically connected to
totaly separated 1d BECs were accessible, and phase locking by
coherent tunneling in the intermediate regime could be demonstrated.

With continued progress on these topics, together with techniques
for reducing dephasing of interferometers using BECs,
interferometers using confined atoms hold the promise to be
employed as highly sensitive devices that will allow exploration
of a large variety of physics questions. These range from
exploring of atom-surface interactions to the intrinsic phase
dynamics in complex interacting (low dimensional) quantum systems
and the influence of the coupling to an external `environment'
(decoherence).

%
%

\section{FUNDAMENTAL STUDIES}  \label{sec:fundamental studies}

In this chapter, we address two questions that lay people often ask
once they have understood the basic ideas of atom interferometry:
``Can you make interferometers with any object, people for example?"
and ``Of what use are atom interferometers?" We discuss the limits
to particle size in section A, experiments that probe the transition
from quantum behavior to classical behavior via the process of
decoherence in B, and how the ideas of single particle coherence can
be extended in D.  The question of utility is first addressed in
section C, where we show that measurable phase shifts arise not only
from potential differences, but from ``weirder" things like the
Aharanov-Bohm effect and topological transport in general.  Then we
describe how atom interference can be used to study four different
features of many-body systems in section E, and finally address
fundamental tests of charge equality for protons and electrons.  The
actual order of the sections does not reflect the answers to these
questions in sequence, however; rather the first three address
single particle questions, section D addresses extensions of
coherence first to extended single particles, and then to
multi-particle systems, and section E is devoted to describing
studies of many particle systems that reveal many-particle coherence
and decoherence processes, or in which atom interference is the tool
that enabled the study of their collective properties.

\subsection{Basic questions: How large a particle can interfere?}

When the first atom interferometers were demonstrated, some of our
colleagues expressed surprise that ``composite" particles would give
such high contrast fringes.  These sentiments are in line with the
idea that there exists a quantum-classical boundary and that somehow
there must be a limit on the number or spacing of internal states
(i.e. the ``complexity") for particles in an interferometer. Perhaps
the mass, the size of a molecule, or the strength of interactions
with the environment can limit or eliminate the interference. In
this section we investigate the limits to coherent manipulation of
the center of mass motion of larger and more complex particles, and
point to some interesting open problems. We shall first consider
practical limits set by particle size, grating size, and
interactions with the grating, and then move on to more fundamental
limits determined by interactions with the surrounding environment.

Experiments with Na$_2$ molecules \cite{CEH95} demonstrate that
particles with many internal states show interference fringes even
if the paths go on opposite sides of a thin conductor.   These
experiments also confirm what the first atom interferometers showed:
interference fringes can be observed when the size of the particle
is considerably larger than both its de Broglie wavelength and its
coherence length. For example in the separated beam interferometer
with Na$_2$ $\lambda_\textrm{dB} \approx $ 10 pm and the coherence
length $l_\textrm{coh} \approx $ 100 pm  are both much smaller then
the size of the molecule ($\sim $400 pm).  For the experiments with
$C_{60}$ or larger molecules the parameters are even more extreme
\cite{AHZ05,ANV99,ANP01,HUH03,BAZ03,Clauser1997a}.

Perhaps a bit more surprising is the observation of fringes in
Talbot-Lau interferometers with hot particles like $C_{60}$, the
surprise being that they have a spontaneous emission rate fast
enough to emit IR photons during the interference process.  But
since the maximum separation of the paths in these experiments
(about a grating period) is much less than the wavelength of the IR
radiation, a few photons of emitted radiation cannot be used to
localize the molecule to one path or the other \cite{HHB04,HOR06}.
Thus the interference is between two spatially separated paths along
which the molecule emitted a photon and changed from internal state
\ket{i} to final state \ket{f}. Interestingly, IR emission would
localize a molecule on one side or the other of a conducting plate,
so hot molecule interference would not occur between paths separated
by a conductor\footnote{Of course a separated path interferometer,
not a Talbot-Lau interferomter would be needed for this
experiment.}. This makes an important point: information left in the
environment is sufficient to destroy the coherence; no actual
measurement by a macroscopic apparatus is necessary.

Even though a particle's size itself poses no fundamental limit to
matter wave interferometry, there are more practical limitations to
interferometry with large particles, such as 1.~the time required to
propagate through an interferometer, 2.~the requirement that the
particles fit through the openings on material gratings without
undue effects from Van der Waals interactions, and 3.~whether
laser-based beamsplitters can work with particles larger than the
laser wavelength.

The time it takes a diffracted particle (with one grating
momentum, $\hbar G$) to move one grating period sets the
characteristic time for interference of a particle of mass $m$,
\beq t_{char}=\frac{d}{\hbar G/m}=\frac{md^2}{h} =  \frac{\hbar}{2
E_G} \label{eq:t_char}\eeq where $E_G$ is defined as in Eq.
\ref{eq:EG}.  For a grating period 100 nm and a flight time of one
second this limits the mass to $\sim$ $10^{-17}$g, or about one
million Na atoms.  Such a cluster would have a size of $\sim 30$
nm and would just fit through the gratings.  For the 0.01s flight
times characteristic of current Talbot-Lau interferometers, this
limit would be around atomic mass $10^5$, about an order of
magnitude heavier than current practice. Increasing the time by an
impractical factor (e.g. to a year, with concomitant inertial
stabilization of the gratings) does not improve the mass limit
proportionately.  The reason is that the grating period has to be
increased to accommodate the diameter of the particle
\cite{BER97,HEK98,HEK00,STS98} which grows as $m^{1/3}$.  Thus a
year-long interferometer can barely interfere a large bacterium as
pointed out by \textcite{BER97}.

While this discussion of size/mass limits applies quite accurately
to Talbot-Lau interferometers, the requirements of a separated
beam interferometer are several times more stringent.  In order to
separate the paths the beam must be collimated to better than the
diffraction momentum, which requires that the beam (and its
transverse coherence length) be several grating periods wide. To
separate these wider beams, the particles must propagate for
several characteristic times.  Even worse, the flux of particles
will be dramatically reduced due to the tight collimation.  In
contrast, Talbot-Lau interferometers have no restriction on their
width. Not surprisingly they are the interferometer of choice for
demonstrating interference of heavy particles.  And even with
them, it will be some time before sentient beings can be sent
through an interferometer and subsequently asked which path they
took.

While equation \ref{eq:t_char} shows that if molecules spend too
little time in the interferometer, they will not exhibit quantum
interference \cite{OBR96}; on the other hand, if particles spend too
long interacting with mechanical gratings, they will interact with
the grating bars, or be diffracted into very high orders.  This is
because of Van der Waals or Casimir-Polder interactions between
molecules and the grating bars \cite{GST99}. To keep half the
diffracted molecules in the central $n$ orders requires \beq \left[
\frac{\partial }{\partial r}V(r) \right|_{r=d/8} < \frac{h n v}{d^2}
\label{eq:vdwn}  \eeq where $V(r)$ is the atom-surface interaction
potential.  Equation \ref{eq:vdwn} assumes a grating with an open
fraction of 50\% and a grating thickness equal to the grating period
($d$) \cite{PCS05}. \textcite{HSA04} and \textcite{BAZ03} discussed
how the useful range of molecular velocities for a TLI gets severely
restricted for large molecules or small gratings. Van der Waals
interactions also set a minimum mechanical grating period for Sagnac
gyroscopes. For a large Sagnac response factor, one would naturally
select small grating periods. However, Van der Waals interactions
cause the uncertainty of a Sagnac rotation sensor to increase if
grating periods smaller than 44 nm are used with 1000 m/s Na atoms.
For helium atoms, which have much weaker vdW interactions, the
optimum grating period for a rotation sensor is 8 nm, about ten
times smaller than current practice. This is discussed for a MZI by
\textcite{CWP05}.

These limitations from grating bars and Van der Waals interactions
have lead to proposals for Talbot-Lau interferometers for large
molecules based on light gratings  \cite{BAZ03}. If the particle's
size is a large fraction of the wavelength, the light forces will
have gradients inside the particle that will excite the collective
oscillations of the particle unless the turn on/off time extends
over many periods of oscillation. For even larger homogeneous
particles the light force averages out to nearly zero.  This can be
overcome by localizing the interaction [e.g. with a color center
\cite{HSA04,NBA01}] or by making particles with periodic structure
on the scale of the wavelength.  Nevertheless the question of how
much internal excitation will occur still remains to be answered.
Finally, it should be possible to impart lots of momentum with long
wavelength photons by using multi-photon processes.

\subsection{Decoherence}

Quantum mechanics makes assertions so at odds with everyday
experience, that the mechanisms by which a quantum mechanical
treatment of macroscopic objects reduce to purely classical
behavior have long been considered a fascinating topic.  Indeed
wrestling with this problem has led a number of scientists to make
radical suggestions for changes in quantum theory itself (e.g.
spontaneous projection, pilot wave, etc.) or the nature of reality
(many worlds, etc.).  Observation of decoherence, and the
suppression, avoidance, control and correction of decoherence
mechanisms is a busy field made especially topical by the fruits
of, and need for, advances in quantum computation and
nanotechnology.

Atom interefrometry is based on coherence and therefore is sensitive
to interactions that upset this coherence.  Relative to neutrons,
atoms have large polarizability, magnetic moment, and scattering
cross sections and are therefore both more sensitive to, and easy to
use as quantitative probes for, decoherence processes.  In this
section we discuss atom interferometry's historical role in gedanken
experiments about quantum uncertainty and its present role in
providing an environment in which clean quantitative tests of
decoherence is possible.

\subsubsection{Interference and `Welcher Weg' information}

Perhaps the first general realization about interference fringes
was that they can easily be destroyed by interactions that, even
in principle, allow one to determine which path an atom took
through the interferometer. This is deeply rooted in Bohr's
principle of \emph{complementarity} which forbids simultaneous
observation of the wave and particle behaviors.  It is best
illustrated in the debate between Einstein and Bohr on the
question `can one know which path the particle took and still
observe the interference of the waves?' \cite{BOH49,WOZ79}.
Einstein proposed the famous recoiling-slit experiment to gently
measure which path the particle took through a two-path
interferometer.  In reply Bohr pointed out that the slit itself
must also obey the laws of quantum mechanics and therefore is
subject to the Heisenberg uncertainty principle. He showed
quantitatively that if the initial momentum of the slit-assembly
is known well enough to permit the recoil measurement of which
path the particle took, then the initial position of the slit must
have been so uncertain that fringes would be unobservable.

According to Feynman, this experiment ``has in it the heart of
quantum mechanics. In reality it contains the only mystery.''
\cite{FLS65}.  (Subsequently Fenyman acknowledged that entanglement
was another mystery.)  In 1960, Feynman proposed a related gedanken
experiment in which a perfect light microscope (i.e. one
fundamentally limited by Heisenberg uncertainty) is used to
determine ``which-way" information in a two-slit electron
interferometer by analyzing a single scattered photon \cite{FLS65}.
In Feynman's analysis of this gedanken experiment, electron
interference (a manifestly wave-like behavior) is destroyed when the
separation of the interfering paths exceeds the wavelength of the
probe (i.e. when it is possible to resolve on which path the
electron traversed). In fact the contrast is lost whether or not
anyone actually looks with the microscope; the ability in principle
to identify the electron's path is enough to destroy the
interference pattern. Feynman concludes,
\begin{quotation}  ``If an apparatus is capable of determining which
hole the electron goes through, it \emph{cannot} be so delicate
that it dos not disturb the pattern in an essential way."
\end{quotation}

More recently, a quantitative  duality relation was derived by
\cite{JSV95} and \cite{ENG96} to quantify how much `which-path'
knowledge ($K$) can be obtained and how much contrast ($C$) can be
observed at the output of an interferometer.
  \beq K^2 +  C^2 \le 1 \eeq
It is based on the analysis of a detector that quantifies how well
the two paths can be distinguished.  The detector could be similar
to Feynman's light microscope, as studied theoretically by
\textcite{TAW93,TEG93,HMM96,WHC97,STP95,GOG96,SAI90} and examined
experimentally by \textcite{CHL95,KCR01,CLL94,MEW01}.
Alternatively the detector could monitor spin polarization or the
internal state of atoms as proposed by \textcite{SEW91}, discussed
by \cite{LUS98,ESW00,BAR00} and examined experimentally by
\textcite{DNR98,DNR98b,DUR00}.  We also note the similarity with
many neutron spin-superposition experiments
\cite{BRS83,SBR83,BOR79,BRS88}.

Modern decoherence theories no longer invoke Bohr's collapse
postulate, and they do not rely on the uncertainty principle.
Instead they treat quantum systems (such as atoms in an
interferometer) as being coupled to their environment (including the
which-way detector) together as one combined (open) quantum system.
In this view, the interaction between the observed quantum system
and its (quantum) environment is a unitary process that causes
entanglement so that the state of the observed quantum system
becomes correlated with the quantum state of the environment.  Then
a measurement made on the environment allows inferences on the
quantum system.  For example, if a photon in the environment allows
an inference of which path the atom took, then a trace over the
environment would reduce the coherence remaining in the atom density
matrix, even if the coupling interaction were now turned off. For
more details we refer the reader to a set of excellent articles by
\textcite{JOZ85,ZUR91,ZUR03,TAW93,TEG93} and books by
\textcite{GJK96} and \textcite{WHZ83}

Since atoms couple strongly to electromagnetic fields in a
well-understood way, atom interferometers provide ideal tools for
studying decoherence.

\subsubsection{Internal state marking}

The simplest way of measuring an atom's `path' through the
interferometer is by marking it with an internal state of the
atom.  This is analogous to an interferometer for light where the
polarization is rotated in one arm. Measuring the internal state
of the atom then determines which path it took, and consequently
destroys the interference.

For example, \textcite{DNR98,DNR98b} studied the complementary
nature of fringe contrast and path information by using atoms
prepared in a superposition of internal states before they pass
through an interferometer for their external (center of mass)
states.  The interferometer was based on Bragg diffraction gratings
that affect the internal states differently so that the
interferometer paths became correlated with internal states. This
caused a controllable amount of contrast loss, based on how well the
internal states labeled which path the atom took.

These experiments are very similar to earlier neutron
interferometer experiments where loss of interference caused by
correlations between spin polarization and interferometer path was
studied \cite{BRS83}. In both the atom and neutron experiments the
coherence can be retrieved \cite{SBR83,DUR00,DUR00b} by projecting
the internal state vector onto a measurement basis that does not
allow one to distinguish the encoded internal states. The path
information is thereby erased and the full interference contrast
regained. This is a nice demonstration that interference will be
lost if the internal states contain which-path information; the
loss of interference occurs without invoking any coupling to an
external environment.

To substantiate that there is no coupling to the environment, note
that the transitions to prepare the internal state label are driven
with microwave fields that are in coherent states with large photon
number uncertainty, and hence one can not use a measurement of the
microwave field itself to get information about whether the atom
absorbed a single photon on the labeled path. Thus no information
about the internal state is transferred to the environment.  The
coherence is not really gone, it is hidden behind the choice of what
to measure (interference or path). One can easily get it back by
rotating the basis for the measurement, so that the `which path
information' is erased, as it was done in the beautiful experiments
by \textcite{BRS83,DUR00,DUR00b}.

This is different from the decoherence described by the recoiling
slit or Feynman's microscope discussed above. There one has to
look into the environment to get the coherence back. One has to
find the other part of the entangled state.

\subsubsection{Coupling to an environment}

We now discuss situations in which the interferometer looses
coherence because of coupling to the environment.  It is closely
related to modern theories of decoherence as will become obvious. As
an example, consider that the initial state involves an atom
traversing an interferometer and a well-collimated photon incident
on the atom; then the final state may involve an atom at the
detector and a photon in the environment traveling toward infinity.
This is a prototypical example of an interferometer that becomes
entangled with an external environment or particle. The interaction
and its strength is well known, but the final state is unknown.

\paragraph{Decoherence in Diffraction}

Several experiments have demonstrated decoherence due to spontaneous
emission of light quanta. \textcite{GMR91,PSK94,KSZ00} used atom
diffraction patterns caused by diffraction from a grating to observe
how the spatial coherence of an atom beam gets reduced by
spontaneous emission of a photon. A good picture is that the recoil
from the spontaneously emitted photon shifts the momentum of each
atom randomly, along with its individual diffraction patterns. Since
the direction of the final photon is random, these experiments
revealed a decrease of contrast of the summed patterns. There was a
transition from diffraction to diffusion with increasing probability
of spontaneous emission. In a similar spirit, the visibility in the
diffraction patterns of fullerenes $C_{60}$ and $C_{70}$ has been
used to bound the amount of decoherence for the molecule waves
caused by emitting thermal photons \cite{HHB04}.

All of these experiments can be perfectly explained by the random
momentum kicks given by the spontaneously emitted photons.
Interestingly the result is the same regardless of the place of
the photon emission, as long as it is at or upstream of the
grating. Consequently the effect is the same as if the incident
beam had a wider transverse momentum distribution, with associated
smaller transverse coherence length.

\paragraph{Decoherence in Talbot Lau interferometer}

In a three grating Talbot Lau interferometer, \textcite{CLL94}
showed that resonant laser light scattered from atoms in the
middle of the interferometer can destroy fringe contrast. This
experiment actually detected the fringes by selectively destroying
the contrast for different velocity classes that were Doppler
shifted into resonance with a laser beam.

More recently, but in a similar spirit, \textcite{MEW01}
demonstrated that photon scattering in a multiple beam Ramsey
interferometer also leads to decoherence for the atoms that
scatter light.  Furthermore, because some of the multiple paths in
this experiment cause fringes that are out of phase with the other
two-path combinations, it was shown that decoherence of one beam
can either increase or decrease the net contrast.

\textcite{HHB04} observed decoherence of internally hot Fullerene
matter waves caused by emission of radiation in a Talbot-Lau
interferometer.  This experiment is remarkable, since the emission
spectrum of the hot Fullerene is very close to thermal radiation,
and in that sense looks more like a (mesoscopic) classical
particle which `cools' internally by emitting photons during the
flight in the TLI.

All of these experiments can again be perfectly explained by the
(classical) random momentum kicks given by the spontaneously
emitted photons.

\paragraph{Photon scattering in an Interferometer}
\textcite{CHL95} studied the loss of coherence in a Mach-Zehnder
interferometer when each atom scattered exactly one photon. Loss
of contrast was observed which depended on the separation between
the two interferometer paths at the point of photon scattering.
This is a close realization of Feynman's gedanken experiment, and
we will discuss it below in detail.

\begin{figure}[t]
\begin{center}
\includegraphics[width = 8cm]{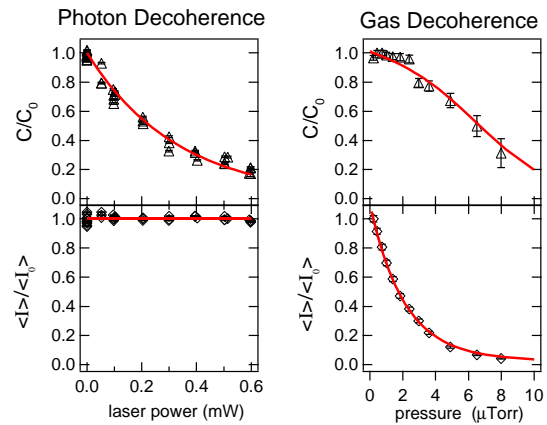}    
\caption{(color online) Comparison of decoherence from photon
scattering (left) to gas particle scattering (right).  Contrast
and atom beam intensity are reported as a function of the resonant
laser beam power or background gas pressure.  The light scattering
occurs where the separation $d=0.16 \lambda_{ph}$, and the gas
scattering occurs throughout the interferometer. The theoretical
curves come from equation \ref{eqn:betat} for the detected atoms
as discussed below. Figure from \cite{UPC05}.} \label{fig:gas
decoherence data}
\end{center}
\end{figure}

\paragraph{Scattering from background Gas in an Interferometer}
Scattering from a background gas of massive atoms or molecules has
also been used to cause a controlled amount of decoherence.
Collisional decoherence was observed by \textcite{HHB03,HUB03}
with Talbot-Lau atom interferometer, and similar work with a
Mach-Zehnder interferometer \cite{UPC05} is shown in Figure
\ref{fig:gas decoherence data}.

It is interesting to note the difference between the decoherence due
to photon scattering and atom scattering. The basic physics
processes are very similar, except that the momentum transfer is
much much larger in the case of the atoms and many of the
`collisions' lead to the atoms being scattered out of the detected
beam. Consequently the loss of contrast in atom collisions is not so
bad, but the overall intensity goes down significantly.  In addition
the atom-atom scattering is a probabilistic process, whereas the
photon scattering can be made deterministic (see \cite{CHL95}).
Additional theory work on collisional decoherence with massive
particles can be found in \cite{HOS03,HSA04,VAC04,FIH03,KLR01}.

Closely related to these atom scattering decoherence experiments
are the studies of stochastic or deterministic absorption and its
effect on coherence in neutron interferometers
\cite{SRT87,SRT88,RAS92,NPR93}.

\subsubsection{Realization of Feynman's gedanken experiment}

\begin{figure}[t]
\begin{center}
\includegraphics[width = 7cm]{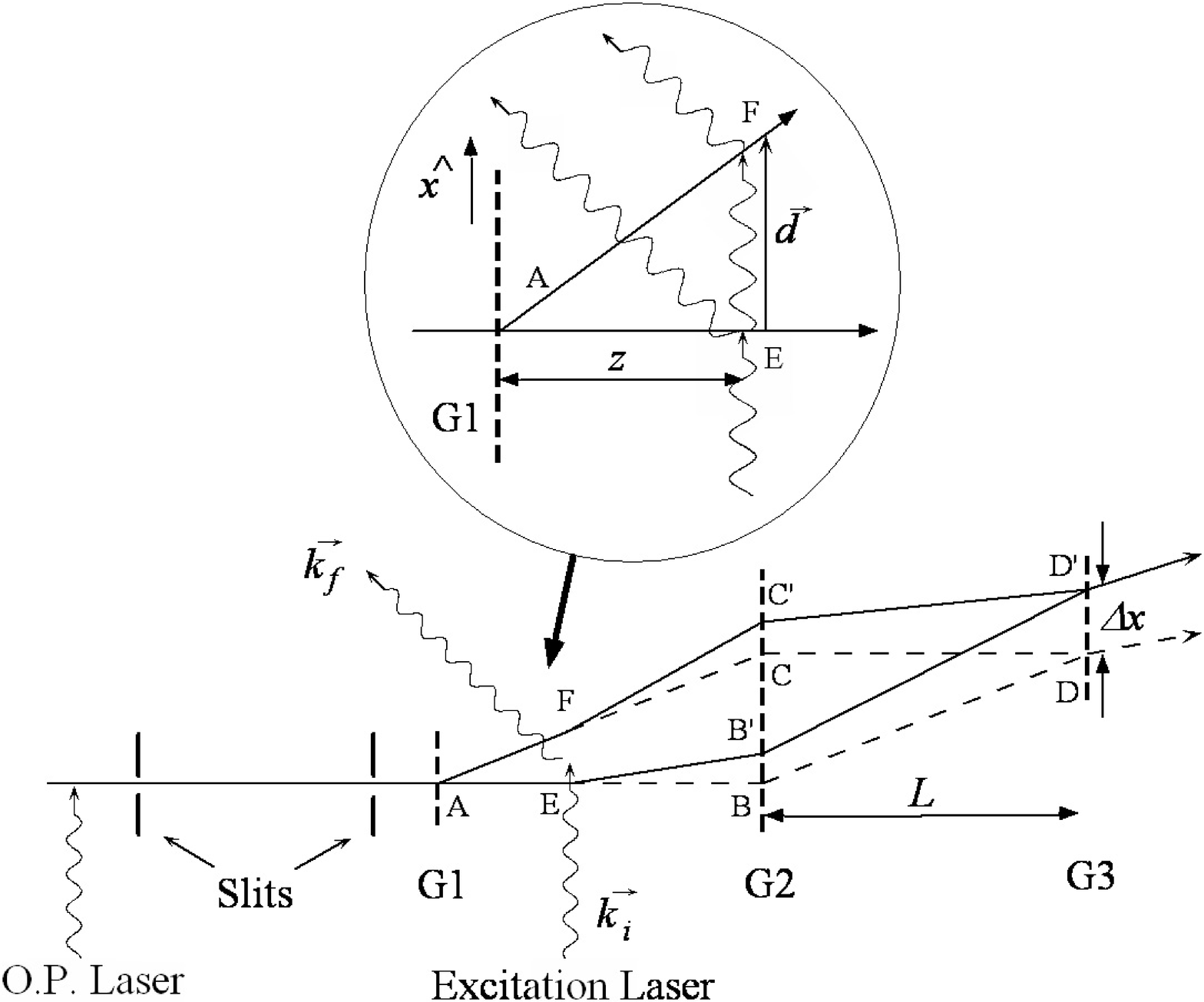}    
\caption{Schematic for the photon scattering decoherence
experiment in \cite{CHL95,KCR01}.  The path separation, $d$, and
the number of photons scattered per atom can both be controlled.}
\label{fig:1photonschem}
\end{center}
\end{figure}

Scattering a single photon from an atom in superposition of two
locations, is one of the icons of decoherence experiments.  It is
directly related to Feynman's gedanken experiment discussed above.
To realize such an experiment \textcite{CHL95} scattered single
photons from atoms within a two-path mach Zehnder atom
interferometer (Fig.~\ref{fig:1photonschem}).  Exactly one photon
was scattered by adjusting a tightly focused laser beam so that each
traversing atom made exactly half a Rabi cycle,  exiting the laser
beam in the excited state. To achieve this the transit time of the
atoms through the excitation laser ($T_\textrm{trans}\sim 5 ns$) was
much shorter then the lifetime of the excited state ($\tau \sim 16
ns$). Translating the laser beam along the interferometer caused
excitations at different locations corresponding to different
spatial separations of the interfering atom waves.

The experimental results are displayed in Fig.\ \ref{fig:1photon
decoherence data}. The contrast (which is a direct measure of
coherence) decreases smoothly towards zero as the distance between
the two paths grows to $d=\lambda/2$. At this point, the separation
between paths is equal to the Heisenberg microscope resolution. The
observed contrast recurrences at $d>\lambda/2$ have their
mathematical origin in the Fourier transform of the dipole pattern
for spontaneous photon scattering \cite{TAW93,HMM96,STP95,GOG96}.
Feynman, who might be surprised at their existence, would be
reassured to note that they occur where the prominent diffraction
rings of a perfect light microscope would lead to path ambiguity.

\begin{figure}[t]
\begin{center}
\includegraphics[width = 8cm]{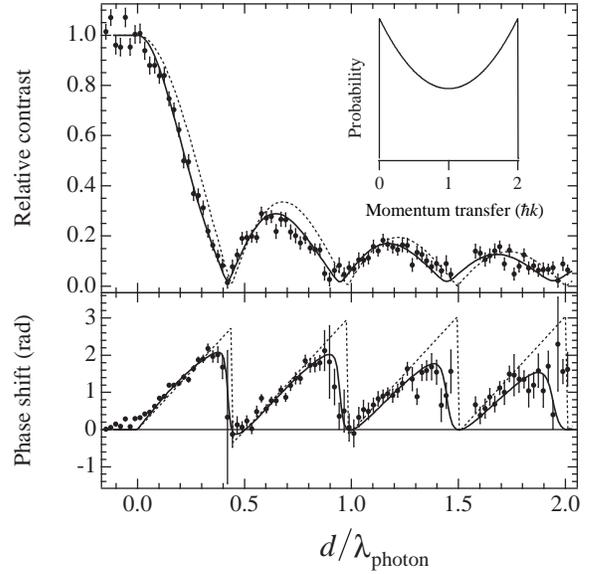}
\caption{Contrast as a function of the path separation, $d$, at
the location of scattering.  Each atom scattered nearly exactly
one photon in this experiment \cite{CHL95}.} \label{fig:1photon
decoherence data}
\end{center}
\end{figure}

The specific arrangement of the experiment allowed separation of the
effects of the (classical) momentum transfer and the entanglement
between the atom at two locations and scattered photon and the
related phase shift.  As seen in Figure~\ref{fig:1photonschem} the
average shift of the pattern at the $3^\textrm{rd}$ grating, and its
random variation from the recoil of the emitted photons is much
larger then the period of the interference pattern at the
$3^\textrm{rd}$ grating ($\sim 30 \mu m$ vs. 200 nm). This
demonstrates that the momentum recoil by itself can not explain the
loss of contrast (as it can in the diffraction experiments), but the
path separation at the point of scattering and the phase shift
imprinted by the entanglement in the scattering process must also be
taken into account.

\begin{figure}[t]
\begin{center}
\includegraphics[width = 7cm]{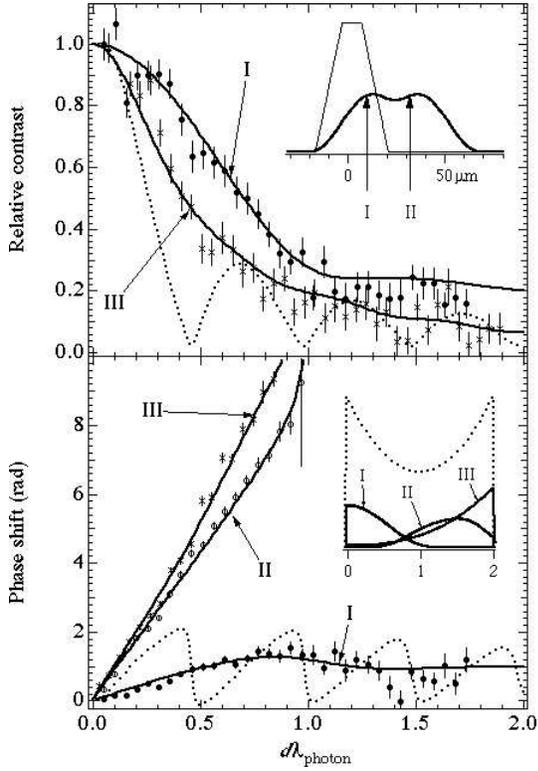}
\caption{Relative contrast and phase shift of the interferometer
as a function of $d$ for the cases in which atoms are correlated
with photons scattered into a limited range of directions. The
solid curves are calculated using the known collimator geometry,
beam velocity, and momentum recoil distribution and are compared
with the uncorrelated case (dashed curves). The upper inset shows
atomic beam profiles at the third grating when the laser is off
(thin line) and when the laser is on (thick line). The arrows
indicate the third grating positions for cases I and II. The lower
inset shows the acceptance of the detector for each case, compared
to the original distribution (dotted line). \cite{CHL95}.}
\label{fig:1photon_recoh}
\end{center}
\end{figure}

The classical recoil shift also allowed a second ``recoherence"
experiment by allowing the experimenters to  \emph{infer the
momentum} of the scattered photon by measuring the atomic recoil.
Interference contrast could be regained
(Fig.~\ref{fig:1photon_recoh}) by selecting atoms within a reduced
range of momentum transfer.  The modern interpretation is that
coherence lost to teh environment because of entanglement can be
regained by learning about the environment. Feynman might say: By
restricting the momentum, the microscope could not use the full 4
$\pi$ acceptance but only a much smaller numerical aperture.
Consequently the maximum obtainable resolution would be degraded, no
`which path' information obtained, and the interference contrast
thereby regained. This experiment demonstrated the importance of
correlations between the recoil momentum and the phase of
interference fringes.

These experiments nicely illustrate how the interaction with an
environment causes decoherence through entanglement with the states
of the environment.  If an atom in the two-path interferometer, with
the paths separated by $d$, scatters a photon the quantum state
evolves into:
\begin{eqnarray}
    \label{eqn:idealmeas}
    \big|\psi\big\rangle_{i} & = &\Big(|x\rangle +
    |x+d\rangle\Big) \otimes|e_{0}\rangle
    \stackrel{\mathrm{interaction}}{\longrightarrow} \nonumber \\
    && \big|x\big\rangle\otimes\big|e_{x}\big\rangle +
    \big|x+d\big\rangle\otimes\big|e_{x+d}\big\rangle,
\end{eqnarray}
where $|e_{0}\rangle$ is the initial wave function of the
environment (photon) and $|e_{x}\rangle$ is the post-interaction
wave function of the environment (photon) given an atom at
position $x$.

If the environment is now observed to be in state $|e_{x}\rangle$,
the (unnormalized) state of the atom becomes:
\begin{equation}
    \big|\psi_{e}\big\rangle = \big|x\big\rangle +
    \beta(d)\big|x+d\big\rangle, \label{conditon}
\end{equation}
where
\begin{equation}
    \beta(d) = \langle e_{x} | e_{x+d} \rangle.
\end{equation}
If the two environment states are nearly identical then
$|\beta(d)|\approx 1$; very little which-way information is
available in the measured state of the environment, and the atom is
left in nearly the original superposition.  If $|\beta(d)| \ll 1$,
significant which-way information about the atom has been left in
the environment, and the atom is highly likely, with probability
$(1+|\beta(d)|^2)^{-1}$, to be found in state $|x\rangle$.

Whereas Equation \ref{conditon}  gives the atomic state
conditioned on an observation of the environment, we often want to
find the final quantum state of the atom when the environment is
not observed.  This requires averaging over all possible
environment states, obtained by taking the trace of the
atom+environment density matrix over environment degrees of
freedom.  Applied to the atom interferometer, this procedure
results in a reduction of contrast by a factor $|\beta(d)|$ for
every photon scattered, and can be directly applied to describe
the results of the Feynman gedanken experiment \cite{CHL95}.

Focusing on the which-way information carried away by the
scattered photons is not the only way decoherence may be
understood.  An alternative, but completely equivalent picture
involves the phase shift between the two components of the atomic
wave function.  We switch to this viewpoint by using the
translation operator for photon momentum states ($\hat{T}(\vec{x})
= e^{i\hat{k} \cdot \vec{x}}$) to identify that the environment
states are related by \beq \langle k_f | e_{x+d} \rangle = \langle
k_f | e^{i(\hat{k}_f - \vec{k}_i) \cdot \vec{d}} |e_{x} \rangle
\eeq where the momentum of the absorbed photon $\vec{k}_i$ was
assumed to be precisely defined by the incident laser beam. Thus,
if one were to measure the momentum of the scattered photon (to be
$\vec{k}_f$) the atom would then be found in a superposition state
with known phase shift between the two components of \beq
\Delta\phi = (\vec{k}_{f} - \vec{k}_{i}) \cdot \vec{d}.
\label{eq:eidkd}\eeq Interference fringe patterns for atoms with
different recoil momentum kicks will then be slightly out of phase
and the ensemble average - the measured interference pattern -
will have a reduced contrast. This point of view is useful to
calculate \beq \beta(d) = \langle e_x | e_{x+d} \rangle \\ = \int
d\vec{k}_f e^{i (\vec{k}_{f} - \vec{k}_{i}) \cdot \vec{d}}
|\langle k_f | e_x \rangle|^2. \eeq This is a scaled Fourier
transform of the probability distribution $P(\Delta k)$.

We have discussed two views (which-way and dephasing) of the
decoherence that accrues when an atom in an interferometer
scatters photons.  They correspond to two different ways to
describe the scattered photon. (position basis vs. momentum
basis). In these two cases, an observer in the environment can
determine either which path the atom took, or else the phase shift
of its fringe pattern. The key point is that when the experimenter
is completely ignorant of the state of the scattered photons,
whether an apparatus has been set up to measure them or not, the
``which-path'' and phase diffusion pictures are equally valid
\cite{SAI90}.  Both predict decoherence, i.e. loss of contrast.

Building upon the simple framework of the single-photon which-way
experiment, we can easily derive the effect of continuous
atom-light interaction involving many scattered photons.  If
successive scattering events are independent, the total
decoherence function includes one factor of $\beta$ for each
scattered photon (with probability $P_n$ of scattering $n$
photons). If the separation does not change ($d$=constant) one
obtains a very simple relation:
\begin{equation}
    \label{eqn:betat}
    \beta_{total}(d)=\sum_{n=0}^{\infty}P_n\beta(d)^{n}.
\end{equation}
Even at small separations each successive photon scattering found
in $| e_{x} \rangle$  ($ | e_{x+d} \rangle$), will reduce by a
small factor the probability that the atom is in state
$|x+d\rangle$ ($|x\rangle$) until only one component of the
superposition has any remaining amplitude; that is, until
``complete'' which-path information has been obtained.

This was nicely demonstrated in the experiment by \textcite{KCR01}
studying scattering multiple photons from each atom inside the
interferometer at a location where the separation is small
compared to the light wavelength.  The contrast vanishes as
information about which path each atom took in the interferometer
gradually becomes available in the photon field as a result of
multiple scattering events. These experiments are also discussed
in \cite{BER97,PCH98,PCG01} and extended to include two separated
environments inside the interferometer by \textcite{CKR03}.

Multiple photon scattering results in a Brownian motion of the
phase of the atomic superposition and can be analyzed as
\emph{phase diffusion}.  It leads again to an exponential decay of
contrast as a function of time (i.e. the average number of
scattered photons $ \bar{n}$). Taking the specifics of the photon
scattering process one finds, in perfect agreement with the
experiment, a Gaussian loss of contrast as a function of the path
separation $d$.
 \beq
 \frac{C}{C_0}= \left\langle e^{i\phi} \right\rangle =
 e^{-\bar{n}(d \cdot \sigma_k)^2/2} \label{eq:closs}
 \eeq
where, $\sigma_k$ is the RMS spread in momentum per scattered
photon.

Contrast loss due to scattering multiple photons makes contact
with more formal theories that describe the dynamics of open
quantum systems.  A modified Heisenberg equation of motion for the
density matrix has been derived for various environments by
\cite{CAL83,TEG93,OMN97,HSA04,GAL93,DEK81,GAF90,JOZ85}. For
example, an environment that causes the probability of scattering
waves with wavelength $\lambda_{eff}$ in an infinitesimal time
interval $dt$ to be $\Lambda dt$ (where $\Lambda= \textrm{Flux}
\times \textrm{cross section}$), makes the master equation
 \beq \frac{\partial \rho(x,x')}{\partial t} = - \frac{i}{\hbar}
 [H,\rho(x,x')]-  \frac{\Lambda (x-x')^2}{\lambda_{eff}^2}  \rho(x,x')
 \label{eq:master equation}\eeq
where the final term on the right causes a damping of the
off-diagonal elements of $\rho$ with a rate expressed by
 \beq
\rho(x,x';t) \approx \rho(x,x';0)e^{-\frac{\Lambda
(x-x')^2}{\lambda_{eff}^2} t}. \label{eq:diffusion}\eeq Here
$(x-x')$ denotes the separation of the superposition states in a
general coordinate basis, and the diffusion constant $\Delta =
\Lambda / \lambda_{eff}^2$ is also referred to as the localization
rate \cite{JOZ85}, or the decoherence rate \cite{TEG93}. Values of
decoherence rates are tabulated in \cite{JOZ85,TEG93,HMM96} for
various systems and scattering environments.  Comparing equations
\ref{eq:closs} and \ref{eq:diffusion} allows one to discuss the
localization rate caused by photon scattering for atoms in an
interferometer.

\subsubsection{Realization of Einstein's recoiling slit experiment}

To implement Bohr's original design of Einstein's recoiling slit
Gedanken experiment, one needs a very light beam splitter, which
shows quantum properties and will allow an experimenter to
distinguish the two possible paths taken. In a Ramsey experiment,
one would need to be able to distinguish the photons in the
microwave or optical field used to change the state in the first
interaction region. As discussed above, classical fields can not do
the job. But if the splitting in the first interaction region is
induced by a vacuum field, or a single photon field (more generally
a field with a definite photon number) then measuring the field will
determine if a transition has happened, and consequently infer the
path the atom took.

In their seminal experiment \textcite{BOR01} implemented a Ramsey
interferometer with Rydberg atoms where the first interaction zone
is a high-Q cavity which allows the superposition between the
$|e\rangle$ and $|g\rangle$ states to be created by the interaction
with the \emph{vacuum field} inside the cavity.  This is the
ultimate light beam splitter. After passing the interaction region,
the atom-cavity system is in an entangled state described in the
\mbox{$|$atom$\rangle|$cavity$\rangle$} basis:
\begin{equation}
    |e\rangle |0\rangle \rightarrow \frac{1}{\sqrt{2}}
    \big[ e^{i \Phi} |e\rangle |0\rangle + |g\rangle |1\rangle \big], \label{eq:RecSlit}
\end{equation}
where $\Phi$ is an phase difference between the two states after
the interaction.

With this interaction the information about the state of the atom is
left in the cavity field, and no interference contrast is observed
when completing the Ramsey interferometer with a classical microwave
pulse and state selective detection.

The cavity can also be filled with a very small coherent state
$|\alpha\rangle$ with a mean photon number of a few
($\bar{n}=|\alpha|^2$. The interaction region creates the
entangled sate:
\begin{equation}
    |e\rangle |\alpha_e\rangle \rightarrow \frac{1}{\sqrt{2}}
    \big[ e^{i \Phi} |e\rangle |\alpha_e\rangle + |g\rangle |\alpha_g\rangle \big],
    \label{eq:RecSlit_alpha}
\end{equation}
with
\begin{eqnarray}
    |\alpha_e\rangle &=&  \sqrt{2} \sum_n C_n \cos(\Omega \sqrt{n+1}t_\alpha) |n\rangle \\
    |\alpha_g\rangle &=&  \sqrt{2} \sum_n C_n \cos(\Omega \sqrt{n+1}t_\alpha) |n+1\rangle
    \label{eq:RecSlit_alpha_eg}
\end{eqnarray}
where $t_\alpha$ is an effective atom-cavity interaction time
adjusted to give a equal superposition between $|g\rangle$ and
$|e\rangle$.

\begin{figure}[t]
\begin{center}
\includegraphics[width = 9cm]{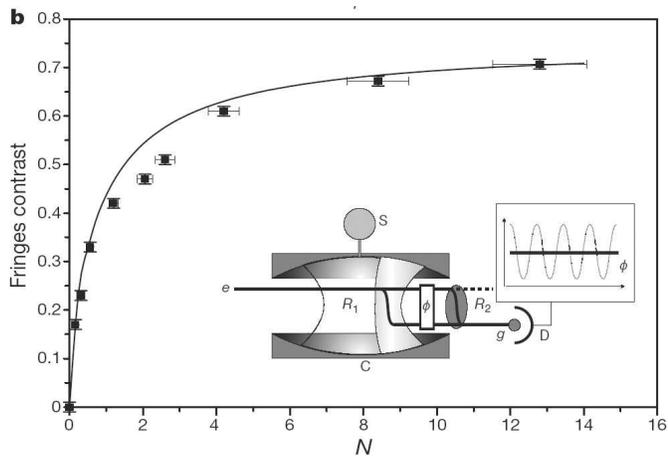}
\caption{Fringe contrast as a function of the mean photon number N
in R1. The points are experimental. The line represents the
theoretical variation of the modulus of the beam-splitter
final-states scalar product. \cite{BOR01}.} \label{fig:RecoilSlit}
\end{center}
\end{figure}

The results of such an experiment are shown in
Figure~\ref{fig:RecoilSlit}.  When employing the lightest beam
splitter, that is the vacuum state with n=0, the contrast in the
Ramsey interferences vanishes completely.  When employing
successively stronger coherent states, the beam splitter becomes
`heavier' in Bohr's argument, and the coherence comes back.  For
$\bar{n} = 12.8$ ($|\alpha|=3.5$) nearly the full interference
contrast is regained.

In a second part of their experiment \textcite{BOR01} employed the
same field twice. Once as first interaction region, and again as
second interaction region.  In this case even for the vacuum field
as a beam splitter no information about the path within the Ramsey
interferometer remains, and full contrast was observed.  This is a
beautiful illustration of an unconditional quantum-eraser
experiment.

As our understanding of quantum mechanics deepens, and in
particular, as we attempt to exploit quantum mechanics to create
more sensitive quantum interferometers, quantum computers, or
perfectly secure communication channels based on quantum
entanglement, we encounter decoherence as a fundamental limit
\cite{UNR95}. Progress relies therefore upon understanding and
correcting for decoherence effects.  Already our increased
understanding of what decoherence means and how to control it has
led to the development of quantum error correction codes
\cite{SHO95,CAS96,STE96} and quantum mechanical systems in which
certain degrees of freedom are intrinsically decoherence-free
\cite{LCW98}.

\subsection{Origins of phase shifts}

Phase shifts for interference fringes (see Section II.~B.~Equation
16) can be induced by photon scattering as discussed in the
previous section (Eq \ref{eq:eidkd}), or by a variety of other
causes such as 1.~different potential energy for atoms in each
path of the interferometer, 2.~transverse or longitudinal forces
on atoms, 3.~inertial displacements such as rotating or
accelerating the interferometer platform, and 4.~geometric and
topological phase shifts such as the Aharonov-Bohm,
Aharonov-Casher, and Berry phase.  In the following section we
discuss the interrelationship between these types of phase shifts.

\subsubsection{Dynamical phase shifts}

Feynman's path-integral formulation \cite{STC94,FEH65} relates
the wave function at $({\bf x},t)$ to the wave function at $({\bf
x}_0,t_0)$ by  \beq \psi({\bf x},t) =
e^{-\frac{i}{\hbar}S_{\Gamma}} \psi({\bf x}_0,t_0)
\label{eq:feynman1} \eeq where the classical action $S_{\Gamma}$
is defined in terms of the Lagrangian \beq S_{\Gamma} \equiv
\int_{\Gamma} \mathcal{L}[\dot{x},x] \textrm{ } dt\eeq and
$\mathcal{L}[\dot{x},x]$ is the Lagrangian and $\Gamma$ is the
classical path from $({\bf x}_0,t_0)$ to $({\bf x},t)$. For
potentials that are only a function of position, the wave function
acquires a phase \emph{shift} due to a potential $U({\bf r})$ of
\beq  \phi_{int} = \int \left[ \sqrt{\frac{2m}{\hbar^2} [E-U({\bf
r})]} - \sqrt{\frac{2m}{\hbar^2} E} \right] dl. \label{eq:phi_int}
\eeq This is analogous to light optics where the wave vector
${\bf{k}} = n({\bf{r}}){\bf{k}}_0$ depends locally on the index of
refraction, and the phase shift due to the index is \beq \phi =
\int (k - k_0 ) dl. \eeq To first order in $U/E$ the interaction
phase shift (\ref{eq:phi_int}) is \beq \phi_{int} \approx
-\frac{1}{\hbar v} \int_{\Gamma} U({\bf r}) dl \label{eq:Udt1}
\eeq where $v$ is the particle's velocity.

This brings up the question, `when does one measure a quantity
described by classical physics like a deflection, and when does one
measure a quantity only observable in an interference experiment?'.
For example, applying a classical force, $\vec{F}$, to change a
particle's motion is identical to applying a phase gradient to the
matter wave. This is because force can be viewed as arising from a
potential gradient [$\vec{F}({\bf r}) = -\vec{\nabla} U({\bf r})$],
and in the same potential $U({\bf r})$ a propagating matter wave
will get a position-dependent phase shift which is exactly the one
needed to account for the deflection.  If there are two paths
through the interferometer, then the fringe phase shift will be
given by \beq \Delta \phi_{int} = \phi_{int 1} - \phi_{int 2}. \eeq
Thus, in a classical apparatus (as in a moir\'{e} deflectometer) or
in an interferometer, forces cause a fringe shift that is
\emph{identical} to the classical deflection, which can be observed
as an \emph{envelope shift} \cite{ZEI86,OBR96}.

On the other hand, there are many cases where the fringe shift is
different from the envelope shift.  A basic example is a constant
potential applied to one arm of an interferometer with separated
beams.  In this case there is no classical deflection, because
neither atom (component) acquires a transverse phase gradient.
Still, there is a different interaction phase $\phi_{int}$ for one
path through the interferometer because of the potential.  For
example, one interferometer path may traverse a capacitor such that the
gradient in potential energy is \emph{along} the atomic path.

In this case \emph{Longitudinal} phase gradients can be caused as
atoms enter and exit the interaction region.  For example, an
attractive potential causes a classical force that first accelerates
then decelerates the atom (component);  If the potential is confined
to one path through the interferometer then the affected atom
component gets displaced ahead of the unperturbed atom component.
Furthermore, if the longitudinal displacement between wave function
components exceeds their coherence length, then contrast is lost. We
prefer to call this `inhomogeneous broadening' (as opposed to
decoherence) because the phase shift is correlated (entangled) with
the atom's own longitudinal velocity.

Another interesting case arises when one applies a time-dependent
potential to one arm of the interferometer so the atom never sees a
gradient in space.  An example is the scalar Aharonov Bohm effect.
Then there will be no change in the classical motion and the
envelope of the atomic probability distribution will remain
stationary as high-contrast fringes (there is no velocity
dispersion) shift underneath.  A similar situation arises when
purely topological phases are involved.  In these cases the full
quantum mechanical properties of an interferometer are in evidence.

\subsubsection{Aharonov-Bohm and Aharonov-Casher effects}

We call a phase shift $\Delta \phi_{int}$ {\em topological} if it
neither depends on the incident $k$-vector (velocity) of the
interfering particle nor on the shape of the particle's path through
the interferometer. Topological phases are characteristic of all
gauge theories, and are related to a singularity enclosed by the
interferometer paths.

\begin{figure}[t]
\begin{center}
\includegraphics[width = 8cm]{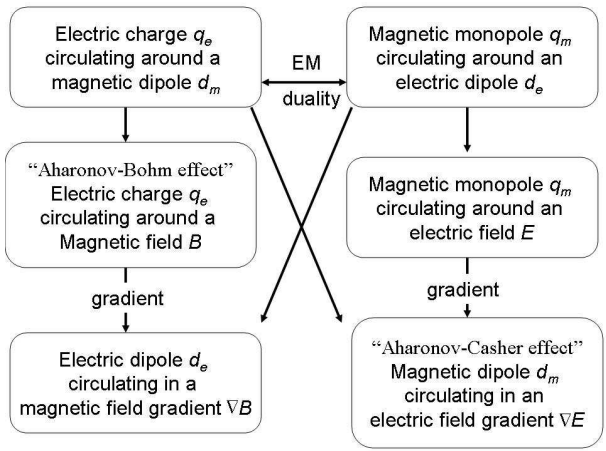}
\caption{The Aharonov-Bohm effect, Aharonov-Casher effect and their
electromagnetic duals.  Figure from Janis Panchos (unpublished); see
also \cite{DWF99}.} \label{fig:em-duality}
\end{center}
\end{figure}

The most widely known topological phase was described by
\textcite{AHB59} for a charged particle passing on either side of a
solenoid.  A related effect was described by \textcite{AHC84} for a
magnetic dipole encircling a line of charge.  To realize a general
framework for the discussion of the quantum interaction between
sources and fields we consider Fig.~\ref{fig:em-duality}. If an
electric charge $q_e$ circulates around a magnetic dipole $d_m$ (or
vice versa) then a quantum phase arises \cite{AHC84}. Particular
configurations of sources can give a variety of contributions.  For
example, a cylinder filled with aligned magnetic dipoles is
equivalent to a solenoid, and creates a homogeneous magnetic field
inside the cylinder but zero magnetic field outside.  When an
electric charge travels around the cylinder it acquires a phase due
to the Aharanov-Bohm effect.  On the other hand, if a cylinder is
filled with electric charges, then a magnetic dipole circulating
around it will obtain a quantum phase due to the Aharonov-Cahser
effect. This can be generalized to the case of a magnetic dipole
moving in the presence of a gradient of electric field.

Employing electromagnetic (EM) duality it is possible to obtain a
series of similar phenomena.  While the charge dual is the
magnetic monopole, $q_m$, which has never been observed, the dual
of the dipoles are well defined.  The interactions between an
electric dipole, $d_e$, and a monopole (or a monopole-like field)
have been extensively studied in the literature
\cite{WIL94,CAS90,SPA06,SPA99,HEM93,DWF99}. This can also be
equivalently viewed as the interactions of an electric dipole,
$d_e$, with an inhomogeneous magnetic field. It should be
understood that this categorization, instructive as it may be, is
not unique nor exhaustive, e.g. quadrupole interactions have not
been considered.

The Aharonov-Bohm phase shift is \beq \Delta \phi _{AB} =
\frac{q_e}{\hbar} \oint \mathbf{A} \cdot d \mathbf{s} \eeq where $A$
is the vector potential that represents the fields. The
Aharonov-Casher effect causes a phase shift \beq \Delta \phi_{AC} =
\frac{1}{\hbar c^2} \oint \mathbf{d_m} \times \mathbf{E} \cdot d
\mathbf{r} \eeq where $d_m$ is the magnetic dipole.

\begin{figure}[t]
\begin{center}
\includegraphics[width = 8cm]{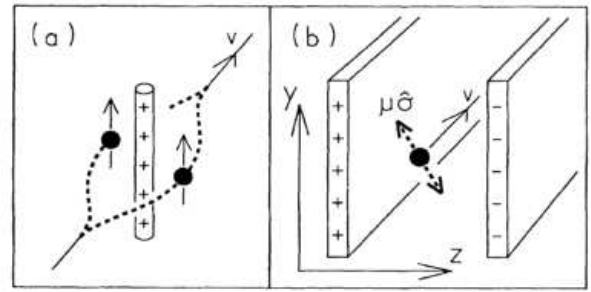}
\caption{Aharonov-Casher effect.  (a)  Geometry of the original
measurement using a neutron interferometer, in which the two
interfering states encircle a charge and have the same magnetic
moments.  (b) Geometry used in \cite{SHB93}.  Particles travel in a
uniform magnetic field in a coherent superposition of opposite
magnetic moments $\pm \mu \mathbf{\hat{\sigma}}$.  the two states
are oppositely shifted by the Aharonov-Cahser phase as they travel
through the field.  Figure and caption reproduced from
\cite{SHB93}.} \label{fig:A-Cgeometry}
\end{center}
\end{figure}

The Aharonov-Casher effect was observed with neutron interferometers
\cite{COK89} using the original geometry proposed by \cite{AHC84},
and the phase shift was 2.19 mrad. With TlF molecules the
Aharonov-Cahser effect has been observed using a geometry where
components of each molecule with different spin states occupy the
same center of mass location \cite{SHB93,SHB95}. This alternative
geometry for the Aharonov-Casher effect, in which the magnetic
dipoles are placed in a superposition of spin orientations (but not
a superposition of center of mass positions) was described by
\cite{CAS90}. The two different geometries are summarized in Figure
\ref{fig:A-Cgeometry}.  With molecules possessing a nuclear magnetic
moment the phase shift was only 3 mrad. Still, this was sufficient
to verify the predicted linear dependance on the electric field and
independence of particle velocity. Atomic sized magnetic moments
were used by \cite{GSW95} to demonstrate a much larger
Aharonov-Caher phase shift of 300 mrad using Rb atoms. An A-C phase
of 150 mrad was observed by \cite{YKM02} using Ca atoms, and related
measurements are found in \cite{ZZR95,ZRZ94}.

The AC phase is a \emph{restricted} topological phase because
although the phase is independent of the speed $|v|$ and the size of
the interferometer loop, the phase does depend on whether
$\mathbf{d_m}$ is perpendicular to both $\mathbf{v}$ and
$\mathbf{E}$. Debate over the topological nature of the AC effect
has stimulated several discussions, among them
\cite{BOY87,APV88,ZGH91,HAK92,LEE01}.  The similarity between the AC
and AB effects has also been discussed in detail by
\cite{HAG90,ORS94}.  One controversy arose over the question of
whether or not a sufficiently large AC phase can lead to
decoherence.  This position was suggested by \cite{BOY87} since the
AC effect can be explained in terms of a classical force due to a
motion-induced magnetic field in the rest frame of the magnetic
dipole. However, as was shown in \cite{ZGH91}, since the classical
force depends on velocity a wave packet envelope does not get
shifted; i.e. $ \partial\phi_{AC}/\partial k_{dB} = 0 $. The AC
effect and AB effect both shift the phase of the wave function, but
do not displace the wave packet envelope (a common miss-impression,
e.g. see Figs 15-7 and 15-8 of the \textcite{FLS65} Lectures in
Physics Vol II).

The scalar Aharonov-Bohm effect (SAB) for neutral particles is a
topological phase that can arise from pulsed magnetic fields
interacting with an atomic magnetic dipole. This has been observed
by \cite{SAM02,AYM03,MBR95} with atoms, and by \cite{ACK92,BWG93}
with neutrons.  It is similar in spirit to the interaction discussed
in the original paper \cite{AHB59} for electrons interacting with
the scalar electrostatic potential.

The electromagnetic dual of the AC effect, in which an electric
dipole moment moves near a line of magnetic monopoles (an idealized
picture of an experiment) was investigated theoretically by
\cite{WIL94}.  The phase shift for polarizable particles moving in
both electric and magnetic fields has been also discussed by
\cite{SHE95,AUS98,ANA89,ANA00}. Furthermore, in the case that
permanent electric dipole moments are used, the electromagnetic dual
to the AC effect can be used to settle any controversy regarding how
the topological nature of the AC effect depends on the dipole moment
being intrinsic and therefore having quantum fluctuations
\cite{LEE01}.

\subsubsection{Berry phase}

Phase effects resulting from parallel transport associated with
adiabatic evolution (Berry phase) can also be topological.
\textcite{BER84} showed that a quantum system in an eigen-state that
is slowly transported round a circuit by varying parameters
$\mathbf{R}$ in its Hamiltonian $H(\mathbf{R})$ will acquire a
geometrical phase factor in addition to the familiar dynamical
phase. For example, the Berry phase of a magnetic moment
adiabatically following a magnetic field will acquire a phase
proportional to the solid angle proscribed by the field during a
closed circuit. Berry also interpreted the Aharonov-Bohm effect as a
geometrical phase factor.

\begin{figure}[t]
\begin{center}
\includegraphics[width = 7cm]{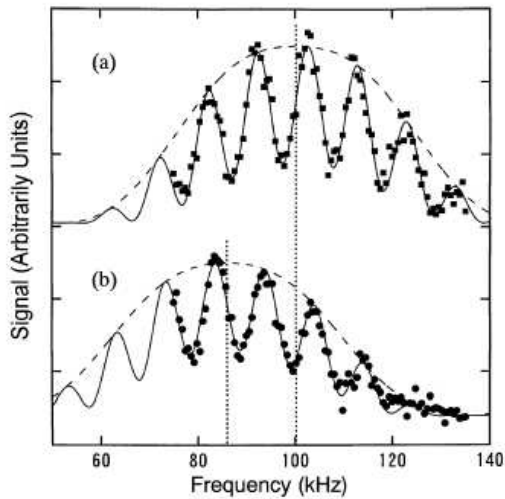}
\caption{Ramsey fringes under (a) a constant magnetic field and
(b) a rotating magnetic field. The rotating angle is $\pi$
radians. The phase difference is observed at the center frequency
of the spectra. Figure and caption reproduced from \cite{YAN05}.}
\label{fig:Berryshift}
\end{center}
\end{figure}

The Berry phase can be studied with many systems in physics. It has
been observed with light in a coiled optical fiber \cite{CHW86},
neutron interferometers \cite{BID87}, nuclear magnetic resonance
experiments \cite{SCH87}, nuclear quadrupole resonance experiments
\cite{TYC87}, and also mesoscopic electronic systems \cite{YPS02}.
Phase shifts due to non-adiabatic circuits \cite{AHA87}, incomplete
circuits \cite{SAB88}, particles in mixed states \cite{SPE00}, and
particles moving relativistically have also been studied
theoretically. For an overview on geometric phases see
\cite{WIS89,ACW97}.

An observation of a Berry phase in atoms in an interferometer for
the polarization states (internal states) is described in
\cite{COM91}. The first observation of a Berry phase in an external
state atom interferometer was accomplished by \cite{MRG92} with a
Stern Gerlach longitudinal interferometer. However in this
experiment the Berry phase was somewhat obscured because the
dynamics were not adiabatic. A Berry phase up to 2$\pi$ radians due
to an atomic state interacting with a laser field was observed by
\cite{WGS99}.  This verified the spin dependence of the Berry phase,
and realized an ``achromatic phase plate for atomic interferometry"
as suggested by \cite{OLS94,RSE93}.  A Berry phase shift for partial
cycles using a time domain atom interferometer was measured by
\cite{YAN05} (see Fig.~\ref{fig:Berryshift}).

\subsubsection{Inertial displacements}

Atom interferometers are impressively sensitive to acceleration and
rotation because the long transit times allow gravity and fictitious
forces due to rotation and acceleration to build up significant
displacements of the interference pattern, which directly influence
the measured phase (introduced in Section III Equation \ref{eq:3g
std phase}).   These are discussed fully in Section V on precision
measurements.

\subsection{Extended Coherence and BEC's}

Bose Einstein Condensates of atomic gasses are very bright sources
for atom optics and atom interferometers.  Additionally in a gas
cooled below $T_c$, a significant fraction of the atoms are in the
condensate, which occupies the lowest translational state of the
trap. Typical BEC's offer a million atoms confined in a cigar
shaped sample 100 microns long and 10 microns across, with
coherence lengths of the same size, and relative velocities around
0.1 mm/sec.  A BEC with its coherence properties (and brightness)
constitutes a source analogous to a laser, whereas the traditional
thermal atom sources are analogous to thermal sources such as
candles or light bulbs in optics.

This ideal source is hindered by the fact that atoms interact which
leads to a mean field interactions (chemical potential). A typical
condensate would have density ~ $10^{14}/$cm$^3$ with associated
mean field energy of $\sim$1 kHz($\times h$), much larger then the
ground state energy of the trap.  If the trap is turned off, and the
BEC released, this mean field energy dominates the expansion and the
condensate atoms will separate with several mm/sec relative velocity
regardless of how small the RMS velocity was inside the trap.
Nevertheless the resulting momentum spread is still an order of
magnitude smaller than the recoil velocity from a resonant photon.
It is therefore easy to separate the momentum states differing by a
photon momentum in atom interferometers based on BEC's as discussed
in section III. (e.g. see Fig.~\ref{fig:rob1}c).

Atom interferometers now offer a powerful tool to study the
properties of a Bose Einstein Condensate.

\begin{figure}[t]
\begin{center}
\includegraphics[width = 8cm]{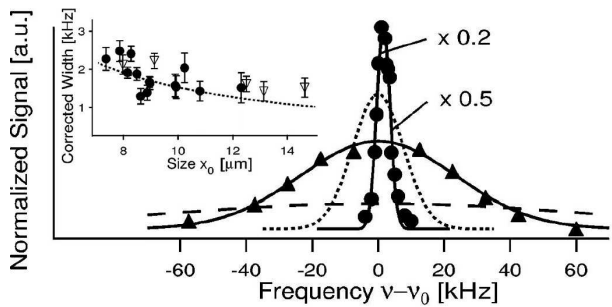}
\caption{Bragg resonances for a trapped condensate (circles) and
after 3 ms time of flight (triangles).  This maps the momentum
distribution in the trapped (or expanding) condensate. For
comparison, the momentum distributions of the ground state of the
trapping potential (dotted curve) and of a 1 mK cold, thermal cloud
(dashed curve) are indicated. (Inset) Bragg peak widths as a
function of condensate size. The plotted Bragg widths have been
corrected by subtracting the contribution of the mean field and the
finite pulse duration. The dashed curve is based on a prediction for
the momentum uncertainty due to the finite size of the condensate
and the uncertainty principle. Figure from \cite{SIC99}.}
\label{fig:Bwidths}
\end{center}
\end{figure}

\subsubsection{Atom Lasers}

Early theoretical studies \cite{BPK87,STO91,MOV94} showed that
making a BEC in a trap is easier than making it in free space
because the critical density had to be reached only at the bottom of
the trap. They showed that the perturbation of the transition
temperature and critical number density due to the s-wave scattering
of the atoms was less than 1\%, encouraging the then-prevalent view
that the condensate is well described as a blob of very cold atoms.
This suggested making a laser like beam of ultra cold atoms simply
by extracting the atoms from the condensate - such a beam would have
an incredibly low temperature, be almost monochromatic, and have an
unprecedented brightness (albeit over a very small cross sectional
area with limited total flux).

Early realizations of atom lasers coupled atoms out from a
condensate with radio frequency (rf) pulses, rf chirps, Raman
pulses, or weak cw rf radiation \cite{MAK97,HDK99,BHE99}.  For
discussions also see \cite{HBG96,kle97}. The out-coupled atoms have
energy given by the out-coupling process plus the mean field energy
they gain when emerging from the condensate.  In addition, they are
accelerated by gravity and any additional potential gradient. The
out-coupling frequency can be adjusted as the condensate number
changes, to account for the changing chemical potential. Moreover,
the number of atoms extractable from the condensate is not limited
because the condensate can be recharged \cite{CSL02} to produce a
continuous atom laser beam.  Although a continuous atom laser is yet
to be demonstrated.

In principle, the output from this type of atom laser can have a
greater coherence length than the condensate simply because it has
the coherence time of the condensate and is traveling.  Using a
stable BEC as a phase reference could enable feedback to perfectly
compensate the changes in chemical potential.  So far, however,
coherence lengths of atom lasers have not exceeded the size of the
condensate.

Phase coherent matter wave amplification, in direct analogy to
laser gain, has been demonstrated \cite{IPG99,KST99} and discussed
early on by \cite{bor_laser95,HBG96,JAW96}.

\subsubsection{Studies of BEC wavefunctions}

In the simplest picture of a BEC, all atoms in the condensate
occupy the quantum ground state of the trap.  This wave function
is modified by the mean field interaction of the atoms. As more
atoms accumulate in the condensate, their mutual interaction
modifies the condensate wave function.  For repulsive interactions
the condensate wave function broadens at the expense of increased
potential energy from the trap in order to minimize the mean field
energy. Each atom in the condensate is coherent across the whole
condensate and a double slit experiment in either space or time
should show interference fringes.

More sophisticated treatments of atoms cooled below the BEC
transition temperature show that they can exist in states called
quasi-condensates that have short range coherence, but not long
range coherence over the whole condensate.  Whether BEC's have long
range coherence was studied in interference experiments on BEC's,
which we now discuss.

Bragg diffraction offers high momentum selectivity.  As discussed in
Section II.C.3, the spread in velocity of atoms that can be
diffracted ($\sigma_v$) is determined by the inverse duration of
interaction with the grating, and can be deduced from the
time-energy uncertainty principle, $\sigma_v = 2/(\tau G)$.
Near-resonant moving standing waves therefore probe a specific
velocity class , creating a high-resolution tool for studying BEC
velocity distribution. By increasing the interaction time to nearly
1 ms, \cite{SIC99} at MIT achieved a velocity selectivity of ~ 0.1
mm/sec, which allowed to study the momentum distribution inside the
trap and in a released condensate Fig.~\ref{fig:Bwidths},
demonstrating the mean field acceleration. The coherence length was
equal to the transverse dimension of the condensate (see
Fig.~\ref{fig:Bwidths} inset).

A similar conclusion was reached independently by \textcite{KDH99}
at NIST using an atom interferometry technique in which KD
out-coupled atom pulses were applied at two closely spaced times.
Each ejected pulse mirrors the condensate itself, so when the
front of the second overlapped the back of the first the
interference observed was indicative of coherence between two
spatially separated places in the condensate.  The decay of the
fringe envelope was as expected for a fully coherent condensate.

\begin{figure}[t]
\begin{center}
\includegraphics[width = 8cm]{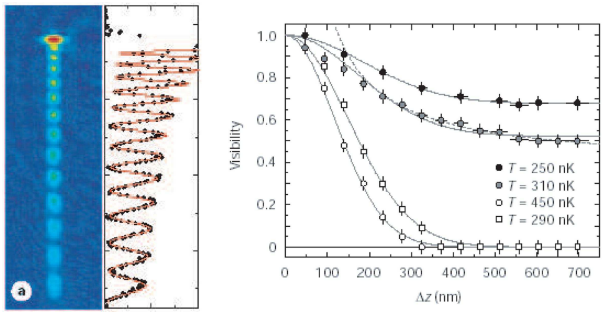}    
\caption{(color online) Spatial correlation function of a trapped
Bose gas as measured by the fringe visibility as a function of slit
separation for temperatures above (white circles  T=450 nK and
squares T=290 nK) and below the critical temperature $T_c$ (grey
T=310 nK and black T=250 nK), where the visibility decays to a
nonzero value due to the long range phase coherence of the BEC. The
data points displayed in the Figure are corrected for the reduction
in visibility which is due to the limited resolution of the imaging
system. Figure reproduced from \cite{BHE00}.}
\label{fig:bec_coherence}
\end{center}
\end{figure}

Experiments studying the coherence of atom laser beams were carried
out by \textcite{BHE00} in T. Haensch's lab in Munich. Two atom
laser beams coupled out from different locations of the trap were
overlapped to interfere.  By changing the separation of the out
coupling locations, and observing the contrast of the interference
between the two out coupled beams they probed the coherence
properties of the condensate wave function on length scales
approaching 1 micron (Figure \ref{fig:bec_coherence}). Measurement
of the temporal coherence of an atom laser has also been used to
give an upper limit for temporal phase fluctuations corresponding to
$\sim$ 700 Hz in the Bose-Einstein condensate \cite{KHE01}.

In a related experiment \textcite{ANK98} observed interference of
atoms from an array of BEC's trapped in an optical lattice.  The
interference between the BEC's at different gravitational potential
leads to a pulsed atom laser beam.  Since many sources contribute,
the pulses are much shorter than the separation between them,
reminiscent of a mode-locked pulsed laser.

\subsubsection{Many particle coherence in BEC's}

The above experiments can all be viewed as looking at single
particle coherence.

BEC's have an even more dramatic coherence than the extended
condensate wave function just discussed.  The atoms in the
condensate are in one macroscopic state with an order parameter, the
phase.  Consequently the phase of one condensate atom is the phase
of all. Therefore, a condensate also exhibits coherence properties
resulting from the interference of different (but indistinguishable)
atoms. This gives it coherence properties like a laser: if the phase
is determined by measuring some of the atoms, other atoms will have
the same phase.

Measuring the phase of condensate atoms requires a coherent and
stable reference. Such a reference can be provided by another
condensate, or by other atoms from the same condensate. This is in
marked contrast to traditional atom interference discussed up to
now, where interference is only that of each atom with itself. The
BEC experiments using Bragg scattering discussed above demonstrate
only the spatial coherence of individual atoms in a BEC.  We now
turn to experiments that show the coherence of different atoms in a
BEC.

The existence of a macroscopic wave functions with an order
parameter means that atoms from different sources can interfere.
If an atom from one interferes with an atom from the other,
subsequent atom pairs will interfere with the same relative phase
and fringes will be built up which reflect the relative phase.
This is similar to interference between two independent lasers
\cite{PFM67,PAU86,KBZ06,CAD97bec}, which also generated
controversy prior to its observation.

\begin{figure}[t]
\begin{center}
\includegraphics[width = 8cm]{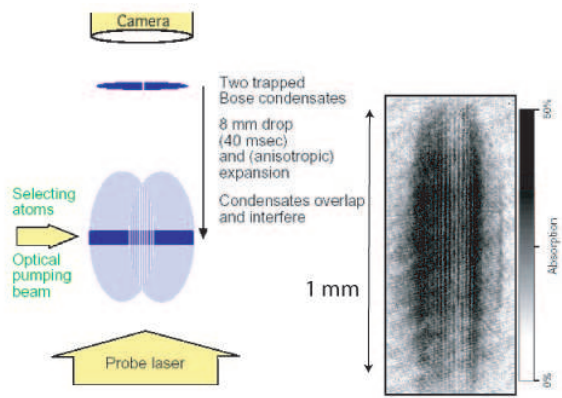}
\caption{(color online) Left: Schematic setup for the observation of
the interference of two independent BEC's separated by a barrier
created by a blue detuned laser beam. After switching off the trap,
the condensates expand ballistically and overlap. In the overlap
region, a high-contrast interference pattern is observed by using
absorption imaging.  Right: Interference pattern  of two expanding
condensates observed after 40  msec time of flight. The width of the
absorption image is 1.1 mm.  The interference fringes have a spacing
of 15 $\mu$m and are conclusive evidence for the multiparticle
coherence of Bose-Einstein condensates. \cite{DUK98} \cite{ATM97}. }
\label{fig:bec_fringes1}
\end{center}
\end{figure}

The first experiment demonstrating this striking behavior was by
\textcite{ATM97} in the Ketterle group at MIT. To demonstrate that
two independent BEC's can interfere, two independent condensates
were produced in a double-trap potential created by dividing a
magnetic trap in half with a focused blue-detuned laser beam.
After two BEC's were created from separate thermal clouds, the
traps were switched off. The atom clouds expanded ballistically
and overlapped.

The atomic density in the overlap region was observed directly with
absorption imaging, and revealed a high contrast interference
pattern extending over a large region of space
(Fig.~\ref{fig:bec_fringes1}). The interference pattern consisted of
straight lines with a spacing of about 15 $\mu$m. This experiment
provided direct evidence for first-order coherence and a macroscopic
wave function with long range order in the BEC, and caused some to
puzzle over why wave packets expanding radially outwards from two
small condensates would produce straight fringes.

\begin{figure}
\begin{center}
\rotatebox{-90}{\includegraphics[width = 4cm] {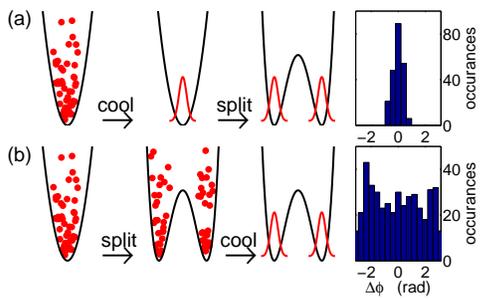}}
 \caption{Comparison of independent and coherently split
BEC's. (a) For the coherent splitting a BEC is produced in the
single well, which is then deformed to a double well. A narrow phase
distribution is observed for many repetitions of an interference
experiment between these two matter waves, showing that there is a
deterministic phase evolution during the splitting. (b) To produce
two independent BEC's, the double well is formed while the atomic
sample is thermal. Condensation is then achieved by evaporative
cooling in the dressed state potential. The observed relative phase
between the two BEC's is completely random, as expected for two
independent matter waves \cite{HLF06}.} \label{fig:cool_split}
\end{center}
\end{figure}

In a related atom chip experiment \textcite{HLF06} compared the
interference of a coherently split BEC with the interference of two
independently created BEC's in identical traps
(Fig.~\ref{fig:cool_split}). The coherently split BEC shows a
well-defined phase, i.e. the same phase for  the fringes each time
the experiment is run. In comparison, the independently formed BEC's
show high contrast interference patterns but with a completely
random phase.

These results are even more surprising than the interference of
independent lasers.  Theories describing laser sources predict
something close to coherent states (for lasers operated well above
threshold), which means that each laser beam may be thought of as
having a well defined (if unknown) phase.  One cannot, however,
assert that the phase of a BEC exists prior to its observation. This
is because a BEC at T=0 can easily contain a known number of atoms
(however many were put in the trap), in which case number-phase
uncertainty prevents the phase from being specified. So the
existence of a well defined relative phase, and hence fringes in the
overlap region seems puzzling.

The resolution to this puzzle is that the phase of the fringes
(i.e. the \emph{relative phase} of the condensates) emerges only
as individual atoms are detected in the overlap region
\cite{CAD97bec}. Since these atoms cannot be attributed to a
particular one of the interfering condensates, an uncertainty
develops in the \emph{relative number} of atoms in the
condensates, and in accord with the relative number-phase
uncertainty principle, they can have a definite relative phase
(even though the total number of atoms in both condensates plus
those detected is known).  Given that neither the phase of either
condensate nor their relative phase existed initially, it should
not be surprising that the fringes in each realization of the
experiment are observed in a different place.  After averaging
over many realizations of this experiment, the fringe contrast
vanishes because the relative phase of each realization is random.

\begin{figure}[t]
\begin{center}
\includegraphics[width = 7cm]{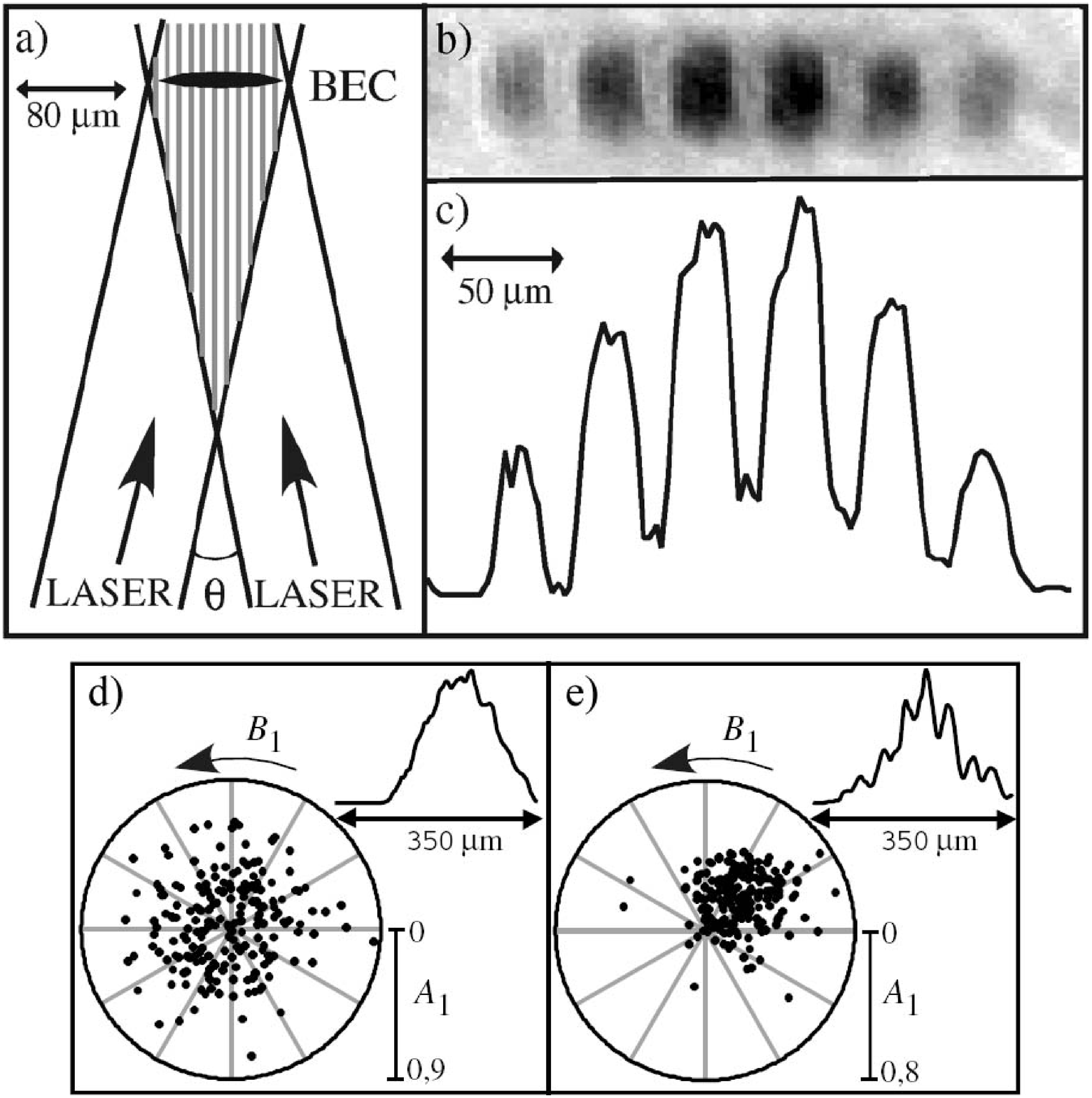}
\caption{Interference of 30 Bose-Einstein condensates each
containing $\sim 10^4$ atoms. (a) A deep 1D optical lattice splits a
cigar shaped condensate into 30 independent BEC's. (b) Absorption
image of the cloud after 22 ms of expansion from the lattice. The
density distribution shows interference fringes. (c) Axial density
profile of the cloud, radially averaged over the central 25 $\mu m$.
(d,e) Polar plots of the fringe amplitudes and phases for 200 images
obtained under the same experimental conditions. (d)
Phase-uncorrelated condensates. (e) Phase correlated condensates.
Insets: Axial density profiles averaged over the 200 images. Figure
reproduced from \cite{HSB04}.} \label{fig:multiBEC}
\end{center}
\end{figure}

Even when many independent condensates interfere, spontaneous
fringes appear.  \textcite{HSB04} observed high-contrast matter wave
interference between 30 Bose-Einstein condensates produced in a
large-period one-dimensional optical lattice. Interference was
studied by releasing the condensates and allowing them to overlap.
High contrast fringes were observed even for independent condensates
with uncorrelated phases as shown in Figure \ref{fig:multiBEC}. This
can be explained the same way as the high-contrast speckles formed
by laser light reflecting off a diffuser.  However, as in the work
with two independent condensates, averaging over many realizations
the experiment causes fringe contrast to vanish because the phase is
random from shot to shot.

\subsubsection{Coupling two BEC's with light}

\textcite{SPS05} have demonstrated a way to make an interferometer
using two BEC's that are never in direct contact and which are
separately trapped at all times.  The key is to use stimulated
light scattering to continuously sample the relative phase of the
two spatially separated BEC's. In fact this sampling creates a
relative phase between the two condensates which in the beginning
had no initial phase relation.

\begin{figure}[t]
    \begin{center}
    \includegraphics[width=7cm]{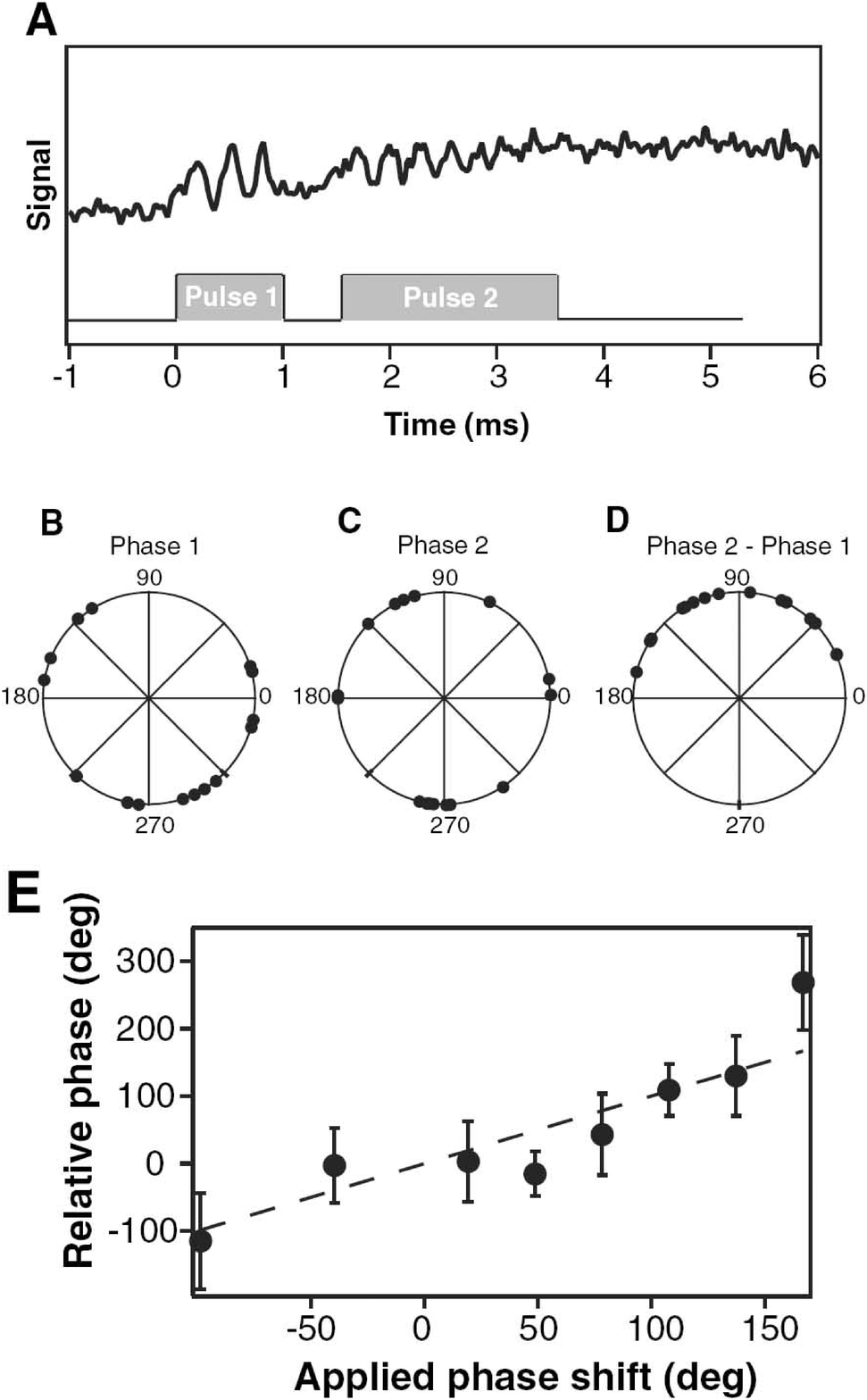}
    \caption{Preparing a relative phase between two independent BEC's
    with no initial phase relation. (A) The temporal trace of the
    Bragg beam intensity shown with the pulse sequence. (B) Phase of
    the oscillations recorded during the first pulse. (C) Phase during
    the second pulse. (D) Phase difference between (B) and (C). (E)
    Phase difference between the oscillations in two pulses as a
    function of the phase shift applied during the evolution time
    between pulses. Each point is the average of several shots
    (between 3 and 10).  Figure and caption reproduced from
    \cite{SPS05}. \label{fig:LightScattBECifm}}
    \end{center}
\end{figure}

The basis of the measurement is the beating of two atom lasers out
coupled from the two condensates by imparting a momentum $\hbar q$.
If the relative phase of the condensates is fixed, the total number
of out coupled atoms oscillates sinusoidally with periodicity $h/d$
as $\hbar q$ is scanned ($d$ is the separation of the condensates).
The experimental tool used to impart a precise momentum to atoms is
Bragg scattering. Two counter-propagating laser beams with wave
vectors $k_1$, $k_2$ hit the atoms so that, by absorbing a photon
from one beam and re-emitting it into the other one, the atoms
acquire recoil momentum $\hbar q=\hbar (k_2-k_1)$ (provided that the
energy difference between photons matches the atom recoil energy).
For each atom out-coupled, a photon is transferred from one beam to
the counter-propagating one. Therefore, all information contained in
the stream of out-coupled atoms is also present in the light
scattered from one beam to the other. Relative phase data were
gathered in real time by monitoring the intensity of the weaker of
the Bragg laser beams, instead of terminating the experiment to
measure the out-coupled atoms using absorption imaging.

Since the relative phase of the condensates can be measured after
scattering only a small fraction of the atoms out of the
condensates, this technique gives a relatively nondestructive
measurement of the relative phase. This technique therefore allows
one to prepare an initial relative phase (by an initial
measurement) of the separated condensates, then to read it out
continuously, and thereby to monitor the phase evolution.  This
way one can realize interferometry between two trapped
Bose-Einstein condensates without ever splitting or recombining
the wavefunction. The condensates can't be too far apart, however,
as the relative atom number uncertainty cannot arise until the
atoms out-coupled from the first condensate have time to reach the
second atom laser beam and create a downstream atom laser whose
atoms could have arisen from either condensate.  (In fact, when
the atom laser beams interfere destructively, the Bragg beams
operating on the second condensate effectively capture atoms from
the first atom laser beam and insert them in the second
condensate!)  The necessity for this process to have occurred
dictates the temporal delay of the buildup of the light fringes in
part A of Fig.~\ref{fig:LightScattBECifm} - it takes about $250$
$\mu sec$ for atoms to make this trip.

This atom interferometer, featuring interference between
always-separated ensembles of interacting atoms is several
significant steps away from the prototypical interferometer in which
uncorrelated non-interacting individual atoms traverse one at a
time. In fact it resembles a gedankenexperiment involving two high Q
L-C circuits resonant with the ac power source in the lab. Suppose
these are both plugged in to different power outlets for a while,
then disconnected.  If some time later these are attached to the
reference and signal ports of a phase detector, it will read a
definite phase. Moreover, this phase will be reproducible shot to
shot.  If one of the L-C circuits is somehow perturbed, then the
phase shift will be systematically modified.  Does this situation,
involving classical L-C circuits constitute an interferometer, or
just classical fields interfering??

In fact it is almost perfectly analogous to the experiment just
described, with the roles of matter and E\&M waves reversed.  The
L-C resonant circuits are classical containers containing coherent
states of low frequency photons; the light traps are classical
containers containing coherent states of atoms.  In either case
phase shifts can be caused by interactions with the container
(squeezing one of the L-C circuit components or light traps), or by
interactions with the quantized medium within (e.g. by non-linear
circuit elements added to the L-C circuit or by a magnetic field
that interacts with the BEC). The initial coherence is induced by
the exchange of photons with the coherent source provided by the
power generation station in one case, and by the mutual exchange of
atoms in the other.  There is a significant interaction among the
atoms in the BEC, whereas the Kerr effect for L-C circuits is small,
but this is not fundamental.  Neither the L-C circuits nor the light
wells are interfering, both function as classical containers for
\textbf{\emph{waves}} that are phased together. The waves undergo
differential interactions, and interfere later to produce a
measurable phase shift.  Ideally this is solely a measure of the
interaction, but in practice small differences between the two
containers cause detrimental phase shifts.

\subsection{Studies with and of BEC's}

Up to now we have reviewed experiments and theories pertaining to
the coherence of BECs.  Now we shift perspective and consider them
as interesting condensed objects in their own right.  Some of the
earliest work that showed this was the study of the frequencies of
the shape oscillations of the condensate.  In this section we
review experiments that were made using the techniques and ideas
of atom optics and interferometry and that allow one to address
other properties of BECs.  In order, we will review the coupling
of two BECs to mimic the physics of Josephson junctions, their
intrinsic decoherence, and two experiments that probe their
structure.

\subsubsection{Josephson oscillations}
 As shown by \textcite{SFG97}, two trapped BEC's that are weakly
coupled (i.e. by tunneling through the barrier) are represented by a
generalization of the equations that apply to a Josephson junction.
The analog is that the sine of the phase difference causes a current
flow between the traps that changes the number difference (and hence
the potential difference that drives the phase change). Given two
trapped BEC's, by adjusting the tunneling rate (i.e. the coupling
strength between the two BEC's), Josephson oscillations between two
weakly linked Bose Einstein condensates can be studied.

\begin{figure}[t]
\begin{center}
\includegraphics[width = \columnwidth]{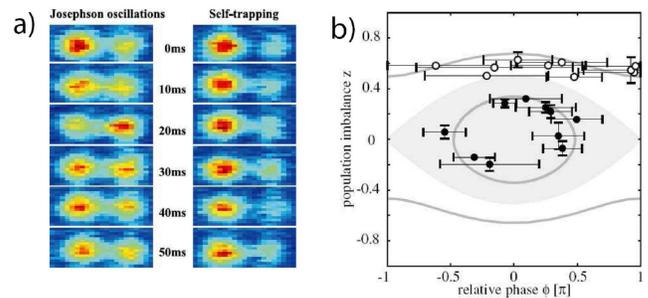}
\caption{(color online) Josephson oscillations. (a) Small population
imbalance causes Josephson oscillations, and large population
imbalance causes self-trapping.   (b) Quantum phase-plane portrait
for the bosonic Josephson junction. In the regime of Josephson
oscillations the experimental data are represented with filled
circles and in the self-trapping regime with open circles. The
shaded region, which indicates the Josephson regime, and the solid
lines are obtained by solving the coupled differential with the
specific experimental parameters. Figure from \cite{AGF05}}
\label{fig:Josephson_PhasePlanePortrait}
\end{center}
\end{figure}

The experiments of \textcite{AGF05} demonstrate both the nonlinear
generalization of tunneling oscillations in Josephson junctions
for small population imbalance $z$, and non linear macroscopic
quantum self-trapping for large population imbalance. The
distinction between the two regimes is very apparent in the
phase-plane portrait of the dynamical variables $z$ and $\Phi$ as
shown in Figure \ref{fig:Josephson_PhasePlanePortrait}. The
successful experimental realization of weakly coupled
Bose-Einstein condensates adds a new tool to both condensed matter
physics and to quantum optics with interacting matter waves.  In
particular, we have to realize that the beamsplitting (and also
the recombination if done at high density) of two BECs must be
discussed in terms of the Josephson effect, or possibly its
generalization.

A detailed study of the phase noise in the interference patterns,
allowed \textcite{GEH06,GAF06,GHF06} to measure the temperature of
the tunnel coupled BEC's. Further examples of tunneling were
investigated with BEC's trapped in optical lattices by
\textcite{ANK98,OTF01} in M. Kasevich's Lab and by
\textcite{CBF01,CFF03} in Florence.

\subsubsection{Spontaneous decoherence and number squeezing}

BECs have an intrinsic decoherence due to fluctuations in the number
of atoms they contain.  If a BEC is prepared in a number (Fock)
state, its phase is indeterminate.  If its phase is determined, for
example by placing the BEC in a coherent state, then it must be in a
superposition of states with different atom number.  (For example, a
coherent state is a (coherent) superposition of states with
different number, with rms variation $\sqrt{N}$.) Since the mean
field energy of a trapped BEC increases with N ($\sim$ $N^{2/5}$ in
a harmonic trap), this means the different components have different
energy, evolve at a different rate, and get out of phase.  The time
for this to happen is typically 25 to 50 ms, severely limiting the
accuracy of BEC interferometers.

Even if a BEC interferometer starts with a definite number of
atoms in the central well, the $\sqrt{N}$ projection noise at the
beam splitter, translates into fluctuations of the chemical
potential which results in fluctuations in the accumulated phase
of the interferometer and consequently in a rapid dephasing of the
split BEC. The phase diffusion rate can then be estimated by:
\begin{equation}
  R_\phi = \frac{1}{\hbar} \frac{d \mu}{d N} \Delta N,
\end{equation}
where $N$ is the number of Atoms in the BEC, $\mu$ its chemical
potential.  With the chemical potential $\mu$ larger then the
trapping frequency $\omega$ ($\mu > \hbar \omega$) for trapped atoms
after typically a few transverse trapping times the phase is random
and the coherence is lost. This phase diffusion caused by the
interactions between the atoms puts stringent limits on the
persistence of coherence in a BEC interferometer.

This interaction-induced dephasing can be reduced in different
ways:

\begin{itemize}
    \item Reduce the effect of interactions by
tuning the scattering length with a Feshbach resonance.  This may
permit setting the scattering length to zero.  This requires
precise control over the magnetic field, and may limit the number
of atoms used in the experiments since the mean field repulsion is
proportional to the scattering length and hence the ground state
condensate will no longer be spread out.

\item If the method of light scattering described above to measure
the phase evolution of the two condensates is applied to two
initially number-squeezed condensates (e.g. if a large condensate
were separated adiabatically), it will add differential number
uncertainty only in proportion to how well the phase is
determined.

\item If the splitting is performed adiabatically, the repulsive
interaction itself will tend to equalize the chemical potentials
of the splitting condensates.  Thus the relative atom number
distribution will be reduced if the splitting is performed slowly.
This will reduce the relative phase diffusion rate of the initial
condensates at the cost of an increased uncertainty in the initial
phase, but this can be increased to the measurement noise level
without penalty.  For interferometers using large condensates this
can lead to significant increases in their sensitivity and
applicability.
\end{itemize}

In fact, dramatic observations of number squeezing have already
been made.  Squeezing between atoms trapped in arrays of traps was
observed by \textcite{OTF01}. Recently \textcite{JSW06} observed a
dramatically reduced phase diffusion in a trapped BEC split with
an RF splitter on an atom chip.

\subsubsection{Structure studies of BEC}

According to theory, a BEC possesses collective modes (e.g. sound
waves) due to the interactions of the atoms.  In a quantum
many-particle description, it's dispersion relation has the
Bogoliubov form \beq \nu = \sqrt{\nu_0^2 + 2\nu_0 \mu/h},
\label{eq:bogoliubov}\eeq where $\mu = n4\pi \hbar^2a/m$ is the
chemical potential, with $a$ and $m$ denoting the scattering
length and the mass, respectively, $n$ is the density of the
condensate, and $h\nu_0 =q^2/2m$ is the free particle dispersion
relation. \cite{SCG01,OKS05}.

\begin{figure}[t]
\includegraphics[angle=0,width=\columnwidth]{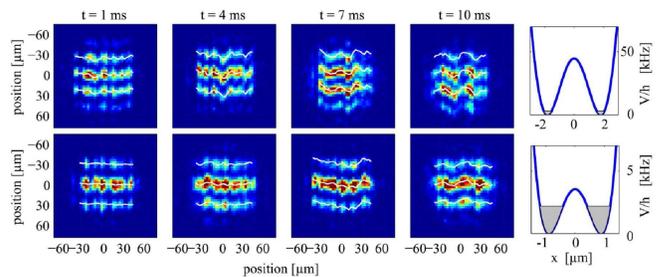}
\caption{\label{Fig:examples} Direct observation of the phase
dynamics through interference. Example images of the observed
interference patterns for hold times $t=1,4,7,10$\,ms \emph{(top)}
in the case of isolated 1d systems and (\emph{bottom)} for finite
tunnel coupling. The different transverse double-well potentials
shown as indicated. (adapted from \textcite{HLF07}}
\end{figure}

In a typical Rb or Na condensate, $\mu/h$ is about a kHz,
corresponding to speeds of $\sim$cm/sec or less.  The Bragg
Spectroscopy discussed previously generates atoms with several times
this speed, which therefore have nearly their free-particle
dispersion relation (the mean field energy term being negligible).
However, by reducing the angle of the Bragg beams from 180 to much
smaller angles, the transferred momentum was correspondingly
reduced, and many fewer atoms are liberated from the condensate
(i.e. the static structure factor is no longer unity), and the
frequency shift relative to a free particle follows
Eq.\ref{eq:bogoliubov}. Studies of BEC structure are in
\cite{KOS04,SKO03}, and theory for these measurements is discussed
by \cite{CAL00,BLB00}.

Physics which goes deeper into the properties of degenerate
quantum gases and their coherence properties is outside the
purview of this review, so we refer the reader to a series of
excellent reviews in the literature which summarize the status of
this still very fast moving field \cite{DGP99,KET02,COW02,KAS02}.

\subsubsection{Dynamics of coherence in 1-D systems}

\begin{figure}[t]
\includegraphics[angle=0,width=.98\columnwidth]{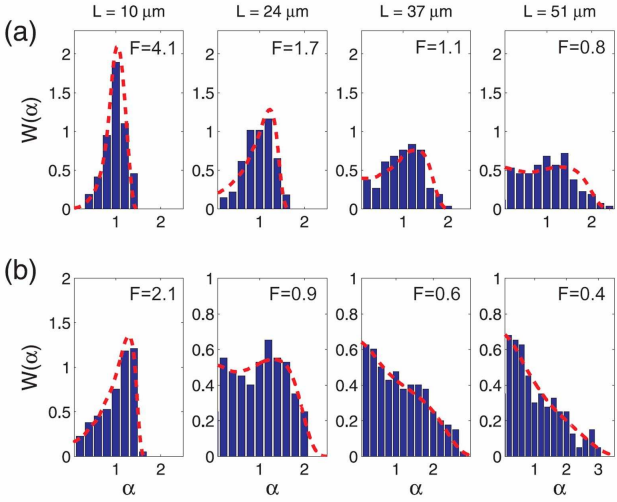}
\caption{\label{Fig:distributions}Distribution functions of the
measured interference contrasts for different lengths $L$ along the
1d Condensate. \textbf{(a)} The length-dependent normalized
interference contrasts $\alpha $ with parameters ($n_{1d}=60
\,\mu$m$^{-1}$, $\nu_\perp=3.0$\,kHz, $K=46$). The red curves show
the corresponding calculated distributions for $T=30$\,nK
($\xi_T=0.9 \,\mu$m). \textbf{(b)} Same parameters as in
\textbf{(a)}, but higher temperature $T=60$\,nK. For both sets
\textcite{HLS07} observe the predicted change of overall shape of
the distribution functions from single peak Gumbel-type
characteristic for quantum noise to Poissonian characteristic for
thermal noise. (adapted from \cite{HLS07})}
\end{figure}

Interference allows to study the dynamics of (de)coherence in
degenerate Bose gases.  This is especially interesting in the
one-dimensional (1D) regime where long-range order is prevented by
the ubiquitous phase-fluctuations.

In their experiments \textcite{HLF07} coherently split a 1d
quasi-condensate, characterized by both the temperature $T$ and
chemical potential $\mu$ fulfilling $k_\text{B} T, \mu < h
\nu_\perp$, along the transverse direction which initializes the
system in a mutually phase coherent state, and phase fluctuation
patterns of the two individual 1d systems being identical. This
highly non-equilibrium state relaxes to equilibrium over time and
the evolution of (de) coherence is revealed in local phase shifts
leading to increased waviness of the interference pattern (Figure
\ref{Fig:examples}).

If the two parts of the system are completely separated, the
equilibrium state consists of two uncorrelated quasi-condensates and
\textcite{HLF07} observe a randomization of the relative phase
$\theta(z,t)$ as expressed in the coherence factor $
\Psi(t)=\frac{1}{L} \left|\int dz\, e^{i\theta(z,t)}\right|$. Most
interestingly $\Psi(t)$ decays sub exponential $\Psi(t)\propto
e^{-(t/t_0)^{2/3}}$ as predicted by \textcite{Burkov2007} based on a
Luttinger liquid approach \cite{Haldane1981}. Qualitatively similar
behavior was recently observed at MIT \cite{Jo2007b} for elongated
condensates with $\mu \sim 2 h \nu_\perp$ and $T \sim 5 h
\nu_\perp$.

For finite tunnel coupling between the two systems, \textcite{HLF07}
observe that the final equilibrium state shows a non-random phase
distribution (Figure \ref{Fig:examples} \emph{(bottom)}). The phase
randomization is counterbalanced by the coherent particle exchange
between the two fractions, equivalent to injection locking of two
matter wave lasers. The final width of the observed phase spread
depends on the strength of the tunnel coupling \cite{GHF06}.

\subsubsection{Measuring noise by interference}

In many-body systems quantum noise can reveal the non-local
correlations of underlying many-body states \textcite{Altman2004}.
Recently it has been suggested that the statistics of the shot to
shot fluctuations in fringe contrast probe higher order correlation
functions \cite{Polkovnikov2006,Gritsev2006}.

This rational was used by \textcite{HLS07} in an experiment
investigating the statistical properties of interference experiments
performed with pairs of independently created one-dimensional atomic
condensates \cite{HLF06}. The shot-to-shot variations of
interference can then be directly related to the full distribution
functions of noise in the system \cite{Polkovnikov2006}. Probing
different system sizes they observe the crossover from quantum noise
to thermal noise, reflected in a characteristic change in the
distribution functions from Gumbel-type to Poissonian
\ref{Fig:distributions}. The results are in excellent agreement with
the predictions of \textcite{Gritsev2006} based on the Luttinger
liquid formalism \cite{Haldane1981}.

These experiments demonstrate the power of quantum noise analysis in
interference patterns as a probe of correlated systems, and the
power of simple ultra cold atom systems to exhibit and illustrate
fundamental quantum processes relevant in many areas of physics.

\subsubsection{Momentum of a photon in a medium}
The momentum of a photon propagating in a medium is a topic fraught
with controversy.  When an electromagnetic wave enters a medium with
index of refraction $n$, its wavelength is reduced, and its
wavenumber is increased, by $n$.  Thus is seems evident that a
single photon in this medium would have momentum $p=n \hbar
k_{vac}$, a conclusion reached by \textcite{MIN08} \cite{MIN10}
using classical physics. On the other hand, if the photon is
considered as a particle, it seems very strange that it should
increase its momentum when entering a medium in which its speed is
reduced! Such a viewpoint is supported by \textcite{ABR09} who found
$p=\hbar k_{vac}/n$. Resolving these two viewpoints has been cited
as one of the challenges of theoretical physics \cite{Peierles91}.

\begin{figure}[t]
\begin{center}
\includegraphics[width = 7cm]{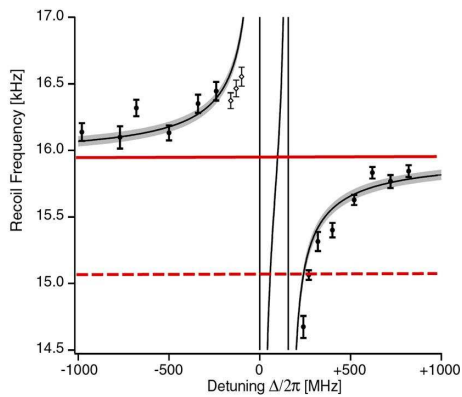}
\caption{Recoil frequency measured by \cite{CLM05} on two sides of
resonance.  The dashed line shows expected result for free atoms,
solid line corrects for chemical potential assuming $p=\hbar
k_{vac}$.  Solid line with error shading shows expectation if $p=n
\hbar k_{vac}$.} \label{fig:campbell}
\end{center}
\end{figure}

When a photon propagating in an atomic gas is absorbed by one of
the atoms in the BEC, what is the momentum of the atom after the
absorption?  This question seems less subject to uncertainty since
it can be settled by a measurement; it is also important in
precision experiments to measure $h/m$ that will be discussed in
the next Section.  For a dilute atomic gas, a third opinion seems
justified: a BEC has only a few atoms per cubic wavelength, and no
obvious mechanism to transfer momentum to/from the atoms not
involved in the absorption - hence the atom will absorb momentum
$p=\hbar k_{vac}$.

In a recent experiment done in a BEC by \textcite{CLM05}, a double
pulse Kakpita-Dirac interferometer was used to measure the recoil
energy of Rb atoms in a BEC for laser frequencies on both sides of
the resonance.  The results showed marked structure near the
resonance, consistent with that predicted from the variation of
$n$ near the resonance only if $p=n \hbar k_{vac}$.

With continued progress on these topics, interferometers with BEC
hold the promise to be employed as highly sensitive devices that
will allow exploration of a large variety of physics questions.
These range from atom-surface interactions to the intrinsic phase
dynamics in interacting (possibly low dimensional) quantum systems
or the influence of the coupling to an external `environment'
(decoherence).

\subsection{Testing the charge neutrality of atoms}

The equality of the electrical charges of the electron and proton,
and the charge neutrality of the neutron are of great significance
in the fundamental theory of particles \cite{CHU87,UNG04}.
Experimental tests of the electrical neutrality of bulk solid
matter and bulk quantities of gas are precise enough at present to
state that $(q_p + q_e)/e < 10^{-21}$ \cite{DYK73,MAM82,MAM84}. An
experiment searching for deflection of a neutron beam has set a
similar limit for the electric charge of the neutron $q_n <
10^{-21}$ \cite{BGK88}. Experiments with individual atoms or
molecules in a beam have only been able to verify the net
electrical charge of $(q_p + q_e)$ is less than $10^{-19}e$
\cite{ZCH63,HFC88}.

A dedicated atom interferometry experiment could detect a phase
shift if $(q_p + q_e)/e = 10^{-22}$. The phase shift would be \beq
\phi = \frac{(q_p + q_e)Z \mathcal{E} \Delta x L }{ v \hbar }
\approx 10^{-4} rad \eeq where we have assumed $Z=55$ is atomic
number, $\mathcal{E} = 10$ kV/mm is the applied field, $\Delta x =
100 \mu$m is the separation of the paths in the interferometer, $L$
is the length of the interaction region, and $v = 100$ m/s is the
atomic velocity. \textcite{CBD01b} and \cite{DCB01} studied this.
The main difficulty will come from the electric polarizability of
atoms which will cause large phase-shifts due to field gradients.
But because these phase-shifts are quadratic in the applied electric
field while the proposed effect is linear, these stray phase-shifts
should mainly limit sensitivity.

\section{PRECISION MEASUREMENTS}

Since their demonstration in 1991, atom interferometers have
become precision measurements tools. The advantages of small de
Broglie wavelengths, long propagation times, and the narrow
frequency response of atoms are responsible for atom
interferometers already having made an impact on many fields of
fundamental science and engineering.  In the present section, we
discuss measurements of acceleration, platform rotation, the Molar
Planck constant ($N_A \times h$) and the fine structure constant
($\alpha$). Although some of the measurements of atomic and
molecular properties are precision measurements by the standards
of those fields, they will all be discussed in Section VI .

\subsection{Gravimeters, Gryroscopes, Gradiometers}

Inertial sensors based on atom interferometers already perform
comparably to the best available sensors based on any technology. At
their current levels of resolution summarized in Table
\ref{tab:inertial sensors} several interesting applications are
within reach.  In fact, development has begun for commercial sensors
and commercial applications using atom interferometers. To explore
the precision, resolution, accuracy, response factor, bandwidth,
dynamic range, and stability achievable with atom interferometers we
begin by looking at the different designs used for gravimeters,
gyroscopes, and gravity gradiometers.

\begin{table}[h]
\caption{\label{tab:inertial sensors}Inertial sensing resolutions
demonstrated with atom interferometers.}
\begin{ruledtabular}
\begin{tabular}{ll}
Sensor&Resolution \\
\hline Gravimeter\footnote{\cite{PCC01}} & $2 \times 10^{-8}$ \
($g$) /$\sqrt{\textrm{Hz}}$ \\
Gravity gradiometer\footnote{\cite{MFF02}} & $4\times 10^{-9}$ \  ($g$/m)/$\sqrt{\textrm{Hz}}$   \\
Gyroscope\footnote{\cite{GLK00}} & $6\times10^{-10}$
(rad/sec)/$\sqrt{\textrm{Hz}}$  \\
\end{tabular}
\end{ruledtabular}
\end{table}

Thermal atom beams for rotation, freely falling atoms for
acceleration, and two clouds of falling atoms with a common laser
beam for the gradiometer have given the best results to date. This
is in part because rotation sensors have a \emph{response factor}
(i.e. phase shift per rotation rate) that increases linearly in
proportion to the atom's time of flight; but accelerometers have a
response factor that increases quadratically with time of flight.
Part of the tradeoff is that fast atom beams offer more atoms per
second than cold atom sources. Larger interferometers will improve
sensitivity and slow atom interferometers can make compact
sensors. In each case, to judge the overall performance one must
also look at systematic errors.

Displacements from an inertial reference frame with a constant
acceleration $\vec{g}$ and a constant rotation $\vec{\Omega}$ causes
a phase shift for a 3-grating interferometer \beq \phi = (\vec{G}
\cdot \vec{g}) \tau^2 + 2 \vec{G} \cdot ( \vec{\Omega} \times
\vec{v}) \tau^2. \label{eq:inertial} \eeq where $\vec{G}$ is the
reciprocal lattice-vector of the gratings, and $\tau$ is the time of
flight for atoms with velocity $\vec{v}$ to travel between gratings
\cite{ANA81,MAL00,DRY79,DUK06,BLK06}. Referring to our previous
section on the origin of phase shifts, this phase is equivalent to
the envelope shift, a classical property. Equation \ref{eq:inertial}
can be derived from the grating phase (introduced in Section III
Equation \ref{eq:3g std phase}) \beq \phi = \vec{G} \cdot
\left[\vec{x}_1(t_1) -2\vec{x}_2(t_2) + \vec{x}_3(t_3) \right] \eeq
where $x_i$ is the transverse position of the $i$th grating (with
respect to an inertial frame) at time $t_i$ (when the atoms interact
with the grating).

Rotation about the center grating in a space-domain interferometer
causes a phase shift \beq \phi_{atom} = 2  \Omega G L \tau = 4 \pi
\Omega\frac{m}{h}A \label{eq:sagnac for atoms} \eeq where $L$ is the
separation between gratings, $\tau$ is the time of flight between
gratings, and $A$ is the area enclosed by the interferometer paths.
For an optical interferometer \beq \phi_{light} = 2  \Omega G L^2
\frac{1}{c} = 4 \pi \Omega \frac{1 }{\lambda_{ph}c}A.
\label{eq:sagnac for light} \eeq The ratio of phase sifts for a
given rotation rate ($\Omega$), assuming equivalent interferometer
areas ($A$), is \beq \frac{\phi_{atom}}{\phi_{light}} =
\frac{mc^2}{\hbar \omega} =
\frac{\lambda_{ph}}{\lambda_{dB}}\frac{c}{v}\approx 10^{10}. \eeq
This famous ratio shows that atom interferometers have a huge Sagnac
response factor compared to optical interferometers.

However, to really gain this large increase in resolution (at the
expense of bandwidth $v/c$), both the enclosed area and the count
rate of the two types of interferometers must be equal.  But a fiber
optic ring gyro can easily have an enclosed area of \mbox{$A= 10^3$
m$^2$} and still have a much better bandwidth compared with the
largest atom interferometers that have \mbox{$A = 10^{-4}$ m$^2$}.
So the response factor is only a few orders larger for today's atom
interferometers. Furthermore, while the count rate for an optical
Watt of power is on the order of $10^{19}$ photons per second,
typical atom interferometers offer only $10^7$ atoms per second.

For acceleration, one can see from Eq.~\ref{eq:inertial} that,
\beq \frac{\phi_{atom}}{ \phi_{light}} = \left( \frac{c}{v}
\right)^2 \eeq if identical gratings are used for light and atom
interferometers.

The presence of velocity $v$ in the Eq.~\ref{eq:inertial} has two
important consequences. For a space-domain interferometer, the
acceleration phase depends on $\tau^2$ while the rotation phase
depends on $\tau$. Therefore, slow atoms are particularly
advantageous for sensing acceleration, but fast atoms (beams) offer
competitive sensitivity for gyroscopes.  That is why the best
gravimeters use cold atoms, and the best gyroscopes use thermal
atomic beams. From the vector notation in Eq. \ref{eq:inertial} one
can see that reversing the atom velocity switches the sign for the
rotation phase but not the acceleration phase. This provides a
method to distinguish $\Omega_x$ from $g_y$ or $g_z$.  Kassevich
used counter-propagating atom beams for this reason \cite{GLK00}.

Instrument \emph{resolution} is given by the response factor times
the precision with which the phase shift can be measured. Since the
noise-limited phase precision increases with the square root of time
(as discussed in Section III Eq. \ref{eq:sigma_phi}), it is
customary to report the resolution per root Hertz. Instrument
\emph{bandwidth} is limited in part by the desired resolution and
also simply by the atom's time of flight. \emph{Dynamic range} can
be limited by dispersion. For example, if there is a velocity spread
in a space domain interferometer, then the resulting spread in
inertial phase decreases the contrast, as discussed in Sections III
and IV.  For a Gaussian distribution in phase with an RMS
$\sigma_{\phi}$, the contrast is reduced by the factor $C/ C_0 =
\langle e^{i\phi} \rangle = e^{-\frac{1}{2}\sigma_{\phi}^2}$.

\begin{figure}[h]
\begin{center}
\includegraphics[width = 8cm]{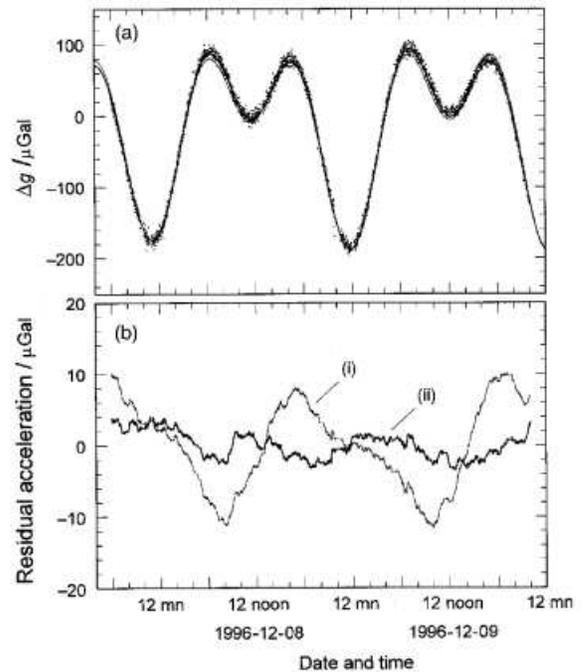}    
\caption{(a) Two days of gravity data.  Each data point represents
a 1 min gravity measurement.  The solid lines represent two
different tidal models.  (b) The residuals of the data with
respect to a tidal model where (i) the earth is modelled as a
solid elastic object and (ii) the effects of ocean loading of the
Earth are taken into account.  Figure and caption reproduced from
\cite{PCC01}.} \label{fig:gravity and tides}
\end{center}
\end{figure}

Measurements of gravitational acceleration in the engineering
literature are often reported in units of $\mu$Gal \mbox{(1
$\mu$Gal = $10^{-8}$ m/s$^2$)} or the more common unit of $g$ ($g
\approx$ 9.8 m/s$^2$). Many applications in geophysics are
currently served with sensors that have $5\times10^{-9}$g
(5$\mu$Gal) precision after averaging for 15 minutes \cite{AGC00}.
The light pulse (Raman) interferometer in \cite{PCC01} (described
in chapter III) attains this precision in less than one minute.
Measurements with this apparatus that show time-variations in
local $g$ due (mostly) to tides are shown in Figure
\ref{fig:gravity and tides}. Some variations in $g$ due to sources
of geophysical interest are shown in Table \ref{tab:inertial
effects}.


\begin{table}
\caption{\label{tab:inertial effects}List of geophysical sources
of change in $g$.\cite{PCC01,AGC00,CSE03}}
\begin{ruledtabular}
\begin{tabular}{lr}
Gravitation Source &Magnitude  \\
\hline
Tides at Stanford, CA & $2 \times 10^{-7}$ $g$   \\
1000 kg  1.5 meters away  & $3 \times 10^{-9}$ $g$   \\
Loaded truck 30 m away & $2 \times 10^{-9}$ g     \\
Elevation variation of 1 cm & $3 \times 10^{-9}$ $g$   \\
Ground water fluctuation of 1 m & $5 \times 10^{-9}$ $g$   \\
10$^8$ kg of oil displacing salt at 1 km & $5\times10^{-7}$ $g$
\end{tabular}
\end{ruledtabular}
\end{table}

\begin{table}
\caption{\label{tab:rotation effects} Rotation rates due to
various causes.}
\begin{ruledtabular}
\begin{tabular}{lr}
Cause & Rotation Rate  (rad/s)  \\
\hline Earth's rotation &  $\Omega_e = 7.2 \times 10^{-5}$ \\
Tidal drag in 1 yr.& $\delta \Omega_e =  10^{-13}$\\
Lense-Thirring & $\Omega_{LT} = 10^{-14}$ \\  Geodetic Effect &
$\Omega_{GD} = 10^{-12}$
\end{tabular}
\end{ruledtabular}
\end{table}

Gravity gradients in the engineering literature are often measured
in units of E (1E = $10^{-9} s^{-2} \approx 10^{-10}$ $g$/m) or
simply $g$/m.  By measuring the differential gravitational
acceleration in two atom interferometers located one meter apart
from each other, \textcite{SMB98} measured the Earth's gravity
gradient ($\nabla g$ = $3\times 10^{-6}$ $g$/m) with an
uncertainty of 5\% and \cite{FFM07} measured the change in gravity
gradient caused by a 540 kg source mass of Pb ($\nabla g$ =
$8\times 10^{-9}$ $g$/m) with an uncertainty of 0.3\%. Related
measurements are also described in \cite{KAS02,FFM02,MFF02}.
Second order phase shifts due earth gravity and gavity-gradients
and centrifugal and Coriolis forces due to Earth rotation have
been identified by \cite{DUK06,BLK06,BLC06}.

\emph{Historical background:} The first measurements of $g$ with a
matter-wave interferometer was done with neutrons by
\textcite{COW75}. An early proposal for atom interferometer
measurements of $g$ by \textcite{CLA88} was followed by several
demonstrations with rapidly improving resolution and accuracy
\cite{KAC92,BER97,BER97-YKC,PYC97,PCC99,PCC01}.  An atom beam
sensor for little $g$ based on the classical moire-effect was also
constructed with three material gratings by \textcite{OBR96}.

In 1913, \textcite{SAG13,SAG13b} made his famous
light-interferometric measurement of platform rotation.
\textcite{MIC25} measured the rotation rate of the earth,
$\Omega_e$, with a large optical interferometer.  The Sagnac effect
with neutron and electron interferometers has also been demonstrated
\cite{WSC79,HAN93}.  Atom interferometer gyroscopes were proposed
early on by \textcite{CLA88}. An atom interferometer Sagnac
gyroscope was first built by \textcite{RKW91}, and huge improvements
in sensitivity were demonstrated by \cite{LHS97,GBK97,GLK00}.

\subsection{Newton's constant $G$}

Newton's constant $G$ is the least accurately known fundamental
constant. The 2005 CODATA value of $G$ has a precision of
$1.4\times10^{-4}$ \cite{MOT05}, although several individual
experiments have recently claimed precision better than this
\cite{GUM00,QPR01}. Atom interferometry is a relatively new method
to measure $G$, and may soon provide comparable precision to the
CODATA value.

The Kasevich group determined $G$ with a precision of $3 \times
10^{-3}$ \cite{FFM07}, and the Tino group reported a value for $G$
with a precision of $1 \times 10^{-2}$ \cite{BLC06}. Both of these
groups use two atom interferometer gravimeters and a movable source
mass of order 500 Kg. The Tino group plans to extend their precision
to the $10^{-4}$ level.  Methods to measure $G$ with atom
interferometry are also discussed in \cite{MFF02,FLP03,SFP03,KAS02}.

\subsection{Tests of Relativity}

In accord with Einstein's principle of equivalence, atomic mass
$m$ does not enter into Equation \ref{eq:inertial}. However,
theories that go beyond Einstein's general relativity motivate the
search for composition-dependent gravitational forces. The
principle of equivalence has been tested accurately enough to
state $\Delta$g/g = 1.2$\pm1.7\times 10^{-7}$ for the two
different Rb isotopes \cite{FDH04} and there are plans based on
current technology to increase the precision of this test to
$\Delta$g/g$\sim 10^{-15}$ (300 times better than current limits
from any method) \cite{DGH06}.

Searches for a breakdown of the $1/r^2$ law are another test of
general relativity, in this case motivated by string theories and
the possibility of compact dimensions. Experiments to detect
non-Newtonian gravitational potentials with multiple atom
interferometers located at different distances from the earth's
center have been discussed in \cite{MDR02,DGH06}. Experiments to
search for a breakdown of the $1/r^2$ law at micrometer length
scales using atom interferometry were discussed in
\cite{DIG03,FPS06}.

The gravitational scalar A-B effect would be an interesting test at
the intersection of quantum mechanics and gravity.  If a 5 cm radius
lead sphere has a small hole in the center, Cs atoms placed there
have a frequency shift of about 7 Hz.  Thus atoms at the top of
their trajectory could easily experience a phase shift $\sim$10
radians, enabling a quantum measurement of the gravitational
potential.  If the lead is assembled around the atoms in one
interferometer arm, or if the lead is moved into position quickly
compared to the atoms' transit time, the effects of the
gravitational field (force) can largely be eliminated, making this a
sensitive measure of gravitational potential.

Atom interferometer rotation sensors in low earth orbit should be
able to measure the geodetic effect and possibly the Lense-Thirring
rotation. According to Special Relativity, freely-falling gyroscopes
orbiting in the vicinity of the Earth will experience the geodetic
effect caused by the motion of the gyroscope in the gravitational
field of the earth \cite{SCH60,JMR04}.  For low earth orbit, the
rotation rate induced by the geodetic effect is $10^{-12}$ rad/sec,
and is independent of the earth's rotation rate.

The Lense-Thirring rotation \textcite{THI18} is a General Relativity
effect that causes a gyroscope to rotate relative to the fixed stars
due to a massive rotating body being nearby. It is also called the
gravito-magnetic effect.  In low earth orbit (700 km altitude), this
can be as large as $10^{-14}$ rad/sec and depends on the orientation
of the earth's spin. Measurements of both the geodetic effect and
the Lense-Thirring effect is the objective of future space bourne
atom interferometer missions \cite{JMR04}.

\subsection{Interferometers in orbit}

In addition to ultra-precise atomic clocks to improve the atomic
clocks already aloft for the GPS system, physics experiments that
could benefit from being in space include measurements of the
gravitational red-shift, tests of Einstein's equivalence
principle, mapping the Lense-Thirring effect close by the Earth,
and measurements of $h/m$.

\begin{figure}[h]
\begin{center}
\includegraphics[width = 6cm]{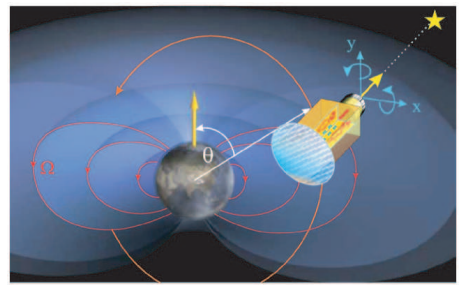}
\caption{(color online) The Mission Scenario: HYPER, which follows
a nearly polar circular orbit, will measure with two atomic
gyroscopes the two characteristic components of the Lense-Thirring
rotation as a function of latitudinal position $\theta$.  Figure
and caption reproduced from \cite{JMR04}. \label{fig:LT2space}}
\end{center}
\end{figure}

NASA works on these goals with the `Laser Cooled Atom Physics'
(LCAP) and ultra-precise primary atomic reference clocks in space
(PARCS) programs planned for the international space station
\cite{LEI03}. The European Space Agency's `HYPER-precision atom
interferometry in space' project is described in several articles in
General Relativity and Gravitation, Vol. 36, No. 10, (2004) starting
with \cite{JMR04} (see Figure \ref{fig:LT2space}).

\subsection{Fine structure constant and $\hbar /$M}

One of the highest precision atom interferometry experiments is the
measurement of $\hbar / m_{atom}$ This leads to a measurement of the
Molar Planck Constant, $N_A \times h=M_{atom}/m_{atom} \times h
\times 1000$ where $M_{atom}$ is the atomic weight of the atom in
grams and the factor of 1000 comes in converting the atomic mass
into kilograms. This was done at Stanford by
\cite{WYC93,WYC94b,WHS02}, and more recently in Paris by Biraben and
cowrokers \cite{CMC06,BCG04}.  Both groups achieved a precision of
$\sim$14 parts per billion by measuring the velocity change of an
atom due to the photon recoil (from emission or absorption). As we
shall discuss, these measurements lead to a value for the fine
structure constant at $\sim$7ppb when combined with other
measurements.

The underlying physics, first exploited using neutrons by
\textcite{KNW95}, is based on the deBroglie wavelength, \beq
\lambda_{dB} = h/mv. \eeq Obviously a simultaneous measurement of
both $\lambda_{dB}$ and $v$ gives $h/m$, where $m$ is the mass of
the neutron of the particular atom used in the experiment. In the
interferometer experiment of Chu, the measured quantity is
essentially the frequency with which an atom with the recoil
velocity (from absorbing a photon of wavevector $k$) \beq v_{rec} =
\hbar k/m_{Cs} \eeq crosses the fringes in a standing wave, \beq
\omega = 2 \pi v_{rec}/ \lambda_{dB} \sim \hbar k^2/m_{Cs} \eeq
where we have replaced $\lambda_{dB}$ with the wavelength of the
light causing the recoil. In the Biraben experiment, the Doppler
shift associated with this recoil is measured: \beq \omega_D = k
v_{rec}=\hbar k^2/m_{Rb}. \eeq These frequencies are both equal to
the recoil frequency (typically 10 kHz) derived earlier from
consideration of the energy of recoil.

In the actual experiments, the measured frequency is several times
the recoil frequency. Measuring the small recoil frequency to ppb
accuracy is impossible given a maximum free fall time for the
atoms of a fraction of a second. Hence both experiments increase
the measured velocity by contriving to add recoil velocities from
the absorption of many photons.  In the Chu experiments these are
added in using up to ~60 Raman pulses or adiabatic (STIRAP)
transfers; in the Biraben experiment by accelerating an optical
lattice into which the atoms are embedded. Although the initial
and final lattice speeds are not quantized, the atoms accelerated
in them always absorb an integral number of lattice momenta (sum
of momenta in the two laser beams forming the moving lattice) - up
to 900 photon momenta in \cite{CMC06}.

Both of these experiments are essentially measurements of
velocity, using the combined techniques of atom optics (to add
velocity) and atom interferometry to detect it.  This is indicated
by the fact that the signal increases \emph{linearly} with the
extra velocity.  An interferometer configuration that uses
contrast interferometry to measure the recoil \emph{energy} has
been proposed and demonstrated \cite{GDH02}.  It shows the
quadratic dependence of phase shift on photon number (velocity)
expected for an energy measurement, and therefore requires that
less additional momentum be added to achieve the same precision.

An important consequence of the $\hbar / m$ measurement is to
provide a very high accuracy route to the determination of the
fine structure constant, $\alpha$.  This is based on the
relationship \beq \alpha^2= \left(\frac{e^2}{\hbar c} \right)^2 =
\frac{2R_{ \infty}}{c}\frac{h}{m_e} \eeq
Combining atom interferometer results with independent
measurements of the optical frequency ($\omega = ck$) \cite{GCT06}
and the mass ratios $m_{Cs}/m_p$ \cite{BPR99,RSZ96}, and $m_p/m_e$
\cite{MOT05}, and the Rydberg $R_{\infty}$ \cite{MOT05}, gives a
value of the fine structure constant.

\begin{figure}[h]
\begin{center}
\includegraphics[width = 7cm]{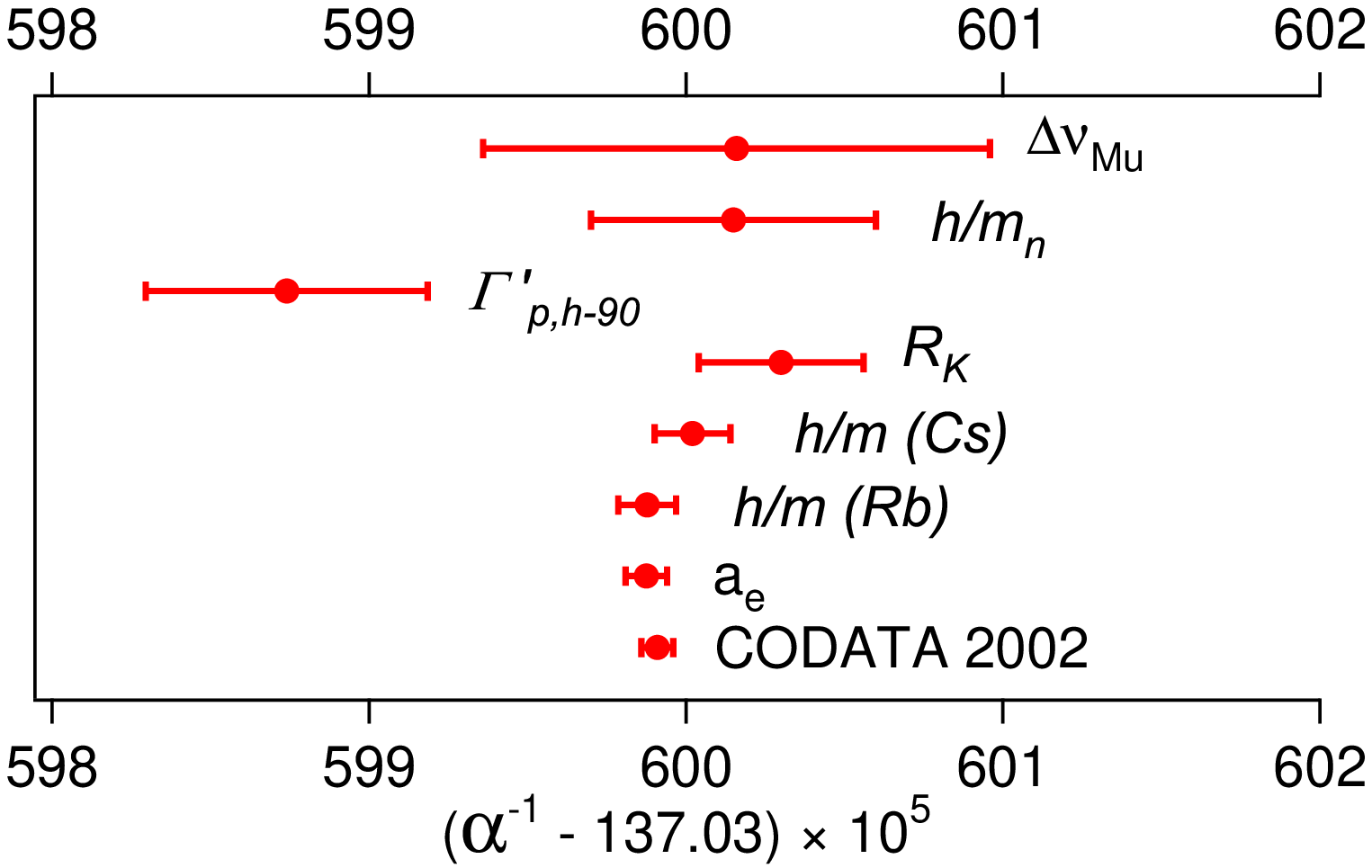}
\caption{Determinations of the fine structure constant, $\alpha$, by
several methods.  The value from $h/m$(Cs) is from atom
interferometry \cite{WHS02} and the $h/m$(Rb) value is determined
with Bloch oscillations \cite{CMC06}.  References are given in
\cite{MOT05} for the values obtained from measuring the muonium
hyperfine splitting ($\Delta \nu_{\textbf{Mu}}$), or measuring the
von Klitzing constant with the quantum Hall effect ($R_K$), or
measuring recoil velocity of neutrons Bragg-reflecting from silicon
crystals ($h/m_n$), or measuring gyromagnetic ratios
($\Gamma'_{p,h-90}$), or measuring electron and positron anomalies
($a_e \equiv g_e/2 - 1$). Figure adapted from \cite{MOT05}.
\label{fig:alpha}}
\end{center}
\end{figure}

The determination of $\alpha$ from $h/m_{Cs}$ has a precision of 7
ppb \cite{WHS02}, and from Rb of 6.7 ppb \cite{CMC06}. Thus this
route already offers the second most accurate value of $\alpha$
(after the measurement of g-2 for the electron), and therefore
allows the most precise comparison across different subfields of
physics, as shown in Figure \ref{fig:alpha}.  It essentially offers
a comparison of QED with such things as the deBroglie wavelength
relationship and calculations of atomic structure in hydrogen. Such
cross-field comparisons are extraordinarily important for the unity
and global understanding of physics, and provide one of the few
routes to discover underlying errors in an isolated subfield.  It is
interesting to note that several of the values appearing in figure
\ref{fig:alpha} have been substantially re-evaluated between 1998
and 2002, which proves that the fine structure constant is not so
well known (it is known mostly due to the electron spin anomaly).

No precision experiment is easy, and the $h/m$ measurements
discussed here experience difficulties from vibration that changes
the velocity of the reference light waves (the scheme in
\cite{GDH02} demonstrated vibrational insensitivity), stray field
gradients, etc.  Other sources of noise and systematic error in
these experiments include the index of refraction for light due to
the atomic ensemble \cite{CLM05,WHS02}, ac-Stark shifts for the
atomic energy levels due to the laser fields \cite{WSH05,WHS02},
beam mis-alignment and wavefront curvature \cite{WHS02,GIB06}, and
mean field shifts for the atomic energy states due to interaction
with nearby atoms\cite{GDH02,LRR06}.

Still, the accuracy of these atom interferometric methods for
measuring $h/m$ is increasing due to the rapid overall progress in
atom interferometry with cold atoms and because sources of error are
being understood and overcome. It is certain that the accuracy of
$h/m$ will soon be improved in both Rb and Cs, which employ
significantly different atom optics methods.  This might mean that
the real limit of confidence in this route to $\alpha$ would be in
the measurements of the atomic masses of Rb and Cs for which there
is only one high precision measurement \cite{BPR99}, and that had
unexplained systematic errors at the 0.2ppb level. We know of no
other experiments planned that could check these heavy masses,
whereas there are two or more measurements of both the Rydberg and
electron mass ratios that are consistent.

\begin{figure}[h]
\begin{center}
\includegraphics[width = 8cm]{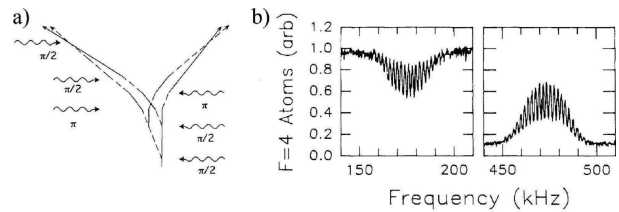}
\caption{(a) A double interferometer where the two interfering pairs
have their velocities shifted with respect to each other by four
photon recoils.  Solid lines indicate atoms in internal state $a$,
and dashed lines represent internal state $b$. (b) Sets of Ramsey
fringes displaced by $2\pi\Delta \omega$ (due to 8 $\pi$ pulses in
the middle of the interferometers). Only the frequency of the final
two   $ \pi/2 $ pulses is scanned. Figure from \cite{WYC94b}.}
\label{fig:recoil}
\end{center}
\end{figure}

Here is more detail on the Chu group experiment, described in
\cite{WYC93,WYC94b,WHS02}.  To determine $h/m$ they measure the
relative frequency of the final $\pi/2$ pulses in two different atom
interferometers (Figure \ref{fig:recoil}). The frequency difference
between the resonances of the two interferometers depends only on
conservation of energy and conservation of momentum. As an example
of the recoil shift, consider a simplified experiment \cite{WYC94b}
where an atom (with mass $m$) in state \ket{a} with zero velocity in
the laboratory frame first absorbs a photon from a leftward
propagating laser beam with frequency $\omega$. The atom recoils by
$\hbar k/m$ and the process has a resonance condition \beq \omega -
\omega_{ab} = \frac{\hbar k^2}{2 m} \eeq  where $\hbar \omega_{ab}$
is the energy difference between atomic states \ket{b} and \ket{a}
at rest. The atom can then be de-excited by a rightward propagating
beam with frequency $\omega'$.  It receives another velocity kick
$\hbar k' / m$ in the same direction and the new resonance condition
is \beq \omega' - \omega_{ab} = -\frac{\hbar k k'}{m} - \frac{\hbar
k'^2}{2 m}. \eeq The two resonances are shifted relative to each
other by \beq \Delta \omega = \omega - \omega'= \hbar(k+k')^2/2m.
\eeq Furthermore, the resonance condition for an atom in \ket{b}
moving with velocity $(N-1)v_{rec}$ towards a laser beam is \beq
\omega' - \omega_{ab} \approx  \frac{\hbar k^2}{2 m} \left[ (N-1)^2
- (N)^2 \right] \eeq  so that \beq \Delta \omega \approx \frac{N
\hbar k^2}{ m} \eeq  where $N$ is the total number of photon recoil
momenta imparted to the atom and the approximation comes from the
fact that $k' \approx k$. This shows why $\Delta \omega$ depends
linearly on $N$.

\section{ATOMIC PHYSICS APPLICATIONS}

A major motivation for atom interference experiments is to learn
more about atoms and molecules themselves.

Atoms in a separated beam interferometer experience a phase shift if
a uniform \emph{but different} potential is applied to each arm.
Thus interferometers offer exquisite sensitivity to the potentials
(not just forces). This sensitivity has been used to measure the
index of refraction due to other atoms and energy shifts due to
electric and magnetic fields. We emphasize that de Broglie wave
phase shift measurements bring spectroscopic precision to
experiments where usually classical methods like beam deflection or
velocity measurement were applied, as discussed in Sections I - IV.

In another application, the nanogratings used as a de Broglie wave
gratings can function as a very gentle spectrometer that diffracts
different molecular species in a molecular beam to different
angles.

The nanostructures themselves also produce potentials due to
atom-surface interactions that have been measured with
interferrometric techniques. For gratings with 50-nm wide slots,
each transmitted atom must pass within 25 nm of a grating bar;
hence the measured intensities are affected by the non-retarded
vdW potential.  With larger gratings, on the other hand, the
Casimir-Polder potential has been probed.

\subsection{Discovery of He$_2$ molecules}

\begin{figure}[b]
\begin{center}
\includegraphics[width = 8cm]{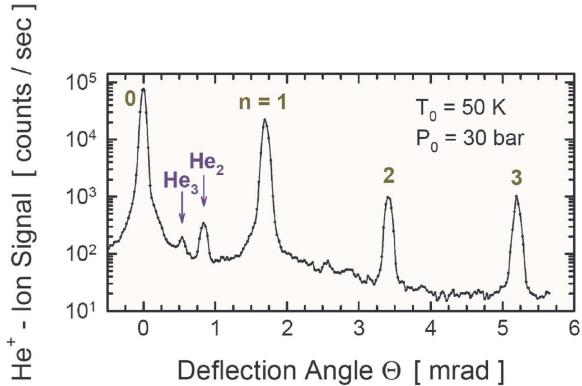}
\caption{Diffraction of helium atoms and helium molecules through
a nano-fabricated grating.  These data, reproduced with permission
from Wieland Schoellkopf, were obtained at the
Max-Planck-Institute in G\"ottingen.} \label{f:He2}
\end{center}
\end{figure}

One stunning application of coherent atom optics laid to rest a long
standing argument concerning whether a stable bound state of the
$^{4}$He$_{2}$ dimer exists. (The attribution of He$^+_2$ to He$_2$
dimers formed in cryogenic expansion by \cite{LMK93} primarily
reopened an old debate). For this a diffraction grating was used to
separate and resolve $^{4}$He$_{2}$ dimers from a helium beam
(Figure~\ref{f:He2}) \cite{SHT96}. Subsequently a grating was used
as a nano-sieve to measure the size of the $^{4}$He$_{2}$ dimers.
They have a size of $\langle r \rangle = 6.2 \pm 1.0 $ nm which
corresponds to a binding energy of $E/k_B$ = 1 mK
\cite{LGG96,GST00b}.  Diffraction has also been used to study the
formation  of more massive clusters \cite{BGK04}, and searches using
this technique are underway for an Efimov-type excited state in
$^4$He$_3$. This would be manifest as a particularly large ($\langle
r \rangle$ = 8.0 nm) excited state helium trimer.
\cite{BKK05,HES05,STK05}.

\subsection{Polarizability measurements}

\subsubsection{Ground state dc scalar polarizability}

\begin{figure}[b]
\begin{center}
\includegraphics[width = 8.5cm]{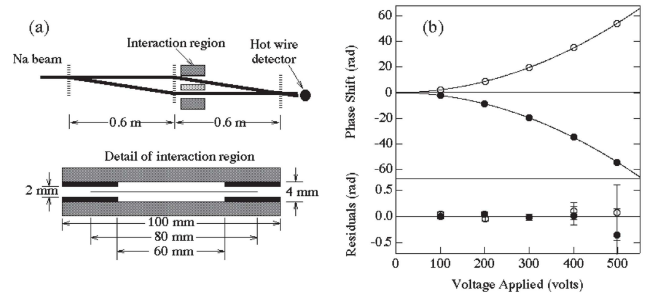}
\caption{Measurement of atomic polarizability. (a) Schematic of
the interaction region installed behind the second grating. (b)
Measured phase shifts vs. applied voltage. The two different signs
of the phase shift stem from the voltage being applied on either
the left (open circles) or the right (filled circles) side of the
interaction region (arm of the interferometer).  The fit is to a
quadratic and the residuals are shown on the lower graph. Figure
from \cite{BER97}.} \label{fig:polarizability}
\end{center}
\end{figure}

By inserting a metal foil between the two separated arms, as shown
in Fig.~\ref{fig:polarizability}, an electric field can be applied
to a single path.  The resulting de Broglie wave phase shift was
used to measure the static ground-state atomic polarizability of
sodium, $\alpha_{Na}$ with a precision of $0.35\%$ \cite{ESC95}.
Similar precision has been demonstrated for $\alpha_{He}$
(Toennies group) and $\alpha_{Li}$ (Vigu\'{e} group) using this
method \cite{TOE01,MJB06,MJB06b}.

In this experiment a uniform electric field $\mathcal{E}$ is
applied to one of the separated atomic beams, shifting its energy
by the Stark potential $U=-\alpha\mathcal{E}^2/2$.  The static
scalar ground-state polarizability $\alpha_{pol}$ is determined
from the phase shift, $\Delta \phi$, of the interference pattern
by \beq \alpha_{pol} = \left(\frac{\Delta \phi} {V^2} \right)
\left( \frac{D^2}{L_{eff}} \right) (2\hbar v), \label{eq:polphase}
\eeq where $V$ is the voltage applied to one electrode in the
interaction region, $D$ is the distance between the electrode and
the septum,  $v$ is the mean velocity of the atomic beam, and
$L_{eff}$ is the effective interaction region length defined as
\beq \left( \frac{V}{D} \right)^2 L_{eff} \equiv \int
\mathcal{E}^2 dz. \eeq

For an accurate determination of electric polarizability, the three
factors in Equation \ref{eq:polphase} must each be determined
precisely.  They are (1) the phase shift as a function of applied
voltage, (2) the geometry and fringing fields of the interaction
region, and (3) the velocity of the atoms.  In \cite{ESC95} the
uncertainty in each term was less than 0.2\%.

Taking all sources of error into account, and adding statistical and
systematic errors in quadrature, the static polarizability of the
ground state of sodium was measured to be $\alpha_{pol}=24.11\times
10^{-24}$ cm$^3$, with a fractional uncertainty of 0.35\%.  This
measurement was a nearly 30 fold improvement on the best previous
direct measurement of the polarizability of sodium \cite{HAZ74}
based on beam deflection.

\begin{figure}[h]
\begin{center}
\includegraphics[width = 8.5cm]{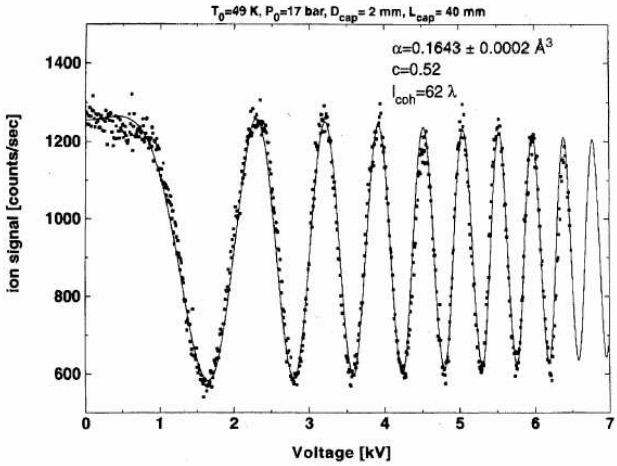}
\caption{Measurement of the electric polarizability of He.  The
gratings are held stationary while the electric field is increased.
The measurement uncertainty is statistical only.  Figure courtesy of
J.P. Toennies and R. Breuhl.} \label{fig:he_pol}
\end{center}
\end{figure}

A similar experiment for He was done with a 3-grating Mach Zehnder
interferometer (with nanogratings) by the Toennies group.  The
phase stability of this interferometer was so good that the
fringes could be observed directly as a function of applied
electric field, while the gratings were not moved. (Figure
\ref{fig:he_pol}). The statistical precision in $\alpha_{He}$ was
0.1\% \cite{TOE01}.

Using three Bragg diffraction gratings for Li atoms and a septum
electrode, the group of Vigu\'{e} measured $\alpha_{Li}$ with a
precision of 0.66\% \cite{MJB06,MJB06b}.

Because atom interferometry gives sub-Hertz precision on the energy
shift of the atomic ground state, \emph{ratios of polarizabilities}
for different species can be very accurately determined with
multi-species atom interferometers. Uncertainty in the interaction
region geometry would then be less significant because the quantity
$(D^2/L_{eff})$ in Eq \ref{eq:polphase} cancels out in a ratio of,
for example, $\alpha_{Rb} / \alpha_{Li}$. The ratio of velocities of
the two species would still need to be measured, or taken into
account. Thus, improved precision in measurements of $\alpha_{pol}$
may come from using an engineered phase shift to cancel the
velocity-dependence of the polarizability phase shift. This is known
as dispersion compensation \cite{RCT04}. Velocity multiplexing
\cite{HPC95} and magnetic re-phasing \cite{SEC94} are other
approaches for dealing with the experimental spread in velocity.
With these improvements it seems feasible to perform polarizability
measurements with uncertainties in the 10$^{-4}$ range.

This precision offers an excellent test of atomic theory, because
theoretical uncertainties in light elements like Li are orders of
magnitude smaller than in heavier alkalis. Polarizability
$\alpha_{pol}$ of an atomic state can be expressed as a sum over
dipole matrix elements: \beq \alpha_{pol} = e^2 \sum_{j\neq i}
\frac{\langle i| \mathbf{r}| j \rangle \langle j| \mathbf{r}| i
\rangle }{E_j - E_i} \eeq where $E_j$ is the energy of state
\ket{j}.  Accurate calculation of static dipole polarizabilities for
heavy atoms still remains a great challenge because electron
correlation and relativistic effects become increasingly important
for heavy atoms. Major theoretical efforts so far have included the
relativistic Hartree Fock approach, many body perturbation theory,
density functional theory, and relativistic coupled-cluster
technique. Several calculations of atomic polarizability all show
the need for precise experimental measurements \cite{BOK94,DEP02,
KVJ01, MAR01, HOH00, SJD99, LPS99, DJS99, RMH98, KBD97,THL05}.

\subsubsection{Transition dc and ac Stark shifts}

\begin{figure}[h]
\begin{center}
\includegraphics[width = 8.5cm]{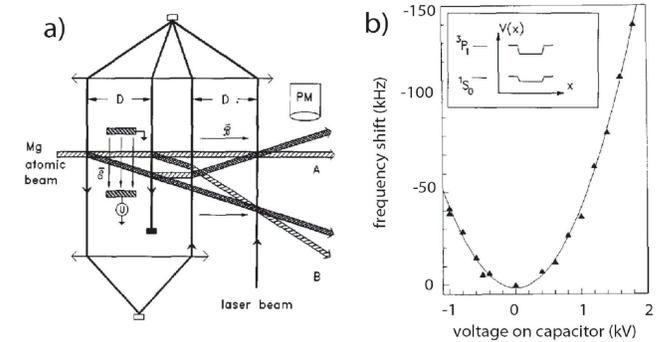}
\caption{Optical Ramsey Bord\'{e} interferometer for measuring
polarizability differences. (a) Schematic of the atom interferometer
with a capacitor. (b) Frequency shift of the interference pattern
versus voltage across the capacitor. The fit is a parabola.  The
inset shows the energy levels as a function of position through the
capacitor. \cite{RSS93}} \label{fig:transition-stark-shift}
\end{center}
\end{figure}

When two paths have different internal states, e.g. in an optical
Ramsey Bord\'{e} interferometer, then a uniform electric field
applied to both paths makes phase shifts proportional to the
\emph{difference} of polarizability of the two states.  (This is
similar to what can be measured with laser spectroscopy.)  For
example, the dc-Stark shift of the magnesium
3s$^2$($1$S$_o$)-3s3p($^3$P$_1$) intercombination line was measured
by subjecting both arms of an atom interferometer to a constant
electric field. The Stark energy perturbation provides two different
potentials in the two arms of the interferometer. The resulting
relative phase shift (Figure \ref{fig:transition-stark-shift})
corresponds to a difference of -(8 $\pm$ 1) kHz (kV/cm)$^{-2}$ in
the polarizabilities of the $^1$S$_o$ and the $^3$P$_1$(m = 1)
states. \cite{RSS93}.

A related approach was used to measure the difference between the
polarizabilities of the $^3$P$_1$ state and the $^1$S$_o$ state of
Ca to be $\alpha$($^3$P$_1$)-$\alpha$($^1$S$_o) =
(13\pm2)\times10^{24}$ cm$^3$ \cite{MNK96}.

The ac Stark shift of the 4s$_2$$^1$S$_o$ – 4s4p$^3$P$_1$ line in
$^{40}$Ca was measured with a time-domain Ramsey-Bord\'{e} atom
interferometer (in a magneto optical trap) for perturbing laser
wavelengths between 780 nm and 1064 nm. \cite{DSS04}. Ac Stark
shifts have also been observed in a double-well interferometer
\cite{SSP04}.

\subsection{Index of refraction due to dilute gasses}

A physical membrane separating the two paths allows insertion of a
gas into one path of the interfering wave, enabling a measurement of
the index of refraction for atom waves traveling through a dilute
gas caused by the collision-induced phase shift. Measurements are
presented in \cite{SCE95,RCK02,GROUP97_KHS,BER97} and these
experiments are discussed in
\cite{BCF03,KHD01,FLK97,CAD97,ADV97,FLK96,VIG95,ADV95}.

Scattering makes a wave function evolve as: \beq \psi
\stackrel{r\rightarrow \infty}{\longrightarrow} e^{i\mathbf{k}r} +
f(\mathbf{k},\mathbf{k}') \frac{e^{i\mathbf{k}'r}}{r}, \eeq where
the scattering amplitude $f$ contains all the information about
the scattering process \cite{SAKURI}. The complex index of
refraction $n$ due to a gas of scattering centers is related to
$f$ by summing the scattered amplitudes in the forward direction
\cite{NEW66}, resulting in \beq n=1+\frac{2 \pi N}{\mathbf{k}^2}
f(\mathbf{k},\mathbf{k}) \label{eq:fkk'}\eeq where $N$ is the gas
density. Atoms propagating through the gas are phase shifted and
attenuated by the index \beq \psi(z) = \psi(0)e^{inkz} =
\psi(0)e^{ikz} e^{i \Delta \phi(N,z)} e^{-\frac{N}{2} \sigma_{tot}
z}. \eeq

\begin{figure}[t]
\begin{center}
\includegraphics[width = 8cm]{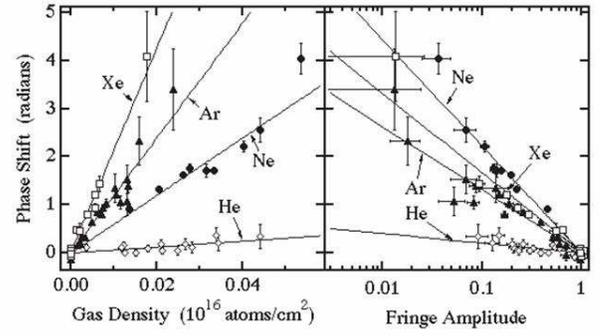}    
\caption{(Left) Phase shift $\Delta \phi$ as a function of gas
density $N$ for different gas samples. (Right) Phase shift vs
Fringe amplitude.  The fringe amplitude is proportional to
$e^{-N\sigma_{tot}z/2}$. Figure from \cite{BER97}.
\label{fig:gas-index}}
\end{center}
\end{figure}

\noindent The phase shift due to the gas, \beq \Delta \phi(N,z) =
(2\pi N k z / k_{cm}) \mathrm{Re}[f(k_{cm})], \eeq is proportional
to the \emph{real} part of the forward scattering amplitude, while
the attenuation is related to the \emph{imaginary} part. Attenuation
is proportional to the total scattering cross section which is
related to $\mathrm{Im}[f]$ by the optical theorem \beq \sigma_{tot}
= \frac{4 \pi}{k_{cm}}
\mathrm{Im}[f(k_{cm})]\label{eq:optical-theorem}.\eeq
Measurements of phase shift as a function of gas density are shown
in Figure \ref{fig:gas-index}.

The ratio of the real and imaginary parts of the forward
scattering amplitude is a natural quantity to measure and compare
with theory.  This ratio, \beq \rho(k) = \frac{\Delta \phi(N)}
{ln[A(N)/A(0)]} = \frac{\mathrm{Re}[f(k)]}{\mathrm{Im}[f(k)]}.\eeq
where $A$ is the fringe amplitude, gives orthogonal information to
the previously studied total scattering cross section.  In
addition it is independent of the absolute pressure in the
scattering region and therefore much better to measure.

\begin{figure}
\begin{center}
\includegraphics[width = 9cm]{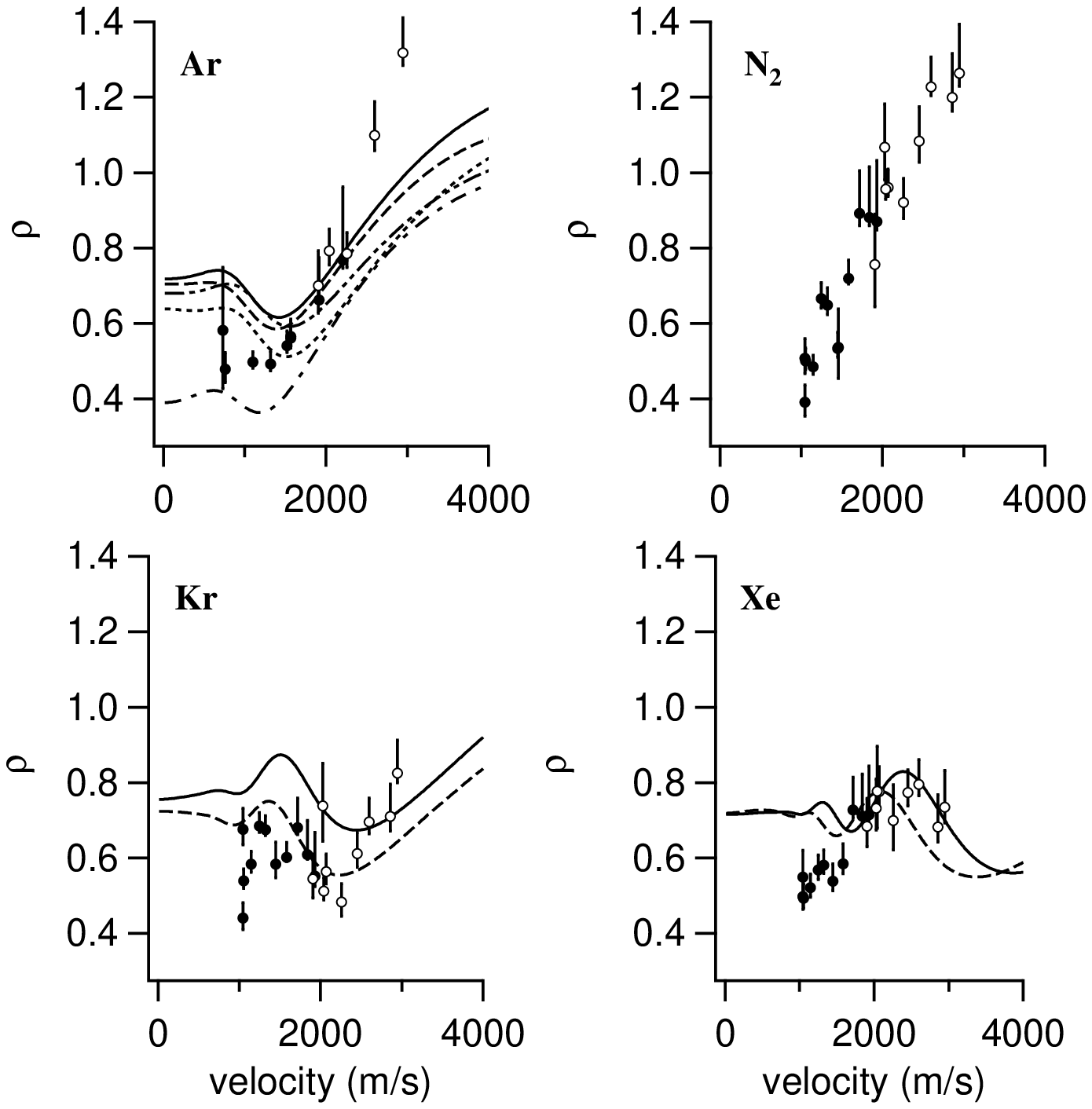}
\caption{ $\rho$ as measured for Na waves in Ar, Kr, Xe, and
N$_2$($\bullet$ using 200 nm gratings, $\circ$ 100 nm), showing
evidence of glory oscillations in comparison to $\rho$ as derived
from predicted potentials: Na-Ar \cite{CAD97}(---),
\cite{DUG78}($---$),
\cite{FLK97}($\cdots$),\cite{TRK79}($\cdot-\cdot-$),
\cite{TAT77}($\cdot\cdot-\cdot\cdot-$); Na-Kr \cite{CAD97}(---),
\cite{DRS68}($---$); and Na-Xe \cite{BZB92}(---),
\cite{DRS68}($---$). Figure from \cite{RCK02}.}\label{fig:glory}
\end{center}
\end{figure}

The ratio $\rho(k)$ shows structure as a function of $k$ known as
\emph{glory oscillations}\footnote{Glory oscillations in the
absorption cross section were first measured by \textcite{ROT62} for
Li and K beams, and related phenomena with light waves have been
studied by \cite{BRC66,CMG98,KHN77,NUS79}.} (Figure
\ref{fig:glory}).  These were predicted in \cite{ADV95,ADV97,FLK97}
and observed in \cite{RCK02}. Measurements of $\rho(k)$ plotted as a
function of Na beam velocity $v$ for target gases of Ar, Kr, Xe, and
N$_2$ are shown in Fig.~\ref{fig:glory}.

To compare these measurements with predictions based on various
potentials $V(r)$, the forward scattering amplitude was computed
using the standard partial wave treatment and the WKB
approximation. Predictions for $\rho$ must also include an average
over the distribution of velocities in the gas sample, and this
damps the glory oscillations as discussed in \cite{CAD97,FLK97}.
Fig.~\ref{fig:glory} shows calculations of $\rho(v)$ based on
predictions of $V(r)$ for Na-Ar, Na-Kr and Na-Xe derived from
spectroscopic measurements and beam scattering experiments.

The motivation for studying the phase shift in collisions is to add
information to long-standing problems such as inversion of the
scattering problem to find the interatomic potential $V(r)$,
interpretation of other data that are sensitive to long-range
interatomic potentials, and description of collective effects in a
weakly interacting gas
\cite{CHS89,BMT93,CMH94,LHP93,WAF94,MOV94,MOS94,STO91,SHV94}. The
glory measurements of $\rho$ are sensitive to the shape of the
potential near the minimum, where the transition from the repulsive
core to the Van der Waals potential is poorly understood. The
measurements of $\rho(k)$ also give information about the rate of
increase of the interatomic potential $V(r)$ for large $r$
independently of the strength of $V(r)$. The real part of $f$ was
inaccessible to measurement before the advent of separated beam atom
interferometers.  Controlled collisions as phase shifting tools are
now widely discussed in the context of quantum computing.

\subsection{Casimir-Polder (atom-surface) potentials}

Atom-surface interactions are important in a wide range of
nanoscale phenomena, including  gas  adsorption, atomic force
microscopy, quantum reflection, AtomChips, and many topics in
biophysics and chemistry. Yet in many situations the forces are
difficult to predict ab initio.  Single atoms passing within 50 nm
of a dielectric surface represent a nice middle ground, where
theoretical calculations are tractable, and precision measurements
are becoming possible.  Here we briefly describe some landmark
theoretical contributions to this field and then survey
measurements done with coherent atom optics.

After J.D van der Waals suggested modifications to the equation of
state for gases to allow for atom-atom interactions (which he did
in 1873), \textcite{lon37} calculated the strength of interactions
between two polarizabile atoms using quantum mechanics, and
similar ideas were used to describe atom-surface interactions
\cite{len32}. \textcite{CAP48} generalized the theory of
atom-surface interactions to include the effect of retardation,
and \textcite{LIF56} modified this theory to allow for a surfaces
with a dielectric permittivity. Since then, hundreds of
theoretical works used quantum electrodynamics to predict the
interaction potential for real atoms near real surfaces.

The Casimir-Polder potential for an ideal surface \cite{CAP48,SBC93}
\beq U(r) = \frac{1}{4\pi\alpha r^4} \int_0^{\infty}
\alpha_{pol}(ix/\alpha r) e^{-2x}[2x^2 + 2x +1 ] dx \label{eq:CP}
\eeq where $\alpha_{pol}$ is atomic polarizability (evaluated as a
function of imaginary frequency), $r$ is the distance to the
surface, and $\alpha$ is the fine structure constant.  This has
well-known limits of the Van der Waals (vdW) regime \beq r
\rightarrow 0 \qquad U(r) = \frac{\hbar}{4\pi r^3} \int_0^{\infty}
\alpha_{pol}(i\omega) d \omega \equiv \frac{C_3}{r^3}
\label{eq:c3calc} \eeq and the retarded regime \beq r \rightarrow
\infty \qquad U(r) = \frac{2hc\alpha_{pol}(0)}{32 \pi \epsilon_0
r^4} \equiv -\frac{K_4}{r^4}. \eeq \textcite{MDB97} evaluated $U(r)$
for sodium atoms at arbitrary distances from a perfectly conducting
half space, using a single electron (Lorenz oscillator) model of the
atom. \textcite{DJS99} calculated $C_3$ for the alkali atoms using
the best available model of frequency-dependent atomic
polarizability. It is noteworthy that 18\% of the interaction
potential between sodium atoms and a perfect mirror is due to
excitations of the core electrons.  The one-electron (Lorenz)
oscillator model yields $ C_3 = \hbar \omega_0 \alpha_{pol}(0) /8 $
with $\alpha_{pol}(0) = e^2/\omega_0^2 m_e [4 \pi \epsilon_0]$ where
$\omega_0$ the resonance frequency and $m_e$ the electron mass. This
one-electron model for sodium atoms and a perfectly conducting
surface gives $C_3$ = 6.3 meV nm$^3$, while the calculation with
many electrons gives $C_3$ = 7.6 meV nm$^3$. The Lifshitz formula
\beq C_3 = \frac{\hbar}{4\pi} \int_0^{\infty}
\alpha_{pol}(i\omega)\frac{\epsilon(i\omega) -1}{\epsilon(i\omega) +
1}d\omega.  \eeq  reduces $C_3$ even further. For sodium and silicon
nitride the Lifshitz formula gives $C_3$ = 3.2 meV nm$^3$.
\textcite{SpT93} and \cite{ZhS95} elaborated on $U(r)$ for arbitrary
$r$ and surfaces composed of multiple layers.

Several experiments can now test these predictions. Atoms
transmitted through a cavity \cite{ande88,SBC93}, atoms
diffracted from a material
grating \cite{GST99,BRU02,GST00,SHI01,CRP04,PCS05},
atoms undergoing quantum
reflection \cite{AHH86,BLS89,SHI01,SHF02a,DRD03,PSJ06},
atoms reflecting from evanescent waves near
surfaces \cite{HBF89,KLL96,WWL98,ESA04}, and atoms
trapped near surfaces \cite{Vul04,MHO04,HOM05} and atoms
in interferometers \cite{BHU02,PEC05,PEC06,NAZ03,KSF03}
have been used to measure atom-surface interaction potentials. For
a review of several such experiments see the CAMS proceedings
\cite{CAMS}.

\subsubsection{VdW-modified diffraction}

\begin{figure}
\begin{center}
\includegraphics[width = 5cm]{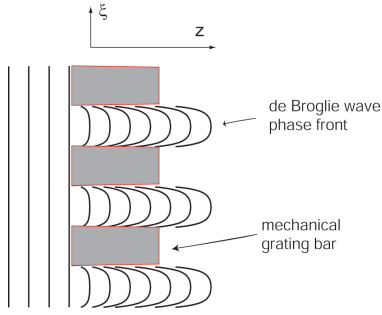}
\caption{Distorted de Broglie waves. Van der Waals interactions with
mechanical grating bars cause near-field phase shifts. This view is
exaggerated: in beam experiments there are typically 10$^4$ wave
fronts in the 100 nm thickness of a nanograting slot
\cite{GST99,PCS05}. \label{fig:phasec}}
\end{center}
\end{figure}

\begin{figure}
\begin{center}
\includegraphics[width = \columnwidth]{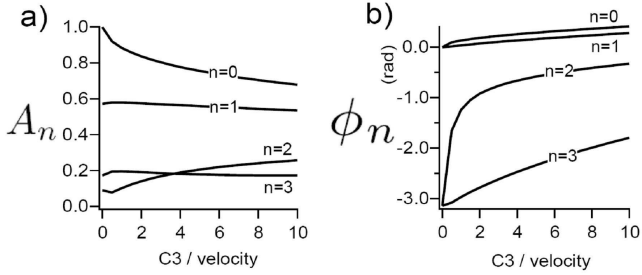}
\caption{The Phase (a) and modulus (b) of far-field diffraction
orders both depend on the vdW coefficient $C_3$ divided by atom
velocity [shown in units of meVnm$^3$ / (km/s)]. Figure adapted from
\cite{PEC06}.} \label{fig:pn}
\end{center}
\end{figure}

Because of van der Waals interactions with mechanical grating bars,
atoms propagating through a nanograting get a phase shift that
depends on position within each slot, as shown in Figure
\ref{fig:phasec}. An analogous structure in light optics is an array
of diverging lenses held between absorbing bars. The index of
refraction in the free-space between material grating bars gives
nanogratings a \emph{complex transmission function} that has been
studied in \cite{GST99,BRU02,GST00,SHI01,CRP04,PCS05,
BHU02,PEC05,PEC06,NAZ03,KSF03}.

\begin{figure}
\begin{center}
\includegraphics[width = \columnwidth]{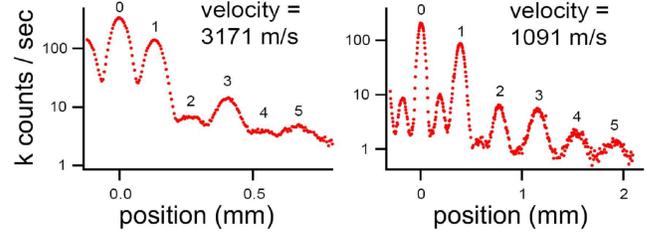}    
\caption{Diffraction intensities used to measure the strength of
$C_3$ for Na-silicon nitride \cite{PCS05}.  Data for two different
velocities show how the 2nd and 3rd order change their relative
intensity (as predicted in Figure \ref{fig:pn}). Diffraction of
Na$_{2}$ molecules is also visible. \label{fig:pn-an}}
\end{center}
\end{figure}

\begin{figure}
\begin{center}
\includegraphics[width = \columnwidth]{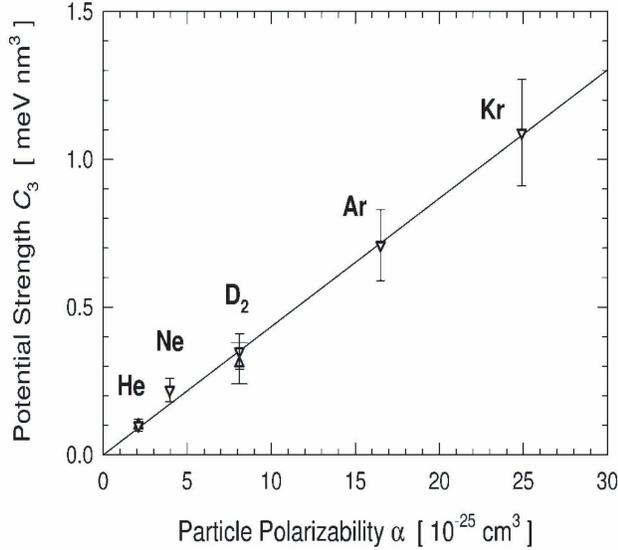}    
\caption{Measurements of $C_3$ for various atoms and a silicon
nitride surface, obtained by studying atom diffraction patterns
\cite{GST99}.\label{fig:c3pol}}
\end{center}
\end{figure}

Figure \ref{fig:phasec} is a cartoon of the de Broglie wave phase
fronts in the near-field immediately after a nanograting.  Far-field
diffraction orders are affected by van der Waals interactions too.
We can describe the $n$th far-field order by \beq \psi_n = A_n
e^{i\phi_n} e^{i \vec{k}_n \cdot \vec{x}} \eeq where the modulus
$A_n$ and phase $\phi_n$ for the $n$th order are given by \beq A_n
e^{i\phi_n} = \int_{-w/2}^{w/2} \exp \left[i\phi(\xi)+ i n G \xi
\right] d \xi \label{eq:Aninteg}. \eeq  Here $w$ is the size of the
windows (or `nano-slots') between grating bars, and $\phi(\xi)$ is
the phase-shift sketched in Figure \ref{fig:phasec} that can be
calculated by putting the atom-surface potential $U(r)$ into the
expression for a phase shift (Equation \ref{eq:phi_int}). Thus the
modulus and a phase of each diffraction order depends on the
strength of the potential ($C_3$ in the vdW regime) and on atomic
velocity as shown in Figure \ref{fig:pn}. Several experiments  have
measured the intensity $|A_n|^2$ in diffraction orders to determine
$C_3$ for various atom-surface combinations, with some results shown
in figures \ref{fig:pn-an} and \ref{fig:c3pol})
\cite{GST99,BRU02,PCS05,CRP04}.

The diffraction intensities $|A_n|^2$ depend on phase
\emph{gradients} induced by $U(r)$. To detect the \emph{diffraction
phases}, $\phi_n$, an atom interferometer can be used as described
in the next Subsection.

\subsubsection{Interferometer VdW and CP measurements}

The complex transmission function of the gratings modifies the
location at which the Talbot effect revivals occur.  This, in
turn, modifies the performance of a Talbot-Lau interferometer, as
discussed in \cite{BAZ03,BHU02} and shown in Figure
\ref{fig:vdw-ifm plan}.  Because gratings in this experiment have
a 1 $\mu$m period, these results probe the retarded Casimir-Polder
regime.

\begin{figure}[t]
\begin{center}
\includegraphics[width = 7cm]{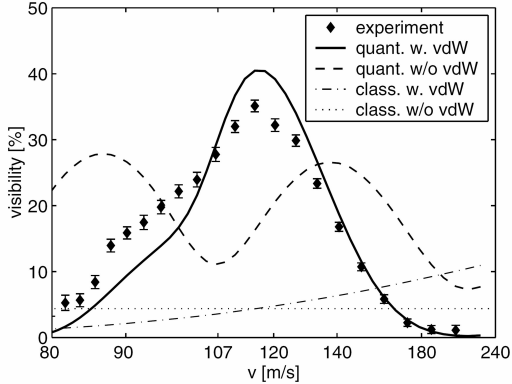}
\caption{Dependence of the interference fringe visibility on the
mean velocity of the molecular beam. Numerical simulation results
are plotted for four models without free parameters: classical or
quantum behavior, with or without consideration of the van der
Waals (vdW) interaction of the molecules with the second grating.
The quantum result including the van der Waals effect is clearly
the only adequate theory \cite{BHU02}. \label{fig:TLI-CP}}
\end{center}
\end{figure}

In a separated-path interferometer, \textcite{PEC05} inserted an
auxiliary \emph{interaction grating} in one path.  This allowed a
measurement of the phase shift $\phi_0$ due to transmission through
the interaction grating as shown in Fig.~ \ref{fig:vdw-ifm plan}. In
a separate experiment the higher-order diffraction phase $\phi_2$
was measured by comparing the output of four different
separated-path interferometers \cite{PEC06}.

\begin{figure}[t]
\begin{center}
\includegraphics[width = 7cm]{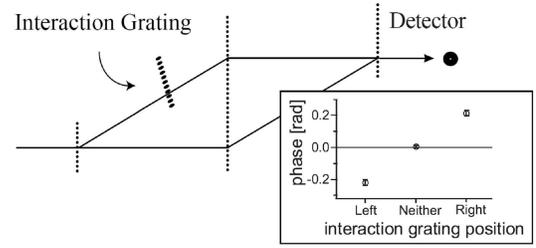}
\caption{An `interaction grating' was inserted and removed from each
path of an interferometer to measure the phase shift $\Phi_0$ due to
Van der Waals interactions \cite{PEC05}. \label{fig:vdw-ifm plan}}
\end{center}
\end{figure}

\begin{figure}
\begin{center}
\includegraphics[width = \columnwidth]{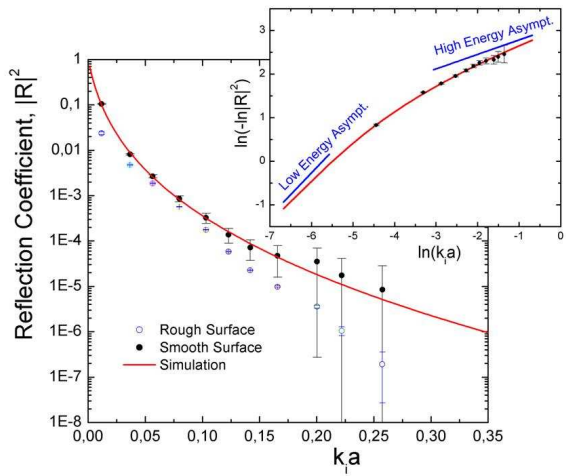}
\caption{Experimental ABSE data for the quantum reflection of
$^3$He atoms from a disordered single crystal quartz surface.
\cite{DRD03} \label{fig:ABSE3}}
\end{center}
\end{figure}

In the Atomic Beam Spin Echo (ABSE) interferometer, discussed in
III.D.2, \textcite{DRD03} observed atoms reflecting from the
attractive part of the atom-surface interaction potential. This
quantum reflection allowed DeKieviet et al to map the van der Waals
potential in an energy range between 1 neV and a sub-meV. Figure
\ref{fig:ABSE3} shows the measured probability of He atoms quantum
reflecting from a quartz surface, as a function of the impinging
wavevector. (Later both metallic and semi-conductor samples were
used.) Deviation of the experimental data from the high-energy
asymptote is attributed to Casimir-Polder retardation. importantly
the spin echo interferometer was used to precisely select the
velocity of the detected atoms. In this regard it complements other
quantum reflection experiments \cite{AHH86,BLS89,SHI01,SHF02a,PSJ06}
that do not explicitly use atom interferometers (though we note that
quantum reflection itself is inherently a wave phenomenon).

\section{OUTLOOK}

In the early 1980's ``Mechanical Effects of Light"  was the name for
the study of light forces on atoms.  See, for example, \cite{CDK85}
and Table \ref{tab:reviews}.  At first these forces were used simply
to change the momentum of atoms.  Then it emerged that, with care in
application, light forces could be conservative.  When atom
diffraction from a standing light wave was demonstrated
\cite{GRP86}, it became appreciated that interactions with classical
light fields can transfer momentum in precise quanta and preserve
the coherence of atomic de Broglie waves.  This led to many papers
contrasting ``Diffraction and Diffusion" and emphasizing that
diffraction of atoms by light was coherent, whereas the occurrence
of some spontaneous decay processes led to diffusion which is not
coherent \cite{TRM83,GMR91,WSM91,RKA98,DEN06}.  It also led workers
in the field to consider other coherent ways to manipulate atoms.
The term ``optics" started to replace ``diffraction" in
conversations. Although some felt that ``atomic optics" was the
preferable phrase (in part to emphasize that atomic physics was the
driving force), we felt that ``Atom Optics" was more closely
analogous to ``electron optics" and decided to make it the title of
our 1991 review in ICAP12 \cite{PRI91}. Perhaps nothing shows the
growth of this field, and the coalescence around this phrase, more
than the fact that there are almost 200,000 Google hits to ``Atom
Optics" plus over 20,000 for atom interferometers (and even fewer
than this for ``atomic optics").

The reviews on Atom Optics that the MIT group wrote in 1990 and 1991
considered Atom Optics as a way to mimic photon optics.  Relative to
a list of standard optical components, it was pointed out that atom
lenses could be made in various ways but that material beamsplitters
were impossible, shifting the burden for coherent beamsplitting and
recombining to diffractive processes using matter and light
gratings.   The observation that light and matter diffraction
gratings would be the beamsplitters has been borne out by the vast
majority of work with atom interferometers over the past 15 years.
However, their refinement has been quite remarkable.  In addition, a
host of new developments in atom optics have greatly lengthened the
list of atom optical components and devices -- see the Atom Optics
Toolkit in Table \ref{tab:hardware}.  The art of atom optics is in
its golden age because the techniques listed in this toolkit are
just beginning to have an impact on scientific questions beyond the
specialty of atom optics. As larger and more controlled atom optical
systems are constructed, opportunities abound to efficiently and
coherently manipulate atoms for scientific gain.

\begin{table*}
\caption{Atom optics tool kit organized by analogy to light optics,
as in \cite{PRI91}. Parentheses () indicate options.}
\label{tab:hardware}
\begin{ruledtabular}
\begin{tabular}{lrll}
& Light Optics  &  $\quad$   & Atom Optics \\
\hline
SOURCES & thermal && (supersonic) beam\\
&&& (moving) molasses, launched (or dropped) MOT, Zeeman slower \\
& coherent LASER && Bose Einstein condensate \\
LENSES & spherical && electrostatic quadrupole or magnetic hexapole \\
  &   cylindrical && Gaussian optical beams   \\
 &    Fresnel&& nano structure zone plates (cyl. or sph.) \\
&achromat & &  combination zone plate + E-M lens\\
&axicon && magnetic quadrupole \\
MIRRORS&&& (giant) quantum reflection  \\
&    && helium from (bent) crystal surfaces\\
&    && evanescent light waves \\
&    && periodically poled magnetic domains (on curved surfaces)\\
GRATINGS&phase && standing waves of light: Bragg, or Kapitza-Dirac (pulses)\\
&amplitude && nano-structure gratings \\
&&& standing waves of resonant radiation\\
& reflection && crystal surfaces \\
&&& quantum reflection from (nano-structured) surfaces\\
&&             &        structured evanescent light \\
& blazed && Bragg scattering \\
&&&two- and three-color standing waves\\
POLARIZING SPLITTERS & && stimulated Raman transitions \\
&& & optical Ramsey $\pi/2$ pulses \\
&& & Stern-Gerlach magnets \\
&& & optical Stern-Gerlach effect\\
PHASE PLATES & glass && $\mathbf{E}$-field \\
&&& $\mathbf{B}$-field \\
&&& dilute gas \\
&&& nearby surface \\
HOLOGRAMS & transmission &&  perforated nano-structures (with $\mathbf{E}$ and $\mathbf{B}$ fields) \\
&reflection && nano-structures (with enhanced quantum reflection) \\
$\lambda$ SHIFTERS &  modulators  && amplitude modulated standing waves \\
&&& gravity\\
&&& bi-chromatic laser fields \\
&&& reflection from a receding rotor \\
INTERFEROMETERS & Young's experiment& & micro (or nano) slits  \\
& Mach-Zehnder&& space domain using (separated) beams (spin entanglement)\\
&&& time domain, with pulsed gratings (spin entanglement) \\
&&& longitudinal (RF or Stern-Gerlach beam splitters) \\
& near field  && Talbot Lau, Lau, and Talbot interferometers \\
& Michelson && atoms confined in a waveguide \\
& Fabry-Perot && atoms confined in a 3-dimensional trap \\
WAVEGUIDES & fiber optics && $\mathbf{B}$ fields from wires (on a chip) \\
 &   && permanent magnets \\
 &   && optical dipole force \\
 &   && evanescent light in hollow fiber \\
DETECTORS & photon counter&& hot wire (or electron bombardment)
ionizer and counter (CEM or MPC)\\
 & state selective && field ionization, laser ionization, metastable detection\\
 &&& polarization spectroscopy \\
 & imaging &&  multichannel plate for ions or (metastable) atoms \\
 &&& (state selective) fluorescence, absorption, or phase contrast imaging\\
 AMPLIFIERS & stimulated emission&& four-wave mixing with BEC [nonlinear quantum optics]\\
\end{tabular}
\end{ruledtabular}
\end{table*}

Not all predictions in \cite{PRI91} were so prescient however;
although coherent atom amplifiers were discussed, they were not
anticipated. Hence the demonstration of coherent atom amplification
(using interferometry to verify its phase coherence) was an
unexpected development, as was non-linear atom optics generally. The
power of, and interest in, non-linear atom optics should lead to
many more advances in atom interferometry such as sub-shot noise
measurements of phase shifts \cite{JSW06,SCD93,SEM03,PES06}, and
coherent oscillations between atomic and molecular BEC's. Non-linear
optics is outside the scope of this paper although techniques of
linear atom optics and interferometry are extremely valuable as
tools in this field \cite{ROP02,and03,bon04,mey01}. Another
unanticipated development is the immense amount of development on
atom chips.

As this review shows, experimental and theoretical understanding of
atomic and molecular matter waves has come a long way since the
first demonstrations of coherent diffraction with laser light and
nanogratings in the early 1990's.  In the MIT group's first paper on
diffraction by a light grating, the rms momentum transfer was far
below predictions; the second paper reported it was low by a factor
of two noting there was ``no explanation for this discrepancy".
Recently this effect was used to measure the standing wave intensity
of a standing wave (depth of optical lattice) to within 1\%
\cite{MMC07}.

This shows the transformation of pioneering scientific work in atom
optics into a high-precision tool for use in cold atom physics.
Similarly, our review shows that atom interferometers are now
routinely used for scientific endeavors ranging from fundamental
investigations of quantum physics to precision metrology. We now
briefly project anticipated progress over the main categories used
in this review (diffraction, interferometry, fundamental studies,
precision measurements, and atomic properties).  We also speculate
on areas that we expect will become more important: e.g. optics with
molecules and ultracold fermions, atom chips and optical lattices,
surface science, fundamental studies of gravitation, new ways to
control atom-atom interactions, entanglement and multi-particle
interferometry, and more formal analogies to condensed matter
phenomena that arise from quantum coherence.

We expect coherent atom optics to become an even more flexible,
powerful, and precise tool for manipulating atoms and molecules,
especially for interferometers, and for applications to other
scientific and technical problems.  The development of techniques
for accelerating (and in the future decelerating) atoms and
especially molecules, both in light crystals and by optimizing the
temporal envelope of light for higher-order beam splitters will
enable coherence to be maintained between wave function components
with relative velocities of meters/sec that are determined with
10$^-10$ accuracy.  This will result in interferometers of far
greater precision with much greater separation of the arms and much
greater enclosed area.

These bigger and better interferometers will be applied to
fundamental problems in gravity and quantum mechanics.  They will
allow one to measure the gravitational potential in experiments
analogous to the scalar Aharonov-Bohm effect in which the potential
has influence in the absence of any gravitational field (such as
when one component of the wave function spends time inside a hollow
massive cylinder).   Placed in orbit around earth, interferometers
with large enclosed area will be useful for fundamental
gravitational measurements such as tests of parallel vector
transport and the Lense-Thirring frame-dragging effect.  As a
byproduct of developing interferometers with larger separation for
heavier particles, more stringent limits will placed on alternative
theories of quantum decoherence that involve spontaneous projection.
It may also be possible to observe some new sources of decoherence
that are hard to shield out \cite{TEG93}. These advances in
interferometer size will also enable better measurements of inertial
effects such as gravitational fields, gravitational gradients, and
in gyroscopes.  These will have application to inertial navigation,
geodesy, and prospecting.

Precision in atom and molecular interference experiments will also
be increased by using higher fluxes and longer interaction times.
This also implies larger instruments in order to reduce the atom
densities, thus reducing the systematic shifts due to atom-atom
interactions.  However, more imaginative approaches are needed since
atom-atom interactions can be a severe problem.  For example, they
are one of the limiting factors for the Cs atomic fountain clocks,
all interferometers using Bose Einstein condensates, and they modify
the index of refraction of near-resonant light passing through even
non-degenerate atom samples.  There are at least two solutions to
the problem:

\begin{itemize}

\item If one uses ultracold fermions in a single atomic state, the Pauli
exclusion principle switches off the s-wave interaction.  Since for
neutral atoms at ultracold energies the higher partial waves can be
ignored, a fermionic ensemble is nearly interaction free, and
therefore ideal for precision measurements.  This was nicely
demonstrated in the Bloch oscillation experiment by \cite{PPS04}.

\item The second solution is to put each atom in a separate potential
well, for example in an optical lattice.  Having only one atom per
well vastly reduces the nonlinear interaction.  The effects of these
additional potential wells can be mitigated by using light that
energy shifts the interfering states equally.

\end{itemize}

Application of atom interferometers to atomic and molecular physics
will benefit from advances in precision and should continue to
provide definitive measurements with higher precision. A key
application will be determination of polarizabilities and stark
shifts for atoms and molecules in applied fields.  These will serve
as benchmark measurements to test and refine atomic theory
calculations as discussed in Section V.

Since atoms are very small, techniques for their manipulation on
small scales will open up many scientific frontiers and technical
possibilities in surface physics, nanophysics, and quantum
information.  The rapid pace of current developments in atom chips,
and the more creative use of focused light beams and light crystals
are both leading to techniques for producing, detecting, and
coherently manipulating atoms on very small spatial scales, e.g.
where tunneling can be carefully studied.

Small interferometers will enable novel applications in surface
science.  Atom interferometry can be used to \textit{(a)} measure
fundamental atom-surface interactions like the Van der Waals and
Casimir potentials or \textit{(b)} study the temporal and spatial
behavior of electro-magnetic fields close to the surface.  This will
allow new probes of surface structure- both magnetic and electric.
The nature of thermally induced time varying fields can be studied,
both for its own sake and because such fields induce decoherence.
This will lead to engineering advances that reduce deleterious
decoherence close to surfaces, advancing quantum information
technology that uses ions and atoms close to surfaces as q-bits.

Coherent atom optics generally, and interferometers in particular,
will be applicable to a central problem in quantum information
science: how to characterize, control and use entanglement and
correlations in atomic ensembles. The challenge here will be to
prepare the ensembles in complex quantum states with high fidelity,
and to develop methods for their characterization - with decoherence
reduced as much as possible (or with its effects reduced by
error-correction methods).  One helpful new interferometric
technique will be the development of powerful homodyne and
heterodyne methods for detecting atoms, in analogy to quantum
optics.  This will be greatly aided by the development of detection
methods with high quantum efficiency, which are also highly
desirable in studying atom-atom correlations, particularly of higher
order.

Having a good understanding of the electromagnetic atom-surface
interaction, and ways to mitigate near-surface decoherence, the
physics community will have a tool to search for fundamental
short-range interactions, as predicted in some unified theories.  In
principle atom interferometry has the potential to improve the
present limits on non-Newtonian gravitational potentials at the
micrometer length scale by many orders of magnitude \cite{DIG03}.
The main challenge here will be to control the systematic effects,
mainly coming from the electro magnetic interactions of the atom
with the close by surfaces, and the atom-atom interactions as
discussed below.  Smaller and more compact atom interferometers also
have application to inertial sensors for commercial applications.

Atom interference will be one of the central tools in the study of
many-atom systems generally and of atoms in lattices that model
condensed matter Hamiltonians in particular.  First, diffraction
peaks are the hallmark of atoms in the regime where tunneling
dominates in a periodic lattice  \cite{BHE00}. As more complex
lattices are studied, higer order interference will play a role.  In
their turn, these lattices can have regions where a particular
number of atoms are confined in each lattice site; this suggests a
way to make a source of atomic number states allowing studies of
degenerate atomic systems. Especially interesting in this arena will
be the study of phase transitions in mesoscopic ensembles, which are
too large to permit full quantum calculations, but too small for the
thermodynamic description to be valid.  This will give us a new and
detailed look at the thermodynamic border.  There are many new
avenues to explore with dense degenerate quantum gases. In the
present review we focused on single particle interference or in the
language of quantum optics: to first order coherence.  One very
fruitful avenue will be extention to multi particle interferometry,
which can give more rapidly varying fringes and sub-shot noise
statistical precision. Detecting higher order coherences requires
measurements of correlations between N atoms. Noise correlation with
bosons and fermions
\cite{MOO06,Altman2004,Polkovnikov2006,Gritsev2006,HLS07,HLF06} are
examples of recent developments in this field.

The field of atom and molecular interference is young but has
already impacted atomic and quantum physics across a broad frontier.
New techniques and the application of previously developed
techniques to new scientific problems promises much future
scientific gain.

\section*{Acknowledgements}
We thank the many research leaders who provided references and
helpful material to make this review more complete. We received a
tremendous number of contributions and could only include a
selection because of the length requirements. ADC and DEP
acknowledge support from the National Science Foundation Grants No.
PHY-0653623 and PHY-0457451 respectively.




\begin{table*}[h]
\caption{Selected books, special journal issues and review articles
germane to atom interferometers\label{tab:reviews} }
\begin{ruledtabular}
\begin{tabular}{l}
\ \ \ \ \ \ Books: \\
\hline
\underline{Atom Interferometry} \cite{BER97book} \\
\underline{Atom Optics} \cite{mey01} \\
\underline{Laser Cooling and Trapping} \cite{MES99} \\
\underline{Neutron Interferometry}  \cite{raw00} \\
\underline{Electron Interferometry} \cite{ton99, ton94} \\
\underline{Molecular Beams} \cite{ram85} \\
\underline{Atomic and Molecular Beam Methods} \cite{SCO88} \\
\underline{Atomic, Molecular, and Optical Physics Handbook}
\cite{dra96} \\
\underline{Atom and Molecular Beams, State of the Art} \cite{camp00} \\
\underline{Encyclopedia of Modern Optics} \cite{GSB04} \\
\\
\hline
\ \ \ \ \ \ Special Journal Issues \\
\hline
JOSA-B: Mechanical effects of Light (1985)\\
JOSA-B: Mechanical effects of Light (1989)\\
JOSA-B: Atom Optics (1992)\\
Applied Physics B {\bf 54} (1992)\\
JOSA-B: Atom Optics (1994)\\
Journal de Physique: Optics and Interferometry with Atoms, {\bf 4}, (11), (1994) \\
SPIE Conference on Atom Optics \cite{SPIE97}\\
Journal of Modern Optics: Quantum State Preparation and Measurement {\bf 44} (1997) \\
Comptes Rendus de L'Academie des sciences Dossier on BEC and atom lasers, t.2 serie IV, (2001) \\
General Relativity and Gravitation, {\bf 36} (10) (2004)\\
Insight Review Articles in Nature {\bf 416} (2002)\\
Applied Physics B: Quantum Mechanics for Space Application {\bf 84},
(4), (2006) \\
\\
\hline
\ \ \ \ \ \ Selected review articles: \\
\hline
Atom Optics \cite{PRI91} \\
Atom Interferometry \cite{SEC93} \\
The Feynman Path-Integral Approach to Atomic Interferometry - a Tutorial \cite{STC94} \\
Atom Optics \cite{ASM94} \\
Atom Interferometry \cite{CAM96} \\
de Broglie Optics \cite{WIL96} \\
Precision atom interferometry \cite{PYC97} \\
Matter-wave index of refraction, inertial sensing, and quantum decoherence in an at. interf. \cite{HCL97} \\
Interferometry with atoms and molecules: a tutorial \cite{PCE97} \\
Atomic interferometry \cite{BMR99} \\
Prospects for atom interferometry  \cite{GAS01} \\
Atom optics: Old ideas, current technology, and new results \cite{PCG01} \\
Miniaturizing atom optics: from wires to atom chips \cite{SCF01} \\
Coherence with atoms \cite{KAS02} \\
Microscopic atom optics: from wires to an atom chip \cite{FKS02} \\
Atom Interferometry \cite{MJB06rev} \\ \\

\hline \ \ \ \ \ \ For nonlinear atom optics see: \\

\hline
Nonlinear and quantum atom optics \cite{ROP02}\\
Nonlinear atom optics \cite{and03}\\
Physics with coherent matter waves \cite{bon04}\\
\end{tabular}
\end{ruledtabular}
\end{table*}

\include*{}  


\end{document}